\documentclass[12pt]{article}
\usepackage{graphicx} % Required for inserting images
\usepackage{mathtools}
\usepackage{setspace}
\usepackage{algorithm}
\usepackage{algorithmic}
\usepackage{float} % Better control of float positions
\usepackage{caption}
\usepackage{subcaption}
\usepackage{xcolor}
\usepackage{tikz}
\usepackage{enumitem}
\usetikzlibrary{positioning,fit,calc}
\usepackage{booktabs}
\usepackage{makecell}
\newcommand{\est}[3]{\makecell{$#1$ \\ \scriptsize(#2, #3)}}
\usepackage{amsmath, amssymb, amsthm}

\usepackage{multirow}
\usepackage[
    backend=biber,
    style=authoryear, 
    giveninits=true,    
    uniquename=false,
    uniquelist=false,
    maxcitenames=2,
    maxbibnames=99,
    natbib=true,
    url=false, 
    doi=true,
    eprint=false
]{biblatex}

\AtEveryBibitem{
  \clearfield{month}
  \clearfield{day}
}

\usepackage{booktabs}
\usepackage{tabularx}
\usepackage{caption}
\usepackage{adjustbox}
\usepackage{authblk}

\usepackage[colorlinks,citecolor=blue,urlcolor=blue,bookmarks=false,hypertexnames=true]{hyperref}
\definecolor{WowColor}{rgb}{.75,0,.75}
\definecolor{SubtleColor}{rgb}{0,0,.50}

\newcounter{margincounter}

\title{Generative Bayesian Filtering and Parameter Learning}

\author{Edoardo Marcelli\thanks{Booth School of Business, University of Chicago}\ ,  Sean O’Hagan\thanks{Department of Statistics, University of Chicago}\ ,  Veronika Ro\v{c}kov\'{a}\protect\footnotemark[1]}

\date{\today}

\begin{document}

\def\spacingset#1{\renewcommand{\baselinestretch}%
{#1}\small\normalsize} \spacingset{1.1}

\maketitle

\begin{abstract}
    \noindent Generative Bayesian Filtering (GBF) provides a powerful and flexible framework for performing posterior inference in complex nonlinear and non-Gaussian state-space models. Our approach extends Generative Bayesian Computation (GBC) to dynamic settings, enabling recursive posterior inference using simulation-based  methods powered by deep neural networks.   
    GBF does not require explicit density evaluations, making it particularly effective when observation or transition distributions are analytically intractable. To address parameter learning, we introduce the Generative-Gibbs sampler, which bypasses explicit density evaluation by iteratively sampling each variable from its \emph{implicit} full conditional distribution. Such technique is broadly applicable and enables inference in 
    hierarchical Bayesian models with intractable densities, including state-space models.
     We assess the performance of the proposed methodologies through both simulated and empirical studies, including the estimation of $\alpha$-stable stochastic volatility models. Our findings indicate that  GBF 
    significantly outperforms existing likelihood-free approaches in accuracy and robustness when dealing with intractable state-space models. 
\end{abstract}

\section{Introduction}

State-space models are a cornerstone of time series analysis in macroeconomics and finance, and are widely used across the physical and social sciences wherever latent dynamic processes must be inferred from noisy or incomplete data. 

Formally, for $t\in \mathbb{N}$, let $ (Y_t)_{t \geq 1}$, $Y_t\in\mathcal{Y}$, denote a sequence of observable outputs and $ (X_t)_{t \geq 0} $, $X_t\in\mathcal{X}$, a sequence of latent (unobservable) states, where $\mathcal{Y}$ and $\mathcal{X}$ are measurable spaces.  
These hidden states may represent, for example, the volatility underlying asset returns, the economy’s potential output or natural rate of unemployment, the true position and velocity of a satellite inferred from radar data, or the migration path of an animal reconstructed from noisy GPS signals. The dynamics of such systems are typically described by the following equations:
\begin{align}
Y_t &= f_t^{\theta}(X_t, \varepsilon_t) \label{eq:obseq}\\
X_t &= g_t^{\theta}(X_{t-1}, \eta_t)\label{eq:stateeq}
\end{align}
Here, $ g_t^{\theta}: \mathcal{X} \times \mathcal{V} \to \mathcal{X} $ denotes the state transition function and $ f_t^{\theta}: \mathcal{X} \times \mathcal{E} \to \mathcal{Y} $ denotes the observation function, both parametrized by $\theta\in\Theta$. The noise sequences $ (\eta_t)_{t\geq 1} $ and $ (\varepsilon_t)_{\geq1} $  follow probability distributions $ p_\eta $ and $ p_\varepsilon $, respectively, and are usually assumed to be independent, although dependence is also possible. As for the initial state $ X_0 $, it is drawn from a prior distribution $ p_\theta(x_0) $. This general formulation encompasses both linear and nonlinear models, and allows for non-Gaussian stochastic dynamics.

The primary goal of inference in state-space models is  to recursively infer the latent signal $ (X_t)_{t\geq 0} $ from the observable data $ (Y_t)_{t\geq 1} $, typically by characterizing the posterior distribution of the current state given the data, a task known as \emph{filtering}. Related problems include \emph{smoothing}, which seeks to estimate past states using both past and future observations, and \emph{prediction}, which involves forecasting future states or observations based on current and historical data.  Another fundamental aspect of state-space modeling is then \emph{parameter learning}, which involves estimating the unknown parameters that govern the system’s dynamics. 

 Except for a very limited number of cases where closed-form solutions are available, state-space inference and learning -- whether pursued jointly or separately -- have traditionally relied on Markov Chain Monte Carlo (MCMC) methods (see, e.g., \cite{CK_1994,F_1994,kim1998stochastic}) or Sequential Monte Carlo (SMC)
 algorithms \citep{doucet2001smc}, including the Bootstrap Particle Filter \citep{Gordon_1993} and its variants. However, these standard techniques are no longer applicable when the state-space model is \emph{intractable}. This occurs when the nature of the stochastic shocks $\varepsilon_t$ and $\eta_t$, or the functional forms of $f_t^\theta$ and $g_t^\theta$ in equations \eqref{eq:obseq} and \eqref{eq:stateeq}, give rise to transition or emission distributions -- i.e. the distributions of $X_t \mid X_{t-1}$ and $Y_t \mid X_t$ -- that do not admit a density with respect to a fixed dominating measure. 
 
Intractable models are encountered across various domains. For instance, in finance, $\alpha$-stable distributions are frequently used to capture asymmetric heavy tails exhibited by asset returns \citep{Mandelbrot_1963,mittnik1993modeling}, yet these distributions lack closed-form density expressions. In macroeconomics, nonlinear DSGE models \citep{fernandez2016solution} rely on numerically solved equilibrium conditions, resulting in transition dynamics that are only implicitly defined and therefore analytically intractable. Similar challenges appear in biology, where mechanistic models such as the Lotka-Volterra predator-prey system \citep{lotka1925elements,volterra1926variazioni} lead to state-space models with intractable transition kernels.  More generally, intractability arises whenever the observation or transition components of a state-space model are defined through a numerical black-box model.

 While the absence of tractable densities precludes standard likelihood-based inference, many intractable state-space models still allow efficient simulation from the generative process. This important feature has motivated the adoption of Approximate Bayesian Computation (ABC) methods in such contexts. In particular,  \citet{Jasra2012} introduce the ABC Particle Filter (ABC-PF) for state inference in state-space models with intractable likelihood, while \citet{jasra2013aliveparticlefilter} propose a Particle MCMC method that employs the ABC-PF to construct an estimator of the likelihood for conducting joint inference on the states and the model's parameters.
 
Although these methods enjoy desirable asymptotic convergence properties -- as we discuss in detail later -- their performance in finite samples remains less well understood. In particular, the accuracy and reliability of the ABC-PF are highly sensitive to several implementation choices. These include the definition of the distance metric used to compare simulated and observed data, the selection of the tolerance threshold that governs acceptance, and the number of particles employed. 
Increasing the number of particles and decreasing the tolerance improves the posterior approximation. However, this combination is often computationally infeasible in practice, and suboptimal tuning can result in biased estimates and highly variable posterior approximations, raising concerns about robustness in empirical applications. Furthermore, these issues are exacerbated by well-documented problems affecting PFs, notably weight degeneracy and sample impoverishment (see, for example, \cite{LI_2014}). 

To address these concerns, several variants of the base ABC-PF have been proposed (as reviewed in Section \ref{sec:ABC-PF}), each aiming to improve either theoretical guarantees or practical performance. Nonetheless, these tools 
remain fundamentally constrained by the structural limitations of both the ABC and SMC approaches.

 Recently, Generative Bayesian Computation (GBC) 
 has emerged as a powerful tool for performing posterior inference in settings where traditional likelihood-based techniques are not applicable. 
In contrast to classical Bayesian methods, GBC employs deep generative models to approximate posterior distributions through simulation rather than explicit likelihood evaluation. By training a deep neural network on a large grid of simulated data generated, GBC learns a mapping from the observed data $y$ to a parameter of interest $\theta$ via a function $F^{-1}_{\theta\mid y}$ that approximates the inverse cumulative distribution function (CDF).
Once trained, posterior samples can be efficiently generated by evaluating this learned inverse Bayes map at the observed data and a random draw from the uniform distribution on the unit interval $u\sim \mathcal{U}(0,1)$. Analogous to classical inverse transform sampling, this procedure yields samples via $\theta \overset{d}{=}F^{-1}_{\theta\mid y}(u)$. Important contributions in this direction include the work of \citet{wang2023adversarialbayesiansimulation}, who learn posterior samplers via generative adversarial networks (GANs); \citet{polson2023generative}, employing conditional quantile learning methods; and \citet{kim2025deepgenerativequantilebayes}, extending the analysis to multivariate settings.

When transitioning to the state-space context, however, GBC presents two structural limitations. First, it is inherently static, being designed for fixed datasets rather than sequentially evolving observations. Second, it assumes a direct link between data and parameters. Both assumptions are  incompatible with state-space modeling, where inferential tasks such as filtering require recursive updating over latent states $X_t$, and where unknown parameters typically influence the observations only indirectly through these latent processes.
Therefore, in this paper, we address both limitations by extending GBC to support recursive inference and to accommodate hierarchical Bayesian dependencies.

\subsection{Contribution and Structure}

We introduce a novel framework for state-space inference and learning that applies to all models expressible in the form presented in equations \eqref{eq:obseq} and \eqref{eq:stateeq}, regardless of  the noise distributions or the functional forms of the transition and observation functions,  provided that simulation from the model is feasible. As such, it also encompasses models with intractable densities. 

Our approach is grounded in GBC and extends the methodology to a dynamic context, where the structure of the problem requires recursive updates of the posterior distribution over the latent state sequence $(X_t)$.  The ultimate goal is to reconstruct key distributions such as the filtering, predictive, and smoothing distributions. 
We consider both the case where the parameter vector is known and the case where $\theta$ is unknown and has to be inferred from the data, starting from a prior distribution $p(\theta)$.

 Our Generative Filter, or more compactly Gen-Filter, is intended as a promising alternative to the existing ABC-PF methods. Both approaches require only the capability to simulate from the data generating process. Yet, in contrast to ABC-PFs, which provide samples from an approximate and inherently biased filtering distribution due to the use of an acceptance threshold, the Gen-Filter allows to sample from the true filtering distribution. This holds under the condition that the training dataset is sufficiently large and the neural network used to approximate the inverse CDF mapping has enough expressive capacity.  
Designing an effective neural network architecture remains a central challenge in our approach. Consistent with the work of \citet{polson2023generative}, we adopt Quantile Neural Networks (QNNs) \citep{dabney2018implicitquantilenetworksdistributional} as our baseline method. We also explore other potential approaches, including the Bayesian alternative proposed by \citet{ohagan2025generativeregressioniqbart}.

Our results indicate that standard deep learning architectures are capable of delivering accurate and reliable performance, provided that the training dataset is sufficiently large. This is typically not a limitation, as generating data from the model is computationally efficient and inexpensive in most cases.

While our Gen-Filter can be naturally employed to construct an estimator of the likelihood, thereby allowing joint inference of the latent trajectory and the unknown parameters in a similar manner of PMCMC algorithms, we also develop an innovative sampling strategy that delivers substantially greater computational efficiency together with remarkable flexibility, which we denote as \emph{Generative Gibbs} sampler. 

With the Generative Gibbs (Gen-Gibbs) sampler, we extend the GBC approach to hierarchical Bayesian modeling, i.e. to settings characterized by multiple levels of latent structure and intricate parameter dependencies.

As in traditional Gibbs sampling, the Gen-Gibbs methodology generates posterior samples by iteratively drawing from the full conditional distributions of the model's parameters. Crucially, and in contrast to the classical approach, all full conditionals are approximated using implicit generative models, thus enabling Gibbs sampling in scenarios where the full conditionals would otherwise be analytically intractable. This renders the Gen-Gibbs sampler broadly applicable and particularly advantageous for intractable state-space models. 
We show that the Gen-Gibbs attains posterior approximations that are consistent with those obtained from traditional MCMC methods. 

The structure of the paper is the following: Section \ref{sec:Background}  review recent advances in state-space modeling using ABC methods, and GBC. In Section \ref{sec:AI-Filter}, we introduce the idea of Generative Bayesian Filtering (GBF) and we present two algorithms: the Gen-Filter and the Pre-Trained Gen-Filter. Their performance is evaluated through simulation studies in Section \ref{sec:sim suty known par}, where we also compare them to established filtering techniques. In Section \ref{sec:Parameter Learning}, we address the setting in which model parameters are unknown and must be inferred jointly with the latent trajectory. To this end, we present the Gen-Gibbs sampler and shows how it can be effectively applied to general state-space models in the form of a  Forward Filtering Backward Sampling (FFBS) strategy. Simulation results for the Gen-Gibbs sampler are reported in Section \ref{sec:sim suty unknown par}. Finally, in Section \ref{sec:Empirical Study}, we present an empirical application using  financial data. A concluding section follows, where we summarize the main findings and discuss potential avenues for future research.

\section{Background}\label{sec:Background}

\subsection{Sequential Inference in State-Space Models}\label{sec:ssm_inference}

Let us consider again the setting introduced in the first section, where the latent process $(X_t)$ evolves as a Markov chain and, conditionally on $(X_t)$, the observations $(Y_t)$ are independent. For illustrative purposes, we henceforth consider the setting in which $X_t$ and $Y_t$ are continuous real-valued random variables, and denote their realizations by $x_t$ and $y_t$. The methods discussed, however, extend naturally to more general settings, including multivariate and non-Euclidean spaces.

Moreover, let us assume for the moment that the parameter vector $\theta$ is known and, to simplify the exposition, omit it from the notation. A detailed discussion of the case where $\theta$ is unknown is deferred to Section \ref{sec:Parameter Learning}.

 The system of equations \eqref{eq:obseq} and \eqref{eq:stateeq} induces a probabilistic model where the state equation defines the transition distribution  $p(x_t\mid x_{t-1})$, and the observation equation specifies the emission distribution $p(y_t\mid x_t)$, which we will also refer to as the likelihood. These two components are central to computing the posterior distribution over the latent states via a sequential updating procedure known as filtering. At each time step $t \in \mathbb{N}$, filtering proceeds in three steps.
Firstly, the one-step-ahead predictive distribution of the latent state is computed by propagating the previous filtering distribution through the transition model:
\begin{equation}\label{eq:predicting}
 p(x_t \mid y_{1:t-1}) = \int p(x_t \mid x_{t-1}) \, p(x_{t-1} \mid y_{1:t-1}) \, dx_{t-1}.   
\end{equation}
Subsequently, the \emph{predictive distribution} of the next observation, also known as the one-step-ahead \emph{forecast density}, is computed as
\begin{equation}\label{eq:forecasting}
p(y_t \mid y_{1:t-1}) = \int p(y_t \mid x_t) \, p(x_t \mid y_{1:t-1}) \, dx_t,
\end{equation}
and, upon receiving the new observation $y_t$, the latent state distribution is  then updated using Bayes’ rule, yielding the \emph{filtering distribution} at time $t$:
\begin{equation}\label{eq:filtering}
p(x_t \mid y_{1:t}) = \frac{p(y_t \mid x_t) \, p(x_t \mid y_{1:t-1})}{p(y_t \mid y_{1:t-1})}.
\end{equation}
The key idea in this recursive procedure is that, at each time $t$, the predictive distribution serves as a prior over future states and observations, which is then refined as new data becomes available through the filtering update. 

Exact and efficient inference in state-space models is feasible only in a limited number of cases such as linear Gaussian models, for which the optimal filtering solution is given by the celebrated Kalman Filter \citep{Kalman_1960}.  However, in practice, many systems exhibit nonlinear dynamics and/or non-Gaussian noise, making exact inference impossible.  As a result, various approximate inference methods have been developed. Among deterministic approaches are the Extended Kalman Filter \citep{maybeck1979stochastic} and the Unscented Kalman Filter \citep{julier1997new}, which attempt to adapt the state-space model to the assumptions of the Kalman Filter by linearizing the dynamics or approximating distributions. While these methods can be effective in some settings, their accuracy deteriorates in the presence of strong nonlinearities or non-Gaussian noise. On the other hand, SMC algorithms have gained prominence in the field, where they are known as PFs. These methods offer a flexible, simulation-based framework for approximating complex posterior distributions that allows to handle general classes of nonlinear and non-Gaussian state-space models.

\subsection{Particle Filters and ABC}\label{sec:ABC-PF}

PFs refer to a class of SMC algorithms that approximate the filtering distribution $p(x_t\mid y_{1:t})$ using a finite collection of particles $\{x_t^{(i)}\}_{i=1}^{N}$ drawn from a proposal distribution $q(\cdot)$. Such algorithms exist in many variants and have been extensively reviewed by \citet{chopin2020introduction}. Since the particles are not sampled directly from the target distrution, importance weights must be assigned and updated sequentially over time to correct for the discrepancy between $q(\cdot)$ and the true posterior. 

Formally, the weight update for particle $i$ at time $t$
is given by the Radon-Nikodym derivative of the target measure $p(x_{0:t} \mid y_{1:t})$ with respect to the proposal $q_t(x_{0:t} \mid y_{1:t})$, i.e.
\[
w_t^{(i)} \propto \frac{p(x_{0:t}^{(i)} \mid y_{1:t})}{q_t(x_{0:t}^{(i)} \mid y_{1:t})}.
\]
In practice, most particle filters assume that the proposal distribution factorizes as
\[
q_t(x_{0:t} \mid y_{1:t}) = q_{t\mid t-1}(x_{0:t-1} \mid y_{1:t-1}) \, q_t(x_t \mid x_{0:t-1}, y_{1:t}),
\]
which allows for a recursive formulation of the weights:
\[
w_t^{(i)} \propto w_{t-1}^{(i)} \cdot \frac{p(y_t \mid x_t^{(i)}) \, p(x_t^{(i)} \mid x_{t-1}^{(i)})}{q_t(x_t^{(i)} \mid x_{0:t-1}^{(i)}, y_{1:t})}.
\]
In the special case where the proposal is chosen as the state transition kernel, i.e. $q_t(x_t \mid x_{t-1}, y_{1:t}) = p(x_t \mid x_{t-1})$, 
the algorithm reduces to the  Bootstrap PF \citep{Gordon_1993}, and the weight update simplifies to
\[
w_t^{(i)} \propto w_{t-1}^{(i)} p(y_t \mid x_t^{(i)}).
\]
A common requirement of PFs is thus that the likelihood $p(y_t\mid x_t)$ must be available analytically, at least up to a normalizing constant -- a condition that may be violated in practice. Nevertheless, despite direct evaluation of the likelihood may be mathematically intractable or computationally prohibitive, simulating from the model is often easy to implement and requires relatively little computational effort. Building on this insights, alternative filtering strategies have been developed, which exploit the generative structure of the model rather than relying on explicit likelihood calculations. 

 One such approach is the Convolution PF \citep{rossi2006nonlinear, rossi2009convolution}, which replaces the intractable likelihood with a kernel-based approximation constructed from pseudo-observations generated conditionally on the latent states. However, this solution can be particularly inefficient, especially in high dimensions, and is sensitive to the choice of the kernel bandwidth, often performing poorly when the bandwidth is not appropriately tuned.
 
 In this context, ABC-PFs \citep{Jasra2012} emerges as a more efficient alternative. These methods, 
also known as Likelihood-Free PFs \citep{LikelihoodfreePF},
update particle weights by evaluating the similarity between simulated and observed data through a distance metric $d(\tilde{y}_t^{(i)},y_t)$, where $y$ denotes the observed data and $\tilde{y}^{(i)}$ is a draw from the emission distribution conditional on the propagated particle $x_t^{(i)}$. The computed distance is then used as the argument of a kernel function $K_\epsilon(\cdot)$, where $\epsilon$ represents the tolerance parameter, and the weights are updated according to
\[
w_t^{(i)} \propto w_{t-1}^{(i)}K_\epsilon\left(d(\tilde{y}_t^{(i)},y_t)\right).
\]
As a specific choice, \citet{Jasra2012} propose to use a uniform kernel $K_\epsilon(d)=\mathbb{I}\{d < \epsilon\}$
together with either the $L_1$ or $L_2$ norm for the distance metric. They further demonstrate that, for fixed $\epsilon$, the ABC-PF converges to a biased posterior distribution as the number of $N\to\infty$. Moreover, the magnitude of the bias becomes negligible as $\epsilon\to0$. In practice, however, the choice of $\epsilon$ and the number of particles $N$ is constrained by computational considerations: decreasing $\epsilon$ 
reduces the bias but leads to a significantly lower acceptance rate, which in turn requires a larger number of particles to maintain particle diversity and reduce variance. However, increasing $N$ raises computational cost. In extreme scenarios where none of the simulated pseudo-observations fall sufficiently close to the observed data, all particle weights are set to zero. When this occurs, the algorithm fails to proceed and, ultimately, the filtering recursion collapses. 

In general, a poor choice of the kernel, distance metric, or tolerance parameter tends to amplify some well-documented challenges inherent to particle filtering methods, in particular weight degeneracy and sample impoverishment. To mitigate this issue, practitioners often rely on adaptive thresholding, smoother kernels, or informative low-dimensional summary statistics to maintain a nonzero acceptance rate and preserve the continuity of the inference procedure. Some relevant examples include the Alive ABC-PF \citep{jasra2013aliveparticlefilter}, which mitigates particle degeneracy by ensuring a fixed number of accepted particles at each time step; the plug-in bandwidth ABC-PF \citep{Calvet_2014}, which is demonstrated to achieve convergence at the optimal decay rate; and the ABC-Auxiliary PF \citep{Vankov_2019}, which improves efficiency by refining the proposal distribution used in the particle filtering step. 

Similarly to the ABC-PF, our Generative Filter avoids direct density evaluations by leveraging the model's underlying data-generating process. However, rather than relying on the acceptance-rejection mechanism typical of ABC methods, it harnesses recent advances in generative modeling to efficiently sample from the filtering distribution $p(x_t \mid y_{1:t})$.  We detail the methodology behind our approach in the following section.

\subsection{Generative Bayesian Computation}\label{sec:GBC}

Generative approaches to Bayesian computation typically rely on \emph{implicit distributions}. These are distributions whose density functions cannot be evaluated directly, yet from which we can readily draw samples through a stochastic generator -- also known as \emph{transport map} -- that converts samples from a reference measure (e.g. a multivariate Gaussian or uniform) into samples from the target probability measure. In modern implementations, the transport map is usually parametrized by a deep neural network \citep{mohamed2016learning}.

This sampling technique proves especially valuable  for addressing key limitations of traditional Bayesian computational methods, particularly the reliance on explicit density evaluations and the substantial computational burden associated with iterative simulation algorithms such as Markov Chain Monte Carlo (MCMC). 
For example, \citet{titsias2019unbiased} employ implicit variational distributions to expand the family of admissible variational approximations, thereby enabling more flexible and expressive posterior representations that go beyond standard parametric forms.

In this article, with GBC we refer more precisely to those approaches that model the posterior itself as an implicit distribution, using a transport map to directly generate samples from the corresponding posterior probability measure. A growing number of studies have recently appeared in the literature, exploring diverse neural architecture for parameterizing the transport map. 
\citet{wang2023adversarialbayesiansimulation} use a conditional Bayesian Generative Adversarial Network (B-GAN) to learn a generative model for the posterior distribution given any observed data vector. \citet{polson2023generative} leverage Implicit Quantile Networks \citep{dabney2018implicitquantilenetworksdistributional} to model the conditional quantile function of a univariate parameter given data, enabling direct posterior sampling, while \citet{kim2025deepgenerativequantilebayes} generalize this idea to multivariate settings, allowing direct sampling from Bayesian credible sets. In a parallel line of research, \citet{sharrock2024sequential} adopt conditional score based diffusion models for posterior sampling.

In developing our GBF framework, we mainly rely on the approach introduced by \citet{polson2023generative}, which we briefly describe. Let $\theta$ denote the parameter of interest and $y$ a generic vector of data.
Based on inverse transform sampling, posterior draws 
$\{\theta^{(i)}\}_{i=1}^{N}$ are obtained by sampling 
$u^{(i)} \sim \mathcal{U}[0,1]$ and setting 
$\theta^{(i)} = F^{-1}_{\theta \mid y}(u^{(i)})$, 
so that $\theta^{(i)} \sim p(\theta \mid y)$. In practice, the exact form of this inverse mapping is generally unknown. Nevertheless, 
we can approximate $F^{-1}_{\theta\mid y}$ with a learnable function  $H:\mathcal{Y}\times [0,1]\to \Theta$,  parameterized by a deep neural network.

The training procedure relies on a large synthetic training dataset composed of parameter–data–base triplets $\{\theta^{(i)},\tilde{y}^{(i)},u^{(i)}\}_{i=1}^{N}$, where $\tilde{y}$ denotes synthetic observations, distinguishing them from the true observations $y$. This dataset is generated by simulating from the model's DGP. Specifically, $\theta$ is drawn from the prior $p(\theta)$ and $\tilde{y}$ from the probabilistic model $p(y\mid \theta)$, while $\{u^{(i)}\}_{i=1}^{N}$ are iid samples from a uniform distribution. The conditional model $p(y\mid \theta)$ represents a forward mapping from parameters to data; therefore, even when an explicit likelihood function is intractable, Bayesian inference via GBC remains feasible. Thus, GBC 
can be regarded as a likelihood-free inference method. 

The function $H$ can be estimated via quantile regression on the simulated data, satisfying the relation
\[
\theta^{(i)} = H(\tilde{y}^{(i)},u^{(i)}), \quad i=1,\dots,N.
\]
This task can be handled by QNNs. In contrast to fixed-quantile estimation methods, QNNs learn a continuous mapping from the quantile level $u\in[0,1]$ and the conditioning variable $\tilde{y}$ to the corresponding quantile value of $\theta$. 

Let $\rho_{u}(z) := u z \mathbb{I}\{z > 0\} - (1 - u)z \mathbb{I}\{z \le 0\}
$ denote the pinball (quantile) loss function and $\mu$ a reference measure, the training objective of QNNs is
\[
H^*\in \underset{H:\mathcal{Y}\times [0,1] \to \mathbb{R}}{\arg\min}
\mathbb{E}_{u \sim \mu}
\Big[\, \mathbb{E}_{\theta\mid y}\big[\rho_{u}(\theta - H(y, u))\big] \,\Big]
\]
where $H^*$ corresponds to the true conditional quantile function.
In practice, this expectation is approximated empirically using the simulated dataset, yielding the optimization problem
\[
\hat{H} \in 
\underset{H \in \mathcal{H}}{\arg\min}
\frac{1}{N}
\sum_{i=1}^{N}
\rho_{u^{(i)}}\!\left(
\theta^{(i)} - H(\tilde{y}^{(i)}, u^{(i)})
\right),
\]
with $\mathcal{H}$ being the hypothesis class defined by the neural network architecture. The parameters of $H$ are optimized using stochastic gradient descent. 

To enhance the model’s ability to capture smooth and nonlinear dependencies across quantile levels, the scalar input $u$ is first projected into a higher-dimensional embedding $\phi(u)$ using a cosine transformation. Similarly, $y$ is encoded into a latent representation $\psi(y)$ of the same dimension of $\phi(u)$. Thus $H$ is expressed as $H(y,u)=h(\psi(y)\circ \phi(u)) $ where $h$ and $\psi$ are feed-forward neural networks, and $\circ$ denotes the elementwise product.

Given the trained model, posterior samples are obtained by evaluating the learned inverse map at the observed data and a uniform random draw as $\theta \overset{d}{\approx} \hat{H}(y,u)$.

In case $y$ is high dimensional, it is possible to compress it into a low-dimensional vector of summary statistics $s=(s_1,\ldots,s_k)$, analogous to the approach used in ABC. 
When the summary statistics are sufficient in Bayesian sense, i.e. $p(\theta\mid y)=p(\theta\mid s)$, then  inference based on $s$ is equivalent to using the entire data vector. Otherwise, $p(\theta\mid s)$ represents a \emph{partial} posterior that may not capture all the information in $y$, but still retains important information about
$\theta$.

As noted by \citet{polson2023generative}, the performance of this approach depends critically on the explicit specification of the neural network architecture, including the choice of layers, activation functions, and optimization routines. QNNs represent only one possible choice among several architectures that can be employed to learn implicit distributions. Other implicit quantile methods, such as distributional regression networks, normalizing flows, or diffusion-based models can likewise be employed for this purpose.

\section{Generative Bayesian Filtering}\label{sec:AI-Filter}

In the context of state-space models, even when the transition or emission distributions are analytically intractable, simulating trajectories of the latent process $(X_t)$ and the observation sequence $(Y_t)$ of a state space model is generally  straightforward. By sampling the initial state from $p(x_0)$ and the noise terms from $p_\varepsilon$ and $p_\eta$, one can recursively obtain a sequence of state-observation pairs $(x_t,y_t)$ via the system of equations \eqref{eq:obseq} and \eqref{eq:stateeq}.

Repeating this procedure $N$ times for sequences of length $T$ produces a collection of time series samples $\{x_{0:T}^{(i)}, \tilde{y}_{1:T}^{(i)}\}_{i=1}^N$, where again we use the notation $\tilde{y}$ to denote synthetic observations, distinguishing them from the actual data $y$. Consistent with the GBC approach, the dataset obtained in this way is then augmented by incorporating $\{u^{(i)}\}_{i=1}^N$, consisting of i.i.d. draws from a base distribution, typically the uniform distribution on the unit interval. This results in an augmented synthetic dataset $\{x_{0:T}^{(i)}, \tilde{y}_{1:T}^{(i)}, u^{(i)}\}_{i=1}^N$ which can be used to train deep learning models and approximate key distributions, such as filtering, smoothing and predictive distributions. 
 For example, by focusing on the subset $\{x_{t}^{(i)}, \tilde{y}_{1:t}^{(i)}, u^{(i)}\}_{i=1}^N$ for a given $t\leq T$, a neural network can be trained to learn the mapping $H_t:\mathcal{Y}^t\times [0,1]\to \mathcal{X}$ approximating the inverse CDF $F_{x_{t}\mid y_{1:t}}^{-1}$. Therefore, by plugging a new base draw $u\sim \mathcal{U}(0,1)$ into $H_t(y_{1:t},u)$, we obtain a sample from $p(x_{t}\mid y_{1:t})$, the filtering distribution. Similarly, we can generate samples from the predictive distribution $p(x_{t+1}\mid y_{1:t})$ and the smoothing distributions $p(x_j\mid y_{1:t}),$ $0 \leq j\leq t-1$. 
 
 While the idea illustrated so far appears conceptually simple, two main challenges arise. Firstly, the series we aim to filter may be very long, with $T$ potentially large or even infinite, requiring us to train a huge collection of models, each tailored to a specific mapping $H_t(y_{1:t},\cdot)$ for $t=1,\dots,T$. One might argue that, although the upfront computational cost of learning many inverse Bayes maps could be substantial, it is in fact amortized over time: once all the maps are estimated, filtering proceeds efficiently by simply evaluating the models $\hat{H}_t(\cdot)$ for $ t=1,\dots,T$ using the observed data matrix and a newly generated base draw.  However, the second and most serious challenge is that, as $T$ increases, the number of predictors $y_{1:T}$ grows too, making the learning task increasingly prone to overfitting and ultimately degrading the performance of the methodology.
 
 We propose two valid strategies addressing this problem, namely the Generative Filter (Gen-Filter) and the Pre-Trained Gen-Filter. The first approach sequentially updates the inverse Bayes map used to sample from the filtering distribution by training a new model at each time step. This model incorporates both the information propagated from the filtering distribution $p(x_{t-1}\mid y_{1:t-1})$ fitted at the previous step and the newly observed data point $y_t$. In contrast, the second strategy uses a summary function to compress the information contained in the full observation vector $y_{1:t}$ into a low-dimensional summary statistic. The choice of this summary function will be discussed later; for now, consider a general $S:\mathcal{Y}^t\to \mathcal{S}^K$.  
 A model $H:\mathcal{S}^K\times [0,1]\to \mathcal{X}$ is then trained only one time ex-ante on a large number of simulated scenarios. Once trained, this model can generate samples from the \emph{partial} posterior $p(x_t \mid S(y_{1:t}))$, which serves as a surrogate for the filtering distribution $p(x_t\mid y_{1:t})$. Since in this second strategy the deep learning model is trained ex-ante and not sequentially, we refer to this method as the Pre-Trained Gen-Filter. Importantly, this approach is only applicable under the assumptions that the underlying process is stationary and the observation model is time-homogeneous. We provide further details on both methodologies in the sections that follow.

\subsection{Generative Filter}

We previously discussed a possible approach for inferring the distribution of $X_t$ given the entire sequence $y_{1:t}$ which involves learning the inverse CDF $F_{x_t\mid y_{1:t}}^{-1}$ using the samples drawn from a prior
$p(x_t) = \int p(x_0)\prod_{j=1}^{t}p(x_j\mid x_{j-1})\,dx_{0:t-1}$ 
and a joint probabilistic model $p(y_{1:t}\mid x_{0:t})=\prod_{j=1}^{t}p(y_j\mid x_j)$. However, we also highlighted that this method suffers from an evident drawback: the curse of dimensionality, which becomes especially severe as $t$ grows. 

A more natural and efficient alternative exploits the recursive structure inherent to the filtering problem, as outlined in Section \ref{sec:ssm_inference}. Specifically, the Gen-Filter reformulates the original learning task into a sequence of local updates, wherein at each time step the inverse CDF  $F_{x_t\mid y_{1:t}}^{-1}$ is learned using the samples drawn from the predictive distribution $p(x_{t}\mid y_{1:t-1})$ and the emission distribution $p(y_t\mid x_t)$. This approach capitalizes on the fact that the predictive distribution propagated from $t-1$ naturally serves as the prior for the Bayesian update once the new observation $y_t$ becomes available.
In other words, the Gen-Filter operates like a traditional filtering method by encapsulating all the relevant information from the process history into the prior distribution $p(x_t\mid y_{1:t-1})$.

In practice, at time $t-1$, a deep learning model $H_{t-1\mid t}$ is trained on a synthetic dataset $\{x_t^{(i)}, \tilde{y}_{t}^{(i)}, u^{(i)}\}_{i=1}^N$, where each $x_t^{(i)}$ is generated from $p(x_{t}\mid y_{1:t-1})$, $\tilde{y}^{(i)}_t\sim p(y_t\mid x_t^{(i)})$ and $u^{(i)}\sim \mathcal{U}(0,1)$. The notation $H_{t-1 \mid t}$ emphasizes that this map is learned at time $t-1$ and is intended to receive $y_t$ as input once observed. Hence, by evaluating $\hat{H}_{t-1\mid t}$ at $y_t$ and a fresh draw from the base distribution, we obtain samples from $p(x_t\mid y_{1:t})$, since $x_t \overset{d}{=} H_{t-1\mid t}(y_t,u)$ and $H_{t-1\mid t}$ approximates the inverse CDF of the filtering distribution. 

Therefore, the Gen-Filter breaks down the sequential inference problem into a sequence of $t$ static updates, each of which can be carried out using GBC as described in Section \ref{sec:GBC}. The only requirement for this method is the ability to sample
 from the transition distribution $p(x_t \mid x_{t-1})$ and the emission distribution $p( y_t \mid x_t) $ via equations \eqref{eq:stateeq} and \eqref{eq:obseq}; there is no need to evaluate these densities explicitly. Therefore, a key strength of the Gen-Filter lies in its minimal assumptions about the DGP, an aspect that we examine further in the following section. 
 However, its computational efficiency heavily depends on the neural network architecture used to approximate the quantile function. This creates a trade-off between accuracy and speed, which may limit the filter’s suitability in scenarios requiring fast response times. To address this problem, we develop an alternative version, namely the Pre-Trained Gen-Filter, which offers significant computational advantages. A full description of the Gen-Filter is presented in Algorithm \ref{alg:Gen-Filter}.
\begin{algorithm}
\caption{Generative Filter}\label{alg:Gen-Filter}
\begin{algorithmic}[1]
\STATE \textbf{Input:} Initial distribution $p(x_{0})$, transition model $p(x_t \mid x_{t-1})$, observation model $p(y_t \mid x_t)$, observed data $y_{1:T}$
\STATE \textbf{Output:} Samples from $p(x_t\mid y_{1:t})$ for $t=1,\dots,T$
\STATE Sample $\{x_{0}^{(i)}\}_{i=1}^{N} \overset{iid}{\sim} p(x_{0})$
\FOR{$t = 1$ to $T$}
    \STATE Sample $x_t \sim p(x_t \mid x_{t-1})$ 
    \STATE Sample $\tilde{y}_t \sim p(y_t \mid x_t)$ 
    \STATE Sample $u_t\sim \mathcal{U}(0,1)$
\STATE Learn the inverse CDF mapping $x_t \overset{d}{=} H_{t-1\mid t}(y_t,u_t)$ using the dataset $\{x_t^{(i)}, \tilde{y}_t^{(i)},u_t^{(i)}\}_{i=1}^{N}$

\STATE Sample new $\{u^{(i)}\}_{i=1}^{N} \overset{\text{iid}}{\sim} \mathcal{U}(0,1)$

    \STATE Get $x_t = \hat{H}_{t-1\mid t}(y_t,u)$
\ENDFOR
\STATE Return $\{x_t^{(i)}\}_{i=1}^{N}$ as samples from $p(x_t\mid y_{1:t})$ for $t=1,\dots,T$
\end{algorithmic}
\end{algorithm}

\subsection{Pre-Trained Generative Filter}

The major strength of the Gen-Filter lies in its versatility, i.e. its ability to handle a wide range of state-space models without requiring strict assumptions about the underlying stochastic processes, particularly stationarity. Nonetheless, when stationarity \emph{is} present, we can take advantage of this property.

The Pre-Trained Gen-Filter exploits the fact that when the latent process in \eqref{eq:stateeq} is stationary and the observation model in \eqref{eq:obseq} is time-homogeneous, the processes $\{x_{1:T}^{(i)}, \tilde{y}_{1:T}^{(i)}, u_{1:T}^{(i)}\}_{i=1}^{N} $ 
generated in the training phase are representative of those observed, as both stem from the same joint stationary distribution. These properties also ensure that the predictive and filtering distribution remain well-behaved over time, neither drifting nor collapsing, and they eventually stabilize. As a result, a general updating rule can be trained once, prior to observing any data, and then reused at each step of the filtering recursion.

Therefore, under the aforementioned assumptions, it is, in principle, possible to learn ex-ante $T$ mappings $\{H_{0\mid t}\}_{t=1}^{T}$, $H_{0\mid t}:\mathcal{Y}^t\times [0,1]\to \mathcal{X}$, each approximating the quantile function $F^{-1}_{x_t\mid y_{1:t}}(u)$, $u\in(0,1)$, which allows the generation of samples from the filtering distributions $p(x_t\mid y_{1:t})$ for any $t\in\{1,\dots,T\}$ by simply feeding the data $y_{1:t}$ in the estimated function $\hat{H}_{0\mid t}$ as soon as they arrive.  However, such strategy is computationally infeasible when $T$ becomes large and potentially enormous. 

A better approach, inspired by ABC, is to replace the full observation vector $y_{1:t}\in\mathcal{Y}^t$ with a lower-dimensional set of summary statistics $s_{1:K}=(s_1,\ldots,s_K)\in \mathcal{S}^K$,
 where $\mathcal{S}^K$ denotes the $K$-dimensional summary space. These summaries are designed to retain the key information needed for inferring the latent states and 
 can be derived directly from the data or, as discussed in the next section, from the previous filtering distribution or other relevant sources.
 Operating in this summary space allows one to learn a single mapping $H_S:\mathcal{S}^K\times [0,1]\to \mathcal{X}$ capable of generating samples from the conditional distribution $p(x_t\mid s_{1:K})$. This \emph{partial} filtering distribution coincides with the full filtering distribution $p(x_t\mid y_{1:t})$ only when the summary statistics are sufficient. This occurs for example when the state-space model is linear and Gaussian. As shown in the Appendix, in this setting the full filtering distribution can be recovered exactly by using the first two moments of the predictive distribution as summaries. In most cases, the Pre-Trained Gen-Filter allows only partial inference. Nevertheless, with an appropriate choice of summaries, it can attain accuracy comparable to full inference while offering significant computational advantages since it no longer requires learning a new inverse CDF mapping at each time step, as in the Gen-Filter algorithm. Once trained, the model $H_s$, indeed, can be efficiently reused upon the arrival of new data to generate samples from the pseudo-filtering distribution, achieving a speed comparable to PFs. 
This filtering strategy is illustrated in its entirety in Algorithm \ref{alg:Pre-Trained_GenFilter}.

\begin{algorithm}
\caption{Pre-Trained Gen-Filter}\label{alg:Pre-Trained_GenFilter}
\begin{algorithmic}[1]

\STATE \textbf{Input:} Initial distribution $p(x_{0})$, transition model $p(x_t \mid x_{t-1})$, observation model $p(y_t \mid x_t)$, a summary function $S:\mathcal{Y}^t\to\mathcal{S}^K$, observed data $y_{1:T}$, a training horizon $\bar{t}$
\STATE \textbf{Output:} Samples from $p(x_t \mid S(y_{1:t}))$ for $t=1,\dots,T$
\STATE Sample $
x_0^{(i)} \!\sim\! p(x_0)$, $x_t^{(i)} \!\sim\! p(x_t \mid x_{t-1}^{(i)})$, $\tilde{y}_t^{(i)} \!\sim\! p(y_t \mid x_t^{(i)})$, $u^{(i)} \!\sim\! \mathcal{U}(0,1)$
for $t = 1,\ldots,\bar{t}$ and $i = 1,\ldots,N$.
\STATE Compute summaries $S(\tilde{y}_{1:\bar{t}}^{(i)})$ for $i = 1,\ldots,N$
\STATE Learn the inverse CDF mapping $x_{\bar{t}}\overset{d}{=}H(S(y_{1:\bar{t}}),u)$ using the synthetic dataset $\{x_{\bar{t}}^{(i)}, \; S(\tilde{y}_{1:\bar{t}}^{(i)}), \; u^{(i)}\}_{i=1}^{N}$
\FOR{$t = 1$ to $T$}
    \STATE Set $x_t \leftarrow \hat{H}(S(y_{1:t}),u)$
\ENDFOR
\STATE Return $\{x_t^{(i)}\}_{i=1}^{N}$ as samples from  the partial filtering distribution $p(x_t \mid S(y_{1:t}))$
\end{algorithmic}
\end{algorithm}

\subsection{Data Compression for State-Space Inference}

 While the use of summary statistics is well established in Bayesian likelihood-free inference, particularly within the ABC framework, extending this paradigm to inference of time-varying quantities -- such as the latent process in a state-space model -- 
  remains largely unexplored. In what follows, we discuss some possible approaches.

  The first approach is actually rooted in the work of \citet{Kalman_1960}, extending the original idea to encompass general classes of state space models. In the linear Gaussian setting, the filtering and the one-step ahead predictive distribution are Gaussian, thus they can be expressed entirely in in terms of their mean and variance which constitute sufficient statistics. It can be formally demonstrated that, starting from an initial distribution $X_0\sim\mathcal{N}(m_0,v_0)$, the Kalman Filter recursively maps 
  $(m_{t-1},v_{t-1})$ to $(m_t,v_t)$ upon observing new data $y_t$, where $m_t$ and $v_t$ denote respectively the mean and variance of the filtering distribution. 
  
 In a similar spirit, a more general update mechanism can be defined by learning a mapping from a set of prior distribution moments and a new data point to the corresponding hidden state. More rigorously, 
 we treat the set of moments $s_{1:K,0}$ associated with an initial distribution $p(x_0)$ as random variables, and generate a large simulated training dataset $\{x_1^{(i)},s_{1:K,0}^{(i)},\tilde{y}_1^{(i)},u^{(i)}\}_{i=1}^{N}$ by drawing $s_{1:K,0}$ from a prior $p(s_{1:K})$, followed by simulating $x_1\sim \int p(x_1\mid x_{0})p(x_{0}\mid s_{1:K})d x_0$, and $\tilde{y}_1\sim p(y_1\mid x_1)$. The resulting dataset is used to learn a map $H_S: \mathcal{Y}\times \mathcal{S}^K\times [0,1]\to \mathcal{X}$ which produces samples from $p(x_1\mid y_1,s_{1:K,0})$. If the set of plausible prior configurations explored during the training of $H_s(\cdot)$ is sufficiently rich, this mapping can be applied recursively. This is particularly the case when the posterior distribution belongs to the same family as the prior, ensuring that the learned update remains consistent across time. Specifically, at each time $t$, one can extract a new set of moments $s_{1:K,t-1}$ from the previous filtering distribution and then feed $(s_{1:K,t-1},y_t,u)$ into the learned map to obtain samples from $p(x_t\mid y_t,s_{1:K,t-1})$ approximating $p(x_t\mid y_{1:t})$. We show in Appendix that, in the linear Gaussian case, the filtering trajectory obtained using this approach is nearly indistinguishable from that of the Kalman filter.

 In practice, the extent to which the posterior distribution deviates from the prior is generally unknown, and constructing summaries in this manner can therefore yield highly variable performance across different classes of state-space models. For this reason, we also propose simpler alternatives, which in our experiments provide comparable or even more stable performance. 
 
 One such alternative is to construct summary statistics directly from the observed data. In this case, we define a summary function $S:\mathcal{Y}^t\to\mathcal{S}^k$ mapping the history of observations to a lower-dimensional summary space.
 A straightforward choice for $S(\cdot)$ is using a truncated window of recent observations, i.e. $S(y_{1:t})=y_{t-l:t}$ for some lag $l$.
More sophisticated summaries can also be designed to account for the structure of the underlying state-space model, such as moving averages or exponentially weighted averages. Importantly, the summary statistics should be selected so as to adequately capture the underlying dynamic structure of the latent process.

In the simulation study that follows, we mainly employ the approach based on using lags of the observed process. Therefore, we approximate the filtering distribution $p(x_t\mid y_{1:t})$ by $p(x_t\mid y_{t-l:t})$. While such approach may initially appear overly simplistic, the approximation often performs remarkably well compared with traditional filtering methods. Nonetheless, selecting the lag length $l$ presents a nontrivial trade-off: a small $l$ may fail to capture sufficient temporal information, whereas a large $l$ introduces many predictors into the regression model, thereby increasing the computational cost of training the deep neural network and raising the risk of overfitting if $N$, the number of training samples, is not sufficiently large. As a consequence, including many lags can potentially degrade the performance of the filtering strategy.

\subsection{Simulation Study}\label{sec:sim suty known par}

\subsubsection{Linear Gaussian Model}

We begin by evaluating our newly developed methodology on a Linear Gaussian (LG) state-space model, which serves as a well-understood and analytically tractable benchmark. As mentioned in Section \ref{sec:ssm_inference}, in this setting the filtering problem can be solved exactly using the Kalman filter. The model is specified as 
\begin{align}
y_t & = x_t + \sigma_y \epsilon_t \label{eq:LG-obs}\\
x_t & = \phi x_{t-1} + \sigma_x \eta_t \label{eq:LG_state}
\end{align}
where $\epsilon_t,\eta_t \overset{iid}{\sim}\mathcal{N}(0,1)$, and we set $\phi = 0.9$, $\sigma_x = 0.2$, and $\sigma_y = 1$. 

To ensure that our findings are not artifacts of simulation randomness, we simulate $100$ independent trajectories $(x_t,y_t)_{t=1}^{T}$, each of length $T=300$. This setup enables a robust comparison between the posterior approximations produced by our novel filtering method, the ground-truth Kalman filter solution, and the existing likelihood–free alternative, namely the ABC-PF.
In particular, we consider two versions of the latter: the original ABC-PF of \citet{Jasra2012}, which uses a uniform kernel, and a variant employing a Gaussian kernel. The latter  provides a smoother weighting scheme that may improve particle diversity and is especially well suited to settings, such as the present one, where the model’s true emission distribution is Gaussian.

For all particle filtering methodologies, we fix the number of particles to $N=1000$ and the tolerance level to $\epsilon=0.1$. Moreover, to mitigate particle degeneracy, resampling is performed whenever the effective sample size falls below 500, and a small regularization term of $10^{-10}$ is added to prevent division by zero in cases where no particles are accepted. 

Regarding our baseline Gen-Filter approach (outlined in Algorithm \ref{alg:Gen-Filter}),  for each time period $1\leq t\leq 300$, we generate a training dataset of size $N=1000$, $\{x_t^{(i)},\tilde{y}_t^{(i)},u_t^{(i)}\}_{i=1}^{N}$, which is used to learn  the inverse CDF mapping corresponding to the filtering distribution. To approximate this mapping, we employ a QNN architecture consisting of three main components: (i) a cosine embedding network that maps sampled quantile levels into a 64-dimensional representation, (ii) a three-layer feedforward network that projects the input variable into the same embedding space, and (iii) a four-layer fusion network with ReLU activations and dropout that processes the element-wise product of these two embeddings to output a quantile estimate. 

To evaluate the accuracy of the learned mapping, we compare samples from the posterior distribution $p(x_t\mid y_{1:t})$ generated by 
all the likelihood-free filtering strategies discussed in the paper, including our Gen-Filter, against those obtained from the exact filtering distribution computed using the Kalman Filter. As a measure of discrepancy, we use the first-order \textit{Wasserstein distance}, $W_1(P,Q):=\int_0^1 |F_P^{-1}(u)-F_Q^{-1}(u)|du$, which captures the overall geometry of the distributions by measuring the cost of transporting mass from one to the other. The \textit{Maximum Mean Discrepancy},  $\mathrm{MMD}^2_k(P, Q): = \mathbb{E}_{X,X' \sim P}\!\left[ k(X,X') \right]
\;+\; \mathbb{E}_{Y,Y' \sim Q}\!\left[ k(Y,Y') \right]
\;-\; 2\,\mathbb{E}_{X \sim P,\, Y \sim Q}\!\left[ k(X,Y) \right]$, which is usually used to detects fine-grained differences in shape (e.g. multimodality and skewness). We consider the MMD with Gaussian kernel \(k(x,y) = \exp\!\left(-\tfrac{(x-y)^2}{2\sigma^2}\right)\) and $\sigma=1$, and the special case with $k(x,y)=-|x-y|$, which is also known as \textit{Energy distance}. Finally we measure the difference in absolute values of the expected values and the standard deviations of the two distributions.

\begin{figure}[ht]
  \centering
  \includegraphics[width=\linewidth]{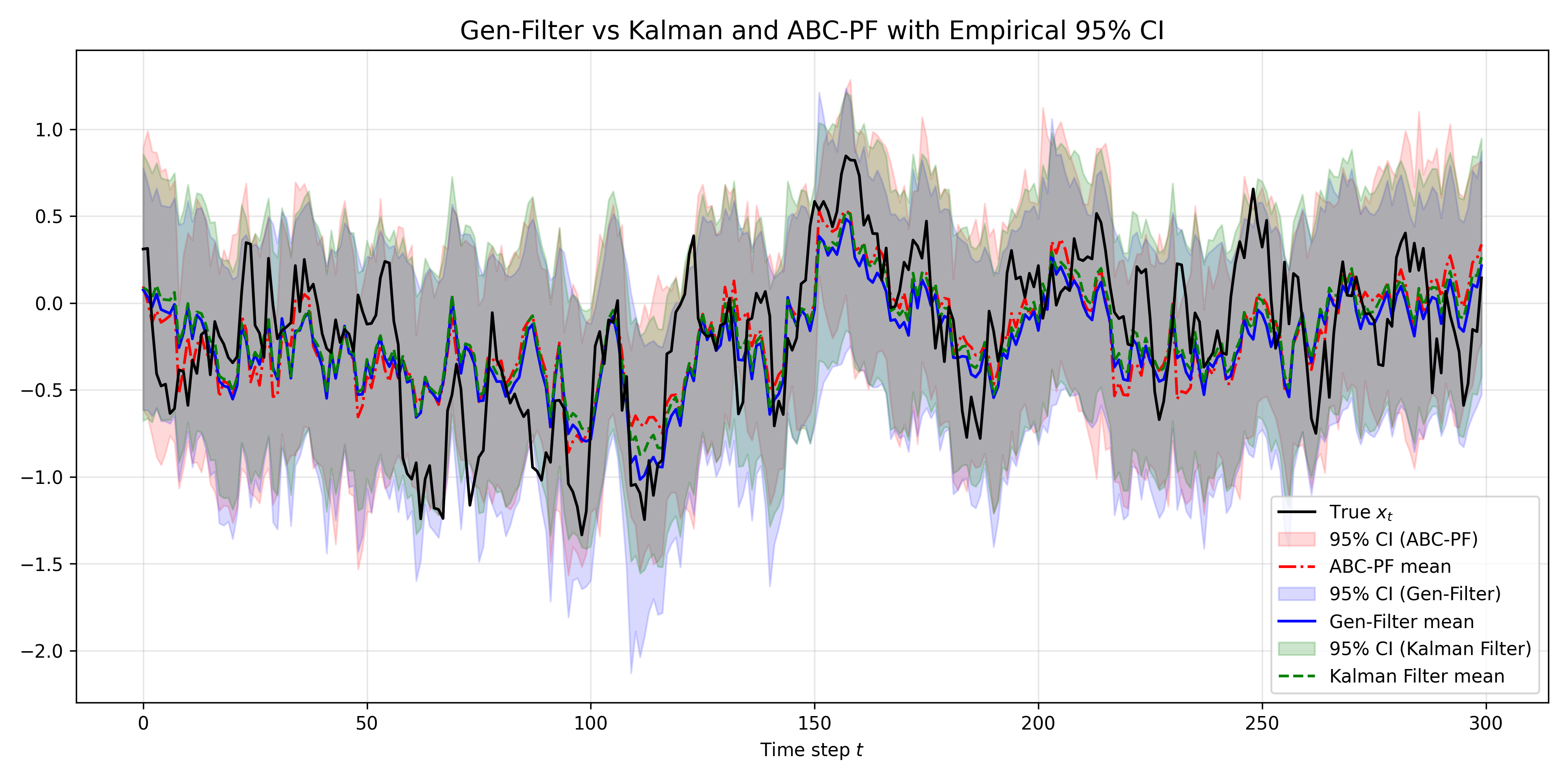}
  \caption{\footnotesize \textbf{LG model}. True latent states (black) with comparison of the Kalman filter (green), our Gen-Filter (blue), and the ABC–PF with a uniform kernel (red). For each method, posterior means are shown together with their corresponding 95\% credible intervals.}
\label{fig:AI_vs_KF_and_ABC_PF_with_Empirical_95CI}
\end{figure}

We report the results as averages over the 
 100 simulated scenarios. In particular, Table \ref{tab:KF_vs_methods} shows that our method provides the closest approximation, across all distance metrics, to the exact filtering distribution. Furthermore, as illustrated in Table \ref{tab:RMSE_coverage_means}, the signal estimated with the Gen-Filter is the closest to that obtained with the Kalman Filter, and the reported coverages at the 75, 90, and 95 levels are also close to the true coverages, a result that ABC-PF methods fail to achieve. This fact is further confirmed in the box plots of Figure \ref{fig:plot_RMSE_coverage}, which display the variability of RMSE and coverage across the simulated scenarios and, in particular, highlight that ABC-PF methods seldom attain the theoretical coverage levels. 
 Finally, Figure \ref{fig:AI_vs_KF_and_ABC_PF_with_Empirical_95CI} presents a visual comparison between the Kalman filter, ABC–PF, and our proposed approach for one simulated process. Although all approaches perform well for this class of state–space models, the latent trajectory estimated using the Gen-Filter appears visibly closer to the one obtained with the Kalman Filter.

The Pre-Trained Gen-Filter (Algorithm \ref{alg:Pre-Trained_GenFilter}) approach also yields promising results. In this case, we train ex-ante a single map $H_S$ over $1$ million synthetic scenarios using the QNN architecture discussed before, and then, at each time step $1\leq t\leq 300$, draw $N=1000$ samples from the approximate filtering distribution $p(x_t\mid S(y_{1:t}))$. For simplicity we use $10$ lags of the observed data as summary statistics, i.e. $S(y_{1:t})=y_{t-l:t}$ with $l=10$. We also report 
results for $l=20$ and $l=30$ in the Appendix.
As shown in Tables \ref{tab:RMSE_coverage_means} and \ref{tab:KF_vs_methods}, and in Figure \ref{fig:plot_RMSE_coverage}, the Pre-Trained Gen-Filter outperforms the ABC-based filters in terms of RMSE, coverage, and proximity to the exact filtering distribution. Interestingly, in this linear Gaussian setting, even using only 10 lags as summary statistics yields a good approximation of the filtering distribution.

The strong performance observed in the linear Gaussian case provides compelling evidence for the effectiveness of the filtering strategies introduced in this paper. To further demonstrate their general applicability, we extend the analysis to a simulated example based on a class of nonlinear, non-Gaussian state-space models commonly known as stochastic volatility models.

\begin{table}[ht]
\centering
\resizebox{\textwidth}{!}{
\begin{tabular}{lrrrr}
\toprule
Method & RMSE & Coverage (0.75) & Coverage (0.90) & Coverage (0.95) \\
\midrule
Kalman Filter & 0.347 & 0.750 & 0.898 & 0.949 \\
Gen-Filter  & 0.354 & 0.761 & 0.898 & 0.940 \\
Pre-Trained Gen-Filter & 0.348 & 0.737 & 0.893 & 0.945 \\
ABC-PF (Gaussian) & 0.360 & 0.709 & 0.859 & 0.911 \\
ABC-PF (Uniform) & 0.370 & 0.691 & 0.831 & 0.883 \\
\bottomrule
\end{tabular}
}
\caption{\footnotesize \textbf{LG model.} RMSE and Coverage at levels 0.75, 0.90, and 0.95 are reported as averages across 100 simulated scenarios. The box plot in Figure \ref{fig:plot_RMSE_coverage} in the Appendix shows the variability of the results across simulations.}
\label{tab:RMSE_coverage_means}
\end{table}

\begin{table}[ht]
\centering
\resizebox{\textwidth}{!}{
\begin{tabular}{lccccc}
\toprule
Distance from Ground Truth & Wasserstein & MMD & Energy & Mean Diff & Std Diff \\
\midrule
Gen-Filter & 0.060 & 0.003 & 0.007 & 0.051 & 0.023 \\
Pre-Trained Gen-Filter & 0.036 & 0.001 & 0.003 & 0.030 & 0.013 \\
ABC-PF (Gaussian) & 0.085 & 0.017 & 0.039 & 0.068 & 0.035 \\
ABC-PF (Uniform) & 0.112 & 0.021 & 0.048 & 0.090 & 0.047 \\
\bottomrule
\end{tabular}
}
\caption{\footnotesize \textbf{LG model.} Average distance metrics between filtering distributions of each method and the Kalman Filter (ground truth), computed over 100 simulations.}
\label{tab:KF_vs_methods}
\end{table}

\subsubsection{Stochastic Volatility Models}\label{sec:SV}

We consider a stochastic volatility model with $\alpha$-stable innovations, which is characterized by the following trajectories for the returns process $y_t$ and log-volatility process $x_t$:
\begin{align*}
    y_t &= \exp\left(\frac{x_t}{2}\right)\varepsilon_t, \\
    x_t &= \mu + \phi(x_{t-1} - \mu) + \sigma_\eta \eta_t,
\end{align*}
where $\varepsilon_t \sim \mathcal{S}(\alpha_y, \beta_y, \gamma_y, \delta_y)$ and $\eta_t \sim \mathcal{N}(0,1)$ are iid $\alpha$-stable random variables, and we assume $x_0\sim \mathcal{N}(\mu,\sigma_\eta^2/(1-\phi^2))$. The notation $\mathcal{S}(\alpha, \beta, \gamma, \delta)$ denotes a stable distribution with tail index $\alpha \in (0,2]$, skewness parameter $\beta \in [-1,1]$, scale $\gamma > 0$, and location $\delta \in \mathbb{R}$. The parameter vector is given by 
\[
\theta = \{\mu, \phi, \sigma_\eta, \alpha_y, \beta_y, \gamma_y, \delta_y\}.
\]
For the latent process we fix $\mu = 0$, $\phi = 0.98$, and $\sigma_\eta = 0.2$.

Despite the Gaussian specification for the return innovations is by far the most common choice, largely because it leads to analytical tractability, well-behaved likelihoods, and straightforward simulation, empirical evidence in finance has consistently shown (e.g. \cite{Cont01022001}, \cite{chakraborti2011econophysics}, and \cite{RevisitingFacts}) that asset returns exhibit excess kurtosis, heavy tails, and skewness, features that the Gaussian law cannot reproduce. Therefore, beginning with the pioneering works of \citet{Mandelbrot_1963}, \citet{Fama1965}, and \citet{mittnik1993modeling}, the $\alpha$-stable distribution gained popularity in the field for its interesting properties. Specifically, stable distributions with $\alpha<2$ naturally accommodate power-law tails, while the skewness parameter $\beta$ allows to model asymmetries. These features make $\alpha$-stable models well-suited for capturing extreme events and asymmetric risk patterns observed in high-frequency and crisis-period financial data, providing a more realistic foundation for risk measurement, option pricing, and portfolio stress testing. 

On the other hand, a notable drawback of using $\alpha$-stable distributions is that the $\alpha$-stable measure $\mathcal{S}(\alpha,\beta,\gamma,\delta)$ only admits a density with respect to Lebesgue measure that can be expressed in terms of elementary functions for three manifolds in the parameter space \citep{nolan2020univariate}. The innovations are distributed as Gaussian when $\alpha=2$, as Cauchy when $\alpha=1$ and $\beta=0$, and as L\'{e}vy when $\alpha=1$ and $\beta=1/2$. This characteristic makes the $\alpha$-stable SV model lends itself to likelihood-free inference whenever $\alpha$ is unknown or fixed at some value $\alpha\not\in\{1,2\}$. 

\begin{figure}[htbp]
    \centering
    \includegraphics[width=\textwidth]{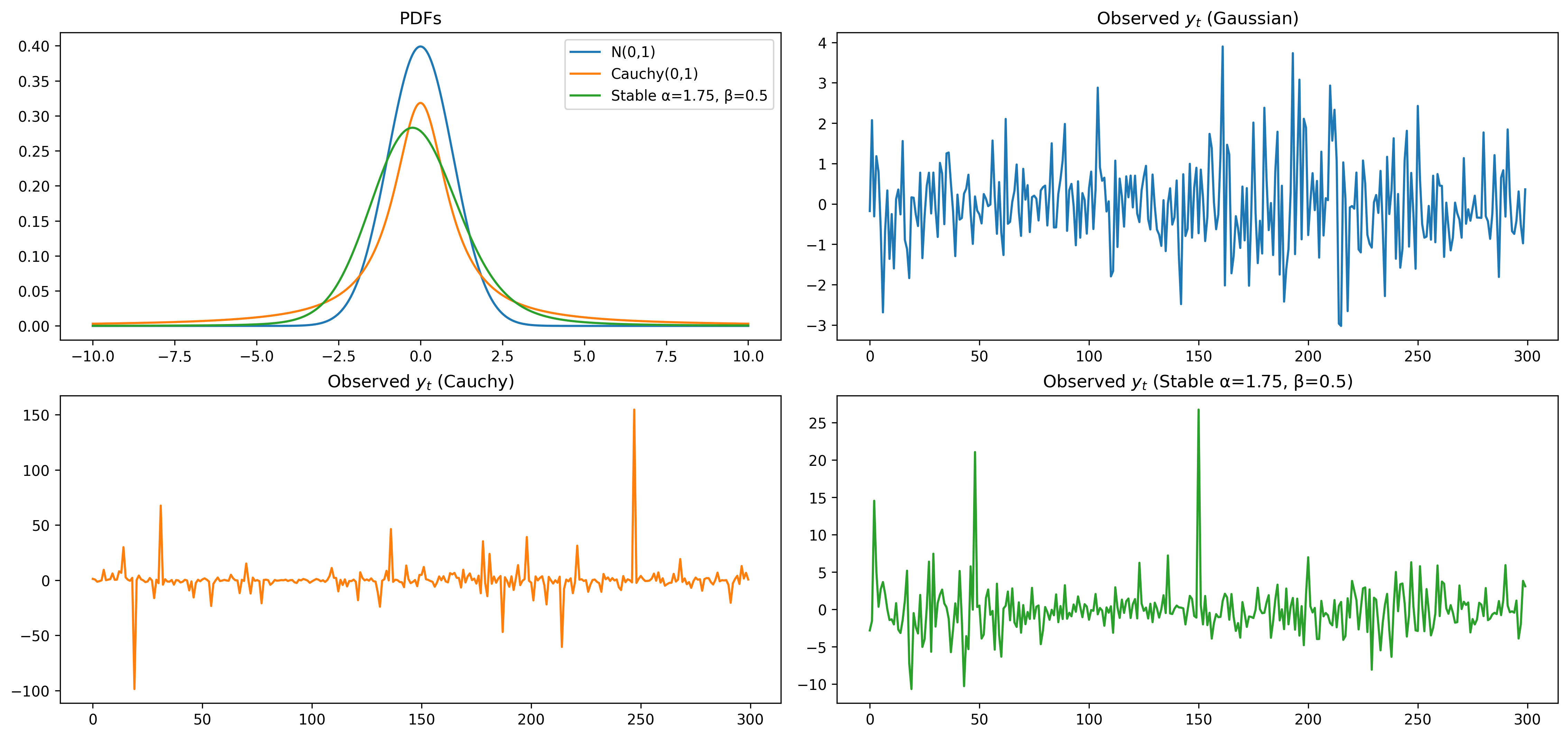} 
    \caption{\footnotesize Comparison of Gaussian, Cauchy, and asymmetric $\alpha$-stable ($\alpha=1.75$, $\beta=0.5$) densities; the $\alpha$-stable PDF is computed by numerically inverting its characteristic function. Panels also display sample observation processes generated under each distributional specification. 
   }
    \label{fig:compare_pdfs}
\end{figure}

In order to perform a principled evaluation of the efficacy of the Gen-Filter in learning the posterior distribution over the log-volatility process $p(x_t\mid y_{1:t})$ for each $t=1\,\ldots,T$, we first consider three sub-cases of the $\alpha$-stable SV model:
\begin{enumerate}
    \item Gaussian SV model: $\varepsilon_t\sim \mathcal{S}(\alpha_y = 2, \beta_y = 0, \gamma_y=1, \delta_y = 0)$
    \item Cauchy SV model: $\varepsilon_t\sim\mathcal{S}(\alpha_y = 1, \beta_y = 0, \gamma_y=1, \delta_y = 0)$
    \item Heavy-tailed asymmetric $\alpha$-stable SV model: $\varepsilon_t\sim\mathcal{S}(\alpha_y = 1.75, \beta_y = 0.5, \gamma_y=1, \delta_y = 0)$
\end{enumerate}
A visual comparison of the three innovation distributions is presented in Figure 
\ref{fig:compare_pdfs}.
The first two cases enable usage of the standard PF to act as a \emph{reference posterior} which we can use to evaluate the Gen-Filter. By contrast, in the third case the likelihood cannot be evaluated in closed form, and thus the standard PF is not applicable. Therefore, we employ an ABC-PF with $N=100,000$ particles as a benchmark for comparing the filtering strategies.

As in the previous section, we simulate 100 scenarios under each innovation distribution specification. For the Cauchy case, we discard draws corresponding to events with probability less than 1 in 10,000. In this distribution, such rare events translate into values that are extremely large in absolute magnitude; besides being unrealistic in practical settings, these extreme outliers can also cause all likelihood-free filtering methods to break down. Similarly, for the heavy-tailed asymmetric $\alpha$-stable case, we remove draws corresponding to events with probability below 1 in 100,000.

 Again, we compare: (i) the baseline Gen-Filter, where at each time $1\leq t \leq 300$ an inverse CDF map corresponding to the filtering distribution is trained on $N=1000$ synthetic samples generated from the predictive distribution $p(x_t\mid y_{1:t-1})$; (ii) the Pre-Trained Gen-Filter, where a single map $H_S$ approximating $F^{-1}_{x_t\mid y_{t-l:t}}$ (we select $l=30$) is fitted ex-ante using $1$ million simulated scenarios; and (iii) the ABC-PF, implemented with both uniform and Gaussian kernels, each using $N=1000$ particles and $\epsilon=0.1$ tolerance level. As shown in Tables \ref{tab:RMSE_coverage_means_panels_SV} and \ref{tab:PF_vs_methods_panels}, both the Gen-Filter and the Pre-Trained Gen-Filter shows superior performance compared to existing likelihood-free filtering methods, both in capturing the true latent trajectory, as reflected by lower RMSE, and in quantifying the associated uncertainty, as indicated by improved coverage. Moreover, the filtering distributions obtained with our methods are consistently closer to the reference filtering distributions across all considered distance metrics.
 
Additional results are provided in the Appendix, including comparisons across different numbers of simulated data points $N$ and lag values $l$, as well as box plots illustrating the variability across scenarios.

\begin{figure}[ht]
  \centering
  \includegraphics[width=\linewidth]{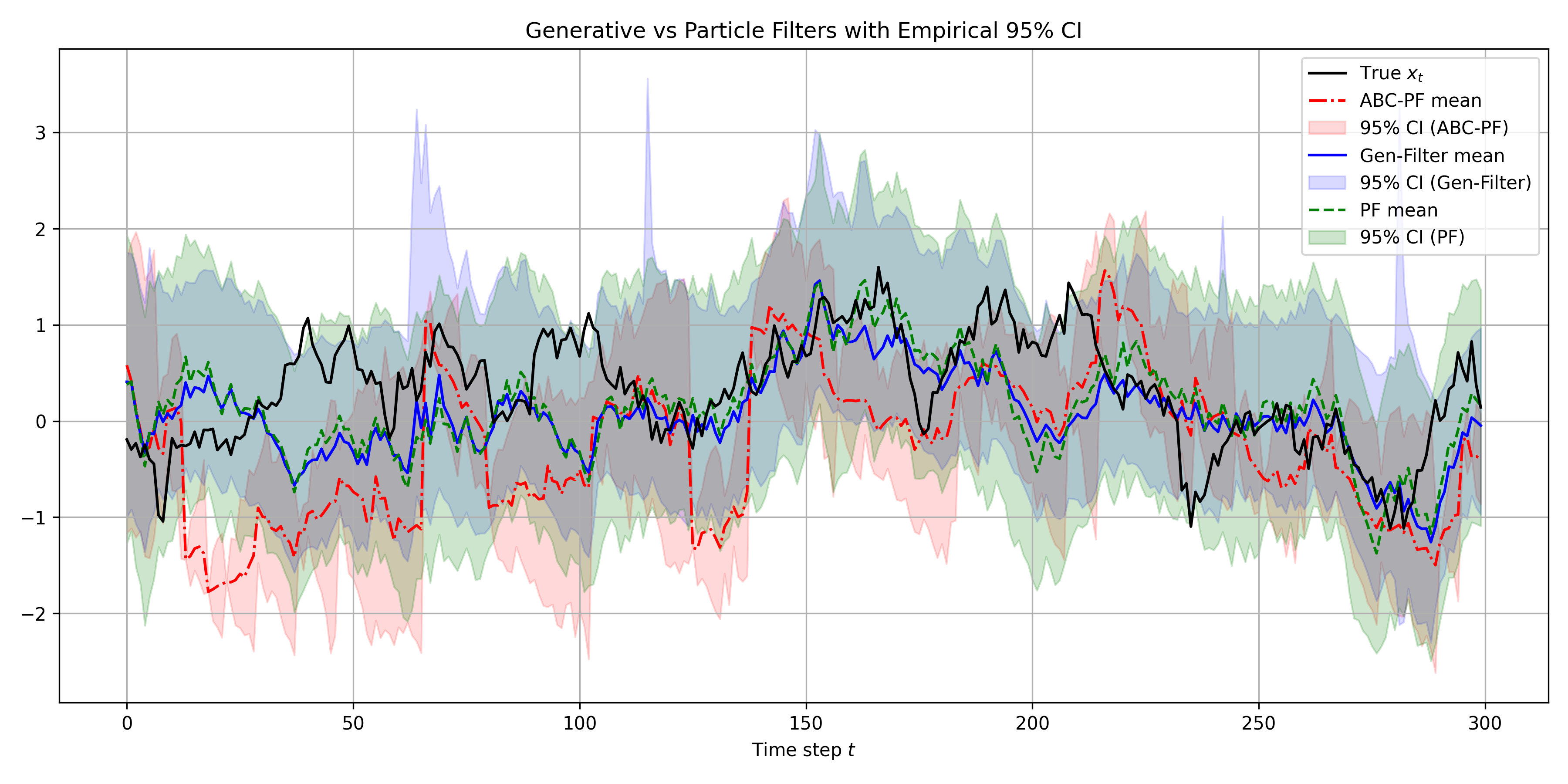}
  \caption{\footnotesize \textbf{Cauchy SV model} True latent states (black) with comparison of the PF (green), our Gen-Filter (blue), and the ABC–PF with a uniform kernel (red). For each method, posterior means are shown together with their corresponding 95\% credible intervals.}
\label{fig:AI_vs_PF_with_Empirical_95CI}
\end{figure}

Figure \ref{fig:AI_vs_PF_with_Empirical_95CI} depicts the estimated latent trajectory for each method for for one simulated scenario under Cauchy innovations. It shows that our strategy remains consistent with the ground-truth estimates provided by the PF, even in this heavy-tailed setting where the ABC-PF struggles.

\begin{table}[ht]
\centering
\resizebox{\textwidth}{!}{
\begin{tabular}{llrrrr}
\toprule
Model & Method & RMSE & Coverage (0.75) & Coverage (0.90) & Coverage (0.95) \\
\midrule
\multirow{4}{*}{GaussianSV}
& PF & 0.494 & 0.753 & 0.902 & 0.951 \\
& Gen-Filter & 0.515 & 0.762 & 0.900 & 0.938 \\
& Pre-Trained Gen-Filter & 0.495 & 0.758 & 0.908 & 0.954 \\
& ABC-PF (Gaussian) & 0.537 & 0.680 & 0.826 & 0.879 \\
& ABC-PF (Uniform) & 0.576 & 0.631 & 0.773 & 0.826 \\
\midrule
\multirow{4}{*}{CauchySV}
& PF & 0.653 & 0.745 & 0.896 & 0.944 \\
& Gen-Filter & 0.687 & 0.589 & 0.783 & 0.854 \\
& Pre-Trained  Gen-Filter & 0.779 & 0.728 & 0.881 & 0.930 \\
& ABC-PF (Gaussian) & 0.871 & 0.438 & 0.559 & 0.611 \\
& ABC-PF (Uniform) & 0.929 & 0.410 & 0.516 & 0.560 \\
\midrule
\multirow{4}{*}{$\alpha$-StableSV}
& Ground Truth &  0.539 & 0.741 & 0.892 & 0.939 \\
& Gen-Filter & 0.576 & 0.650 & 0.845 & 0.917 \\
& Pre-Trained  Gen-Filter & 0.564 & 0.715 & 0.878 & 0.934 \\
& ABC-PF (Gaussian) & 0.610 & 0.616 & 0.758 & 0.815 \\
& ABC-PF (Uniform) & 0.660 & 0.556 & 0.689 & 0.743 \\
\bottomrule
\end{tabular}
}
\label{tab:RMSE_coverage_means_SV}
\caption{\footnotesize \textbf{SV model}. RMSE and Coverage at levels 0.75, 0.90, and 0.95 are reported as averages across 100 simulations.}
\end{table}

\begin{table}[ht]
\centering
\resizebox{\textwidth}{!}{
\begin{tabular}{llccccc}
\toprule
Model & Method & Wasserstein & MMD & Energy & Mean Diff & Std Diff \\
\midrule
\multirow{3}{*}{GaussianSV} 
& Gen-Filter & 0.143 & 0.068 & 0.144 & 0.125 & 0.043 \\
& Pre-Trained Gen-Filter & 0.076 & 0.047 & 0.098 & 0.067 & 0.025 \\
& ABC-PF (Gaussian) & 0.173 & 0.047 & 0.100 & 0.147 & 0.069 \\
& ABC-PF (Uniform) & 0.242 & 0.068 & 0.146 & 0.211 & 0.090 \\ 
\midrule
\multirow{3}{*}{CauchySV} 
& Gen-Filter & 0.224 & 0.064 & 0.132 & 0.168 & 0.142 \\
& Pre-Trained Gen-Filter & 0.330 & 0.063 & 0.152 & 0.316 & 0.100 \\
& ABC-PF (Gaussian) & 0.526 & 0.161 & 0.355 & 0.467 & 0.240 \\
& ABC-PF (Uniform) & 0.581 & 0.183 & 0.409 & 0.522 & 0.252 \\
\midrule
\multirow{3}{*}{$\alpha$-StableSV}
& Gen-Filter & 0.203 & 0.032 & 0.070 & 0.177 & 0.076 \\
& Pre-Trained Gen-Filter & 0.177 & 0.026 & 0.055 & 0.170 & 0.035 \\
& ABC-PF (Gaussian) & 0.265 & 0.069 & 0.148 & 0.230 & 0.107 \\
& ABC-PF (Uniform) & 0.333 & 0.096 & 0.206 & 0.294 & 0.134 \\
\bottomrule
\end{tabular}
}
\label{tab:PF_vs_methods}
\caption{\footnotesize \textbf{SV model}. Average distance metrics between filtering distributions of each method and the PF (ground truth), computed over 100 simulations. For the $\alpha$-stable SV model, the ground truth is given by the ABC-PF with $10^5$ particles.}
\end{table}

\section{Parameter Learning}\label{sec:Parameter Learning}

Thus far, we have assumed $\theta\in\Theta$ to be known; however, this is rarely the case in practical applications. Therefore, we now propose strategies for conducting joint inference on both the latent trajectory and the unknown parameters within the Gen-Filter framework.

In the likelihood-free setting, most work has focused on parameter inference independently of trajectory estimation. For example, \citet{dean2014parameter},  \citet{martin2014approximatebayesiancomputationstate}, \citet{Yıldırım03072015}, and \citet{Martin03072019} develop ABC-based methodologies that primarily target inference on the static parameters  $\theta$, while treating the latent trajectory as secondary or implicitly marginalized. More focused attempts to address both parameter and trajectory inference include \citet{jasra2013aliveparticlefilter} and, subsequently, \citet{Vankov_2019}, who propose the use of PMCMC  algorithms to approximate the joint posterior distribution $p(\theta, x_{0:T} \mid y_{1:T})$, with the ABC particle filter employed as an estimator of the likelihood. 

In principle, a PMCMC-type algorithm could also be formulated within the GBF framework, where the Gen-Filter serves as an estimator of the likelihood in place of the PF. Nevertheless, we pursue more favorable computational approaches.
Specifically, we propose two efficient, fully density-free methodologies that remains applicable across a broad class of state-space models.

The first approach factorizes the joint posterior distribution of the parameters and the latent trajectory as
\[
p(\theta, x_{0:T} \mid y_{1:T})= p(x_{0:t}\mid y_{1:T},\theta)p(\theta\mid y_{1:T}),
\]
thus decomposing the inference task into two steps: parameter inference and conditional state estimation. In the first step, we learn an approximation to the inverse posterior transform
$\hat{F}^{-1}_{\theta\mid y_{1:T}}(\cdot)$ using an appropriate deep learning methodology. This map provides a generator for obtaining samples from the posterior distribution of the state-space parameters.
In the second step, we use the generated samples from the parameters' posterior to recover the latent trajectory $p(\theta\mid y_{1:t})$:
\[
p(x_{0:T}\mid y_{1:T})=\int p(x_{0:T}\mid y_{1:T},\theta)p(\theta\mid y_{1:T})d\theta.
\]
In practice, $y_{1:T}$ may be high-dimensional, making direct use of the raw data impractical. It is therefore recommended to employ summary statistics that compress the relevant information into a lower-dimensional representation. The resulting posterior approximation $p(\theta\mid S(y_{1:T}))$ coincides with the full Bayesian posterior when $S(\cdot)$ is sufficient in the Bayesian sense; otherwise, it constitutes a partial posterior, which may nonetheless yield important inferential results.  

Identifying suitable summary statistics from the observed process can be challenging. 
For example \citet{Martin03072019} explores auxiliary likelihood-based approaches, in which summary statistics are obtained from an auxiliary model that is easier to estimate than the true model, while \citet{maneesoonthorn2024probabilisticpredictionsoptionprices} exploits summaries derived from multiple data sources.

A more robust and intuitive strategy would be to employ summary statistics that also incorporate information from the latent trajectory. However, implementing this approach in practice is challenging because the state sequence is unobserved, and hence such summaries cannot be computed directly in the same way as those based on the observed data $y_{1:T}$. To address this issue, we develop a novel Bayesian computational method, denoted \emph{Gen-Gibbs} sampler. Such approach is broadly applicable to Bayesian inference and is particularly well-suited to hierarchical models, where the presence of multiple levels makes it difficult to construct informative summaries for each latent variable from the observations alone. In such cases, the hierarchical structure itself can be exploited to design more effective summaries and improve the computational efficiency.
We discuss the methodology in details in the next section.

\section{Generative Gibbs Sampling }\label{sec:Gen-Gibbs}

With Gen-Gibbs, we refer to a broadly applicable sampling strategy that integrates the rigorous properties of MCMC algorithms with recent advances in generative modeling, thereby harnessing advanced machine learning techniques within a principled computational Bayesian approach. 

Analogous to Gibbs sampling, the Gen-Gibbs algorithm approximates the posterior distribution by iteratively sampling from the full conditional distributions of the parameters. In contrast to the classical approach, which requires analytical derivation of the conditionals, Gen-Gibbs leverages a deep learning model to approximate their quantile functions. This representation permits direct sampling from the full conditionals via independent draws from $\mathcal{U}(0,1)$ distributions.

Formally, let $\theta_1,\dots,\theta_B$ denote $B$ parameters, each associated with a prior $p(\theta_b)$, $b=1,\dots,B$. Given a probabilistic model $p(y\mid \theta_1,\dots,\theta_B)$ from which sampling is possible, one can (in parallel) learn the inverse CDF maps $F^{-1}_{\theta_b\mid \theta_{-b,y}}:\Theta_{-b}\times \mathcal{Y}^t \times [0,1]\to \Theta_b$ for $b=1,\dots,B$, where $\theta_{-b}$ denotes the vector of all parameter block excluding $\theta_b$, $\Theta_{b}$ is the domain of block $\theta_b$ and $\Theta_{-b}$ is the domain of the remaining blocks. In other words, following the same procedure described in the previous sections, one can generate a large synthetic dataset $\{\theta_b^{(i)}, \theta_{-b}^{(i)}, \tilde{y}^{(i)}, u_b^{(i)}\}_{i=1}^{N}$ by simulating from the prior, the model, and i.i.d. uniform distributions. This dataset can then be used to train a deep learning model of the form $\theta_b^{(i)}=H_{b\mid -b}(\theta_{-b}^{(i)},\tilde{y}^{(i)},u_b^{(i)})$ which can be used as a generator from the full conditional distribution.
Once learned, posterior samples from $p(\theta_1,\dots,\theta_B\mid y_{1:t})$ are obtained by sequentially drawing from the base distribution $\mathcal{U}(0,1)$ and providing these draws, together with the observations $y_{1:T}$ and the most recently updated parameter values, as inputs to the trained models. The entire structure of the Gen-Gibbs sampler is reported in Algorithm \ref{alg:gibbs-training} in Appendix.

The sampling strategy described is particularly useful in hierarchical Bayes models, including state-space models, where some of the parameters are not directly tied to the data. To illustrate, suppose that $\theta$ governs the latent dynamics and therefore depends directly only on the unobserved states rather than the observations themselves. In this case, a state-space model can be written as a two-level hierarchical model
\begin{align*}
    y_t \mid x_t & \sim p(y_t\mid x_t),\\
    x_t  \mid x_{t-1},\theta & \sim p(x_t\mid x_{t-1},\theta),\\
    \theta & \sim p(\theta).
\end{align*}
Here, the conditional distribution of $\theta$ involves only the latent states, while the influence of $y_{1:t}$ is transmitted indirectly through the updates of $x_{0:t}$. Therefore, since the latent states are the true carriers of information about the transition parameters, it becomes natural to extract summaries from the trajectory via $S_x:\mathcal{X}^{t+1}\to\mathcal{S}^K$, and use them to update $\theta$. This approach isolates the signal contained in the dynamics from the noise in the measurements, leading to more consistent and computationally efficient inferential procedure.

As a result, the application of the Gen-Gibbs sampler in the context of state-space models reduces to a sort of FFBS strategy for the latent states, combined with Gibbs updates for the unknown parameters. At each iteration we draw
\begin{align*}
    x_T & \sim p(x_T \mid S_y(y_{1:T}),\theta)\\
    x_{t} & \sim p(x_{t}\mid x_{t+1},S_y(y_{1:t}),\theta)\\
    \theta & \sim p(\theta \mid S_x(x_{0:T}))   
\end{align*}
where $S_y$ denotes the  statistics extracted from the observations, and $S_x$ denotes the summaries from the latent trajectory. 

We provide more details about the Gen-Gibbs sampler for state-space models in Algorithm \ref{alg:ffbs-training}. In practice, different components of $\theta$ may depend exclusively on the observations, exclusively on the latent states, or on both. To account for this, we introduce a general summary function $S_\theta:\mathcal{Y}^T\times \mathcal{X}^{T+1}\to \mathcal{S}^K$ with arguments that can involve either the observed data, the latent trajectory, or a combination of the two.

We will show in the simulation study that, as long as the summaries statistics are enough informative -- and, in the best cases, sufficient -- and the deep learning architecture is adequately expressive, the estimated maps yield accurate  approximations of the true full conditional distributions. Consequently, the Gen-Gibbs sampler produces a chain of draws for the parameters and the latent states, $\{\theta^{(i)},x_{0:t}^{(i)}\}_{i=1}^{M}$, with $M$ being the number of the iterations, which results to be a reliable approximation of the joint posterior distribution, a
conclusion also supported by comparisons with traditional MCMC methods.

\begin{algorithm}[H]
\caption{Pre-Training for Gen-Gibbs Sampler}
\label{alg:ffbs-training}
\begin{algorithmic}[1]
\STATE \textbf{Input:}
Prior $p(\theta)$,
transition model $p(x_t\mid x_{t-1})$, observation model $p(y_t\mid x_t)$, summary functions $S_y:\mathcal{Y}^t\to\mathcal{S}^{K_y}$ and $S_\theta:\mathcal{Y}^T\times \mathcal{X}^{T+1}\to \mathcal{S}^{K_{\theta}}$
\STATE \textbf{Output:} 
$B$ parameters' maps $\{\hat{H}_{b\mid -b}(\cdot)\}_{b=1}^{B}$, a  filtering map $\hat{H}_{t\mid t-1}(\cdot)$, a smoothing map $\hat{H}_{t\mid t+1}(\cdot)$
\STATE Sample $\theta^{(i)}\!\sim\! p(\theta)$, $
x_0^{(i)} \!\sim\! p(x_0\mid \theta^{(i)})$, $x_t^{(i)} \!\sim\! p(x_t \mid x_{t-1}^{(i)},\theta^{(i)})$, $\tilde{y}_t^{(i)} \!\sim\! p(y_t \mid x_t^{(i)},\theta^{(i)})$, $u^{(i)} \!\sim\! \mathcal{U}(0,1)$
for $t = 1,\ldots,T$ and $i = 1,\ldots,N$
\STATE Use the synthetic dataset $\{\theta_b^{(i)},\theta_{-b}^{(i)},S_\theta(x_{0:t}^{(i)},\tilde{y}_{1:t}^{(i)}), S_y(\tilde{y}_{1:t}^{(i)}),u^{(i)}\}_{1=1}^{N}$ to learn:
\\ $\theta_b\overset{d}{=}H_{b\mid -b}(\theta_{-b},S_\theta(x_{0:T},y_{1:T}), u)$ for $b=1,\dots,B$
\\ $x_t\overset{d}{=}H_{t\mid t-1}(S_y(y_{1:t}),\theta, u)$
\\ $x_t\overset{d}{=}H_{t\mid t+1}(x_{t+1}, S_y(y_{1:t}), \theta, u)$
\STATE \textbf{return} $\{\hat{H}_{b\mid -b}(\cdot)\}_{b=1}^{B}$, $\hat{H}_{t-1\mid t}(\cdot)$, and $\hat{H}_{t\mid t+1}(\cdot)$
\end{algorithmic}
\end{algorithm}

\begin{algorithm}[H]
\caption{Gen-Gibbs Sampler for State-space Models}
\label{alg:ffbs-testing}
\begin{algorithmic}[1]
\STATE \textbf{Input:}
Data $y_{1:T}$, $B$ parameters maps $\{\hat{H}_{b\mid -b}(\cdot)\}_{b=1}^{B}$, a filtering $\hat{H}_{t-1\mid t}(\cdot)$, a smoothing $\hat{H}_{t\mid t+1}(\cdot)$, summary functions $S_y:\mathcal{Y}^t\to\mathcal{S}^{K_y}$ and $S_\theta:\mathcal{Y}^T\times \mathcal{X}^{T+1}\to \mathcal{S}^{K_{\theta}}$
\STATE \textbf{Output:} Chain of draws $\{\theta^{(i)},x_{0:t}^{(i)}\}_{i=1}^{M}$ from the joint posterior distribution
\STATE Initialize $\theta^{(0)}$ 
\FOR{$i = 1$ to $M$}
\STATE Sample $u^{(i)}\sim\mathcal{U}(0,1)$
\STATE Set $x_T^{(i)}\leftarrow \hat{H}_{t-1\mid t}(S_y(y_{1:T}),u^{(i)})$
\FOR{$t=T-1$ to $0$}
\STATE Sample $u^{(i)}\sim\mathcal{U}(0,1)$
\STATE Set $x_t^{(i)}\leftarrow \hat{H}_{t\mid t+1}(x_{t+1}^{(i)},S_y(y_{1:t}),u^{(i)})$
\ENDFOR

\STATE Set $\theta^{(i)}\leftarrow \theta^{(i-1)}$
  \FOR{$b = 1$ to $B$}
    \STATE Sample
    $u^{(b)}\sim \mathcal{U}(0,1)$
    \STATE Set $      \theta_b^{(i)} \leftarrow \hat{H}_{b\mid -b}( \theta_{-b}^{(i)},S_\theta(x_{0:T}^{(i)},y_{1:T}),u^{(b)}) $
  \ENDFOR
\ENDFOR

\STATE \textbf{return} $\{\theta^{(i)},x_{0:T}^{(i)}\}_{i=1}^{M}$
\end{algorithmic}
\end{algorithm}

\subsection{Simulation Study}\label{sec:sim suty unknown par}

\subsubsection{Linear Gaussian Model}

Consider again the LG state-space model introduced in equations \ref{eq:LG-obs} and \ref{eq:LG_state}. Here, we are still assuming that $\phi$ is known and fixed to $0.9$ to ensure stationariety, and, in addition to the latent trajectory, the primary unknown quantities of interest are the state noise variance $\sigma_x^2$ and the observation noise variance $\sigma_y^2$. We reparameterize them in terms of their precisions, $\psi_y = 1/\sigma_y^2$ and $\psi_x = 1/\sigma_x^2$ and assign them independent Gamma priors, $\mathcal{G}(a_0, b_0)$, with hyperparameters $a_0 = b_0 = 2$, chosen to provide weakly informative priors. Eliciting Inverse Gamma priors for the variances in a LG model preserves conjugacy, which implies that all full conditional distribution are available in closed form. Therefore, Gibbs sampling can be directly employed to jointly infer the latent trajectory and the state-space parameters. This setup enables a direct comparison between posterior samples produced by our approach and those obtained using the classical FFBS method of \citet{CK_1994} and \citet{F_1994}, as reported in the Appendix. 

For training the Gen-Gibbs sampling, we use again a QNN architecture to approximate the inverse CDF maps of all the full conditionals distributions. As summary statistics for the filtering and smoothing distributions, we use the last $50$ observations, i.e. $S_y(y_{1:t})=y_{t-50:t}$. For the full conditionals of the parameters, the following sufficient statistics are available:
\[
S_{\psi_x}(x_{0:T})=\sum_{t=1}^T (x_t - \phi x_{t-1})^2, \quad S_{\psi_y}(y_{1:t},x_{0:t}) = \sum_{t=1}^T (y_t - x_t)^2.
\]
We train the deep learners on a synthetic dataset generated from the model of size $N=10^7$, and then test the procedure on $100$ simulated processes from the LG model with true parameters $\psi_y=1$ and $\psi_x=5$, which correspond to $\sigma_y^2=1$ and $\sigma_x^2=0.2$. For both the Gibbs and Gen-Gibbs sampling, we produce a total of $1000$ draws, discarding the first $500$ as burn-in. 

\begin{figure}[htbp]
    \centering
    % First plot
    \begin{subfigure}{\textwidth}
        \centering
        \includegraphics[width=\textwidth]{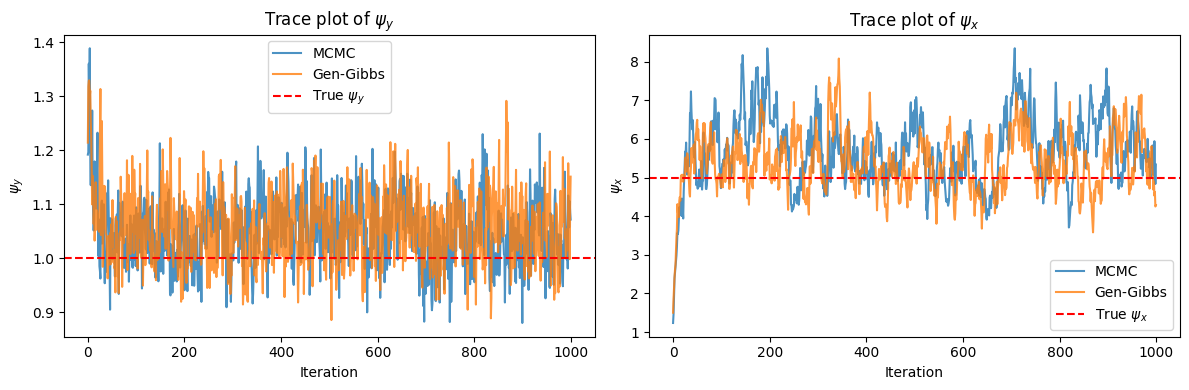}
        \caption{\footnotesize Trace plots of traditional Gibbs (MCMC) and Gen-Gibbs samples for $\psi_y$ and $\psi_x$.}\label{fig:LG-single-a}
    \end{subfigure}
    \vskip\baselineskip
    
    % Second plot
    \begin{subfigure}{\textwidth}
        \centering
        \includegraphics[width=\textwidth]{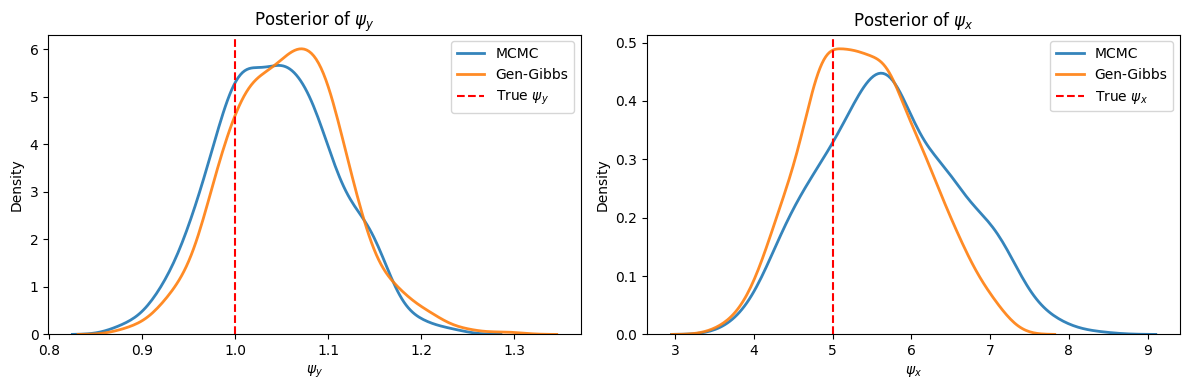}
        \caption{\footnotesize Kernel density estimates of the posterior distributions of $\psi_y$ and $\psi_x$.}\label{fig:LG-single-b}
    \end{subfigure}
    \vskip\baselineskip
    
    % Third plot
    \begin{subfigure}{\textwidth}
        \centering
        \includegraphics[width=\textwidth]{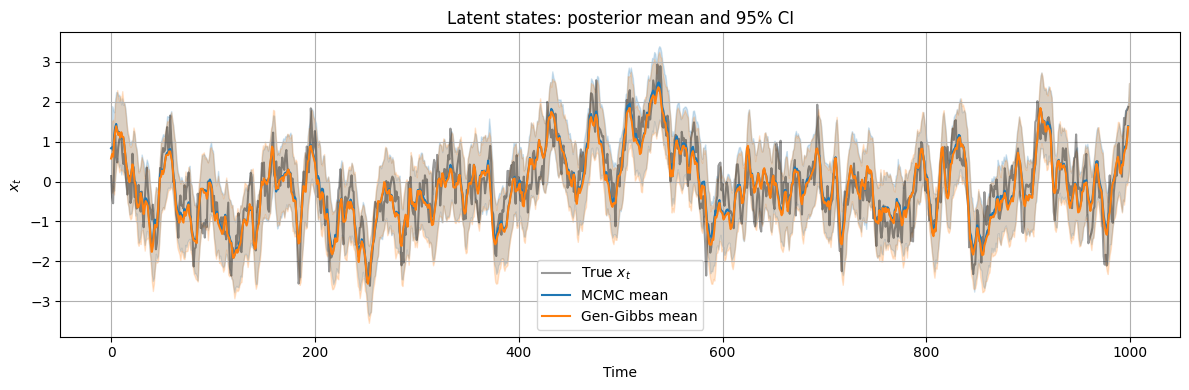}
        \caption{\footnotesize Posterior mean estimate of the latent trajectory $(x_t)_t$ with shaded areas indicating the 95\% credible intervals }\label{fig:LG-single-c}
    \end{subfigure}
    
    \caption{\footnotesize \textbf{LG model} with unknown parameters $\psi_y$ and $\psi_x$.  }\label{fig:LG-single}
\end{figure}

 For a single simulated process, Figures \ref{fig:LG-single-b} and \ref{fig:LG-single-c} illustrates that the Gen-Gibbs posteriors closely match those obtained with the traditional Gibbs sampling. Of particular interest is the mixing and convergence behavior displayed in Figure \ref{fig:LG-single-a}, which demonstrates that the Gen-Gibbs chains achieve rapid mixing and stable convergence, comparable to the classical approach. Notably, Figures \ref{fig:LG-single-b} and \ref{fig:LG-single-c} also show that, when initialized with the same parameter values, both methods converge within approximately the same number of steps. This indicates that the proposed method not only reproduces the posterior distributions with high accuracy but also retains desirable sampling properties, making it a viable alternative to traditional Gibbs sampling, when the latter is not directly available. 

\begin{figure}[h!]
    \centering
\includegraphics[width=\textwidth]{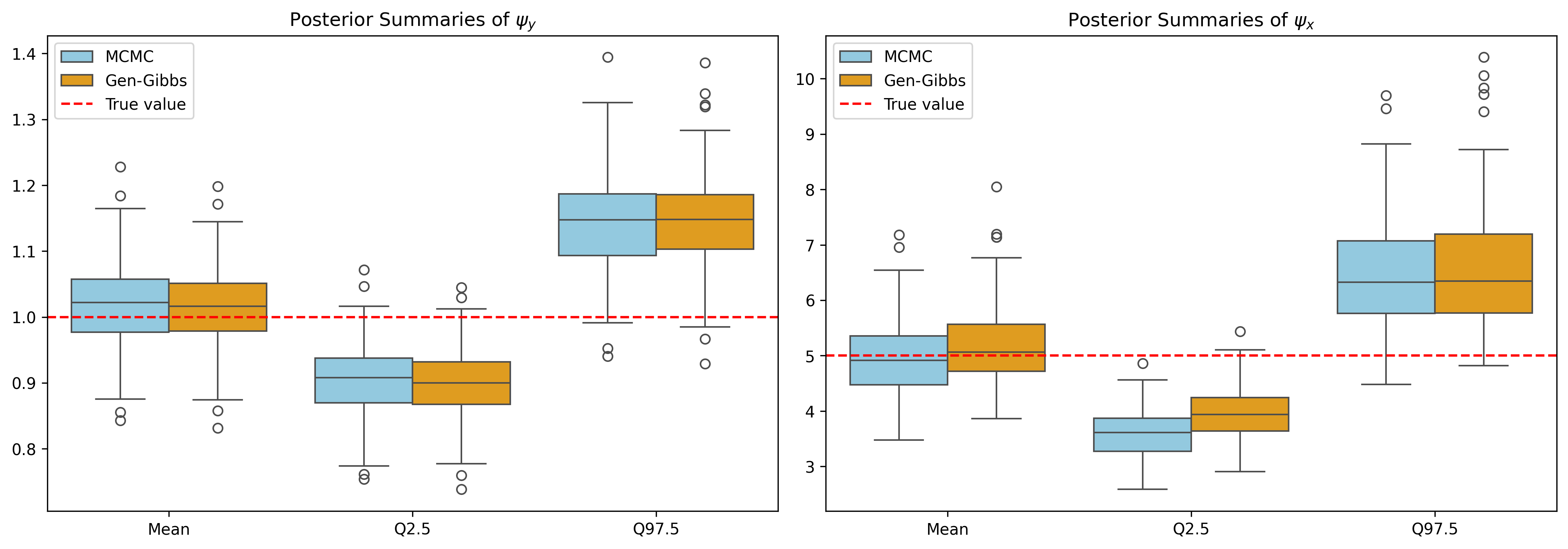}
    \caption{\footnotesize \textbf{LG model}. Estimated posterior means and 95\% credible intervals (0.025 and 0.975 quantiles) for $\psi_y$ (left panel) and $\psi_x$ (right panel), based on $100$ simulated process. Results from traditional Gibbs (MCMC) are displayed in blue, and those from Gen-Gibbs in orange.}
    \label{fig:Comparing_means_quantiles_across_sim}
\end{figure}

\begin{figure}[h!]
    \centering
\includegraphics[width=\textwidth]{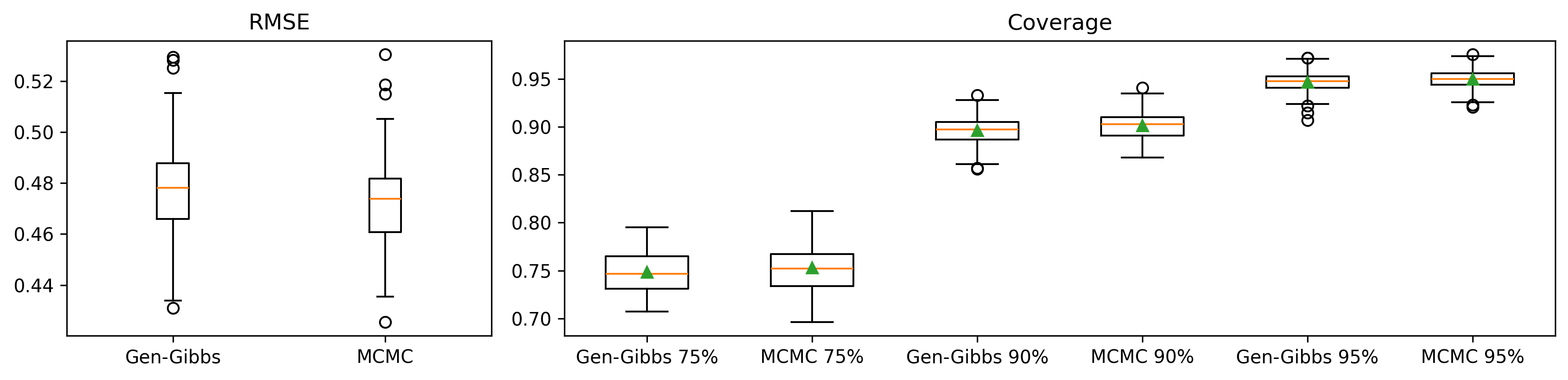}
    \caption{\footnotesize \textbf{LG model}. Performance across $100$ simulated scenarios summarized by boxplots: RMSE of the latent trajectory estimates with traditional Gibbs (MCMC) and Gen-Gibbs (left panel), and empirical coverage of the true trajectory at the 75\%, 90\% and 95\% credible interval levels (right panel). Mean coverages are also indicated in green.
}    \label{fig:Comparing_means_quantiles_across_sim_x}
\end{figure}

Repeating the analysis across all 100 simulated processes, we find that the posterior means and quantiles of the unknown parameters are highly consistent between the two methods, as shown in Figure \ref{fig:Comparing_means_quantiles_across_sim}, as well as the coverage values reported in Table \ref{tab:Gibbs_AI_LG}. Moreover, the latent trajectories estimated by both methods exhibit comparable RMSE and coverage values, as displayed in Figure \ref{fig:Comparing_means_quantiles_across_sim_x}.

Overall, these results are highly promising and motivate extending the analysis to more challenging settings, in particular a non-linear, non-Gaussian example as we have done for the case with known parameters.

\begin{table}
\centering
\begin{tabular}{llrrr}
\toprule
 Parameter & Method & Coverage (0.75) & Coverage (0.90) & Coverage (0.95) \\
\midrule
\multirow[t]{2}{*}{$\psi_y$} & MCMC & 0.74 & 0.89 & 0.91 \\
 & Gen-Gibbs & 0.77 & 0.89 & 0.92 \\
\midrule
\multirow[t]{2}{*}{$\psi_x$} & MCMC & 0.77 & 0.87 & 0.93 \\
 & Gen-Gibbs & 0.76 & 0.91 & 0.96 \\
\bottomrule
\end{tabular}
\caption{\footnotesize \textbf{LG model}. Coverages for $\psi_y$ and $\psi_x$ using traditional Gibbs (MCMC) and Gen-Gibbs sampling.}
\label{tab:Gibbs_AI_LG}
\end{table}

\subsubsection{Stochastic Volatility Model}
We now reconsider the $\alpha$-stable SV model introduced in Section \ref{sec:SV}. As a first step, we focus on the Gaussian SV model specification with unknown parameters $(\mu, \phi, \sigma^2_\eta)$. This setting has been extensively studied in the literature \citep{Jaquier_2002} and 
a wide range of inference and computational strategies have been developed to enable joint estimation of both the latent volatility trajectory and the model’s structural parameters. The non-linear nature of the problem, indeed, does not allow one to employ a traditional Gibbs sampling strategy, since the conditional distributions are not available in closed form. To address this challenge, \citet{kim1998stochastic} proposed an elegant data augmentation scheme based on a finite normal mixture approximation to the distribution of $\log(y_t^2)$. This approach effectively linearizes the observation equation, enabling conditional Gaussian structures that make Gibbs steps feasible for the latent volatilities. Because of its popularity in the field, we adopt it as the benchmark against which we compare our approach. We refer to this method in tables and figures 
as MCMC.

We compare the traditional MCMC strategy with our Gen-Gibbs approach. Both  models are specified with weakly informative priors: $\mu \sim \mathcal{N}(0,1)$, $\phi \sim \mathcal{B}(20,1.5)$, and $\sigma_\eta^2 \sim \mathcal{IG}(2.5,0.025)$. These priors are standard choices in the Bayesian SV literature and are discussed in \citet{kim1998stochastic}. As an additional step to improve numerical stability, we reparametrize the model using
$\tilde{x}_t=x/\sigma^2$ and $\gamma = \mu/\sigma$.
This transformation reduces posterior dependence among parameters and improves mixing in the MCMC sampler, a strategy that has been shown to be particularly effective in stochastic volatility settings (\cite{Robert_2004}, \cite{KASTNER2014408}). 
 As summary statistics for the filtering and smoothing distributions, we use the most recent $50$ observations, i.e. $S_y(y_{1:t})=y_{t-50:t}$. For the full conditional distributions of the parameters, we construct a set of summaries based on both the observed data and the latent trajectories. A detailed description of these summaries is provided in the Appendix.

\begin{figure}[ht]
    \centering
    \includegraphics[width=\textwidth]{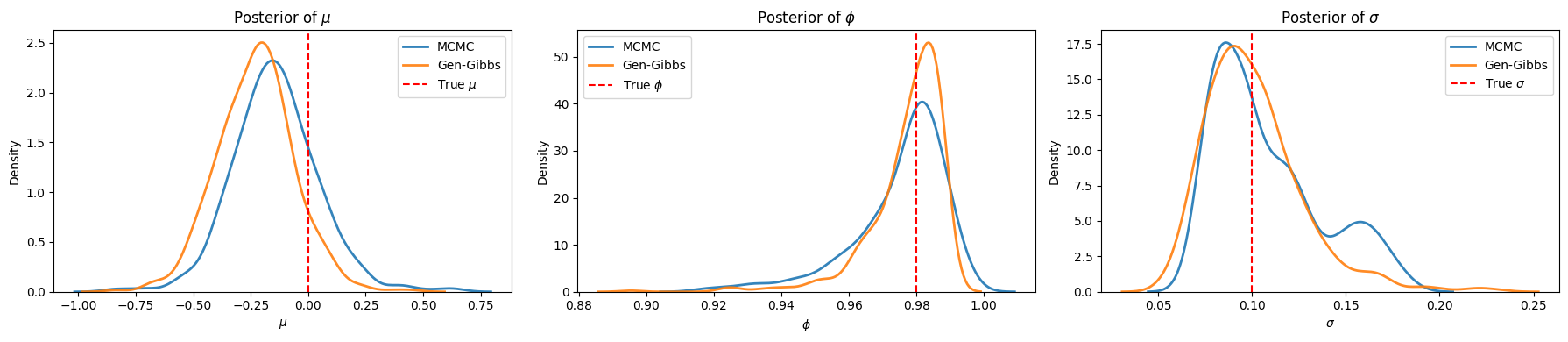}
    \caption{\footnotesize \textbf{Gaussian-SV model}. Kernel density estimates of the posterior distributions of $\mu$, $\phi$, and $\sigma_\eta$ for one simulated scenario.}
    \label{fig:one_process_SV}
\end{figure}

\begin{figure}[ht]
    \centering
\includegraphics[width=\textwidth]{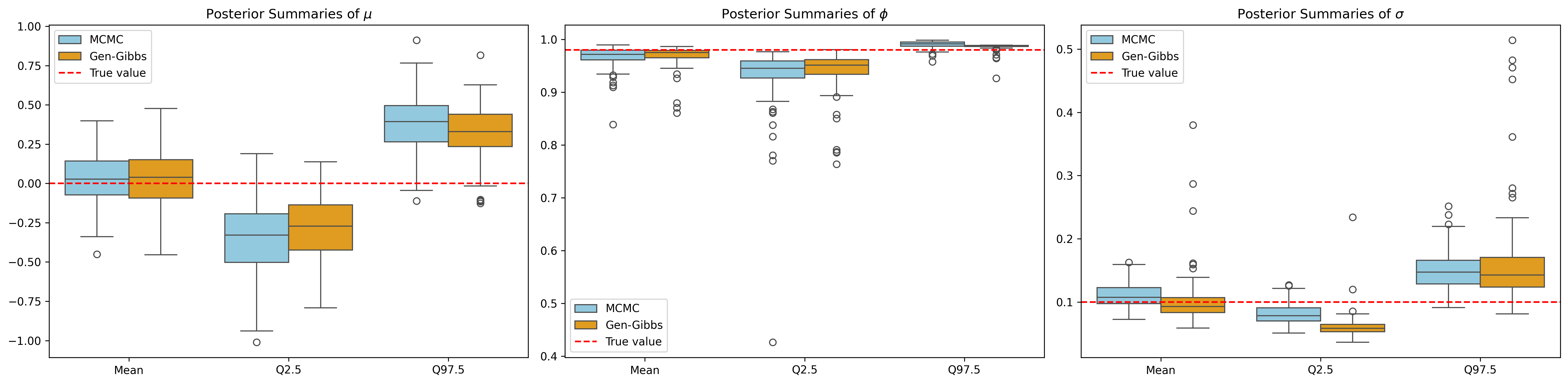}
    \caption{\footnotesize \textbf{Gaussian SV model}. Estimated posterior means and 95\% credible intervals (0.025 and 0.975 quantiles) for 
    $\mu$ (left panel), $\phi$ (central panel), and $\sigma_\eta$ (right panel)  based on $100$ simulated process. Results from the MCMC algorithm of \citet{kim1998stochastic} are displayed in blue, and those from Gen-Gibbs in orange.}
    \label{fig:Comparing_means_quantiles_across_sim_GaussianSV}
\end{figure}

\begin{figure}[ht]
    \centering
\includegraphics[width=\textwidth]{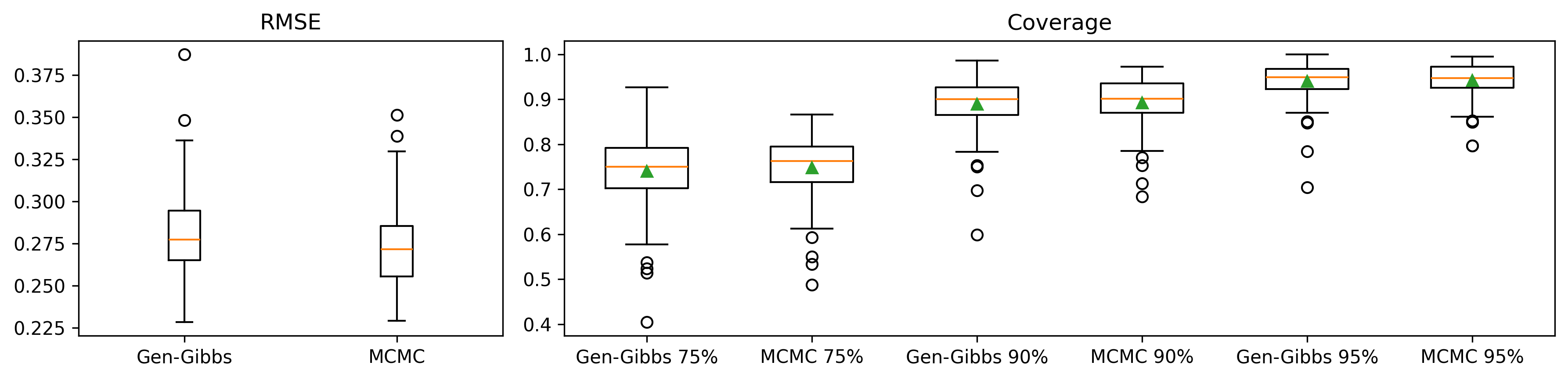}
    \caption{\footnotesize \textbf{Gaussian-SV model}. Performance across $100$ simulated scenarios summarized by box plots: RMSE of the latent trajectory estimates with Gen-Gibbs and traditional MCMC (left panel), and empirical coverage of the true trajectory at the 75\%, 90\% and 95\% credible interval levels (right panel). Mean coverages are also indicated in green.
}    \label{fig:RMSE_COVERAGES_GaussianSV}
\end{figure}

The QNNs for Gen-Gibbs sampling approach are trained on a large synthetic dataset consisting of $20$ million draws from the Gaussian-SV model. During the construction of the training dataset, we excluded samples with $\phi> 0.99$ to avoid issues arising from near-unit-root behavior. To demonstrate the robustness of the proposed procedure, the Gen-Gibbs sampler is subsequently evaluated on 100 independently simulated Gaussian-SV processes with true parameters $\mu = 0$, $\phi = 0.98$, and $\sigma_\eta = 0.1$. This particular parameterization is widely adopted in the empirical stochastic volatility literature, as it captures a realistic degree of persistence and moderate volatility dynamics characteristic of financial time series. For a single simulated process, the approximated posterior distributions of the model parameters are presented in Figure \ref{fig:one_process_SV}. The kernel density estimates based on the generated samples indicate that the Gen-Gibbs sampler yields parameter estimates closely aligned with those obtained using the MCMC algorithm of \citet{kim1998stochastic}. Such findings are further corroborated by Table \ref{tab:coverage_SV_par}, reporting the coverages, and Figure \ref{fig:Comparing_means_quantiles_across_sim_GaussianSV}, which displays box plots summarizing the distribution of parameter estimates across the simulated processes. Furthermore, the estimates of the latent volatility trajectories are highly consistent between the two methods, both in terms of their proximity to the true trajectories and their empirical coverage, as shown in Figure 
\ref{fig:RMSE_COVERAGES_GaussianSV}.

\begin{table}
\centering
\begin{tabular}{llrrr}
\toprule
 Parameter & Method & Coverage (0.75) & Coverage (0.90) & Coverage (0.95) \\
\midrule
\multirow[t]{2}{*}{$\mu$} & MCMC & 0.67 & 0.81 & 0.86 \\
 & Gen-Gibbs & 0.60 & 0.70 & 0.78 \\
\midrule
\multirow[t]{2}{*}{$\phi$} & MCMC & 0.68 & 0.83 & 0.88 \\
 & Gen-Gibbs & 0.73 & 0.90 & 0.94 \\
\midrule
\multirow[t]{2}{*}{$\sigma$} & MCMC & 0.67 & 0.81 & 0.84 \\
 & Gen-Gibbs & 0.73 & 0.90 & 0.94 \\
\bottomrule
\end{tabular}
\caption{\footnotesize \textbf{Gaussian SV model}. Coverages for $\mu$, $\phi$, and $\sigma_\eta$ using the MCMC strategy developed by \citet{kim1998stochastic} and Gen-Gibbs.}
\label{tab:coverage_SV_par}
\end{table}

In addition, we conduct a simulation study based on a heavy-tailed and asymmetric $\alpha$-stable SV model with unknown parameters $\{\mu,\phi,\alpha_y,\beta_y\}$. The new parameters controlling tail thickness and skewness are assigned uniform priors, $\alpha_y \sim \mathcal{U}(1, 2)$ and $\beta_y \sim \mathcal{U}(-1, 1)$.
This setting is particularly relevant as it illustrates how the proposed Gen-Gibbs algorithm can be easily embedded within a traditional MCMC framework to yield an amortized inference procedure. Because the latent process remains Gaussian, samples for $\mu$ and $\phi$ can be obtained by leveraging their full conditional distributions and a Metropolis–Hastings step, following the approach used by \citet{Vankov_2019}. In contrast, the parameters $\alpha_y$ and $\beta_y$ cannot be updated via standard Gibbs steps, since their full conditionals involve an intractable likelihood; hence, we employ a Gen-Gibbs update instead. The volatility of the latent process is fixed at $\sigma_\eta = 0.3$, capturing a more pronounced level of volatility typical of highly fluctuating markets. This simplification avoids the considerable computational burden of estimating this additional parameter, which offers little practical benefit in this context. Details on the MCMC steps and the choice of summaries for $\alpha_y$ and $\beta_y$ are reported in Appendix. 

Table \ref{tab:true_vs_est_with_x} reports the estimation results for the model parameters and the state sequence, averaged over $100$ simulated stochastic volatility processes under several parameter configurations. The results reported in the table indicate that the proposed approach delivers very accurate estimates, and this robust performance is consistently maintained across all configurations. In particular, the method performs well even in challenging scenarios characterized by heavy-tailed and strongly asymmetric, as well as in cases where $\alpha$ is close to 2 and $\beta$ becomes difficult to identify. 

A particularly appealing feature of the Gen-Gibbs approach, which we want to emphasize, is its flexibility. Once the pre-training phase described in Algorithm \ref{alg:ffbs-training} has been completed, the learned mapping functions can be reused within the Gen-Gibbs sampler (Algorithm \ref{alg:ffbs-testing}) to estimate models belonging to the same class of state-space systems at virtually no additional computational cost. In other words, both state and parameter estimation can be carried out rapidly by simply providing a new sequence of observed data. This property offers a substantial computational advantage over conventional ABC methods, which require a full re-estimation procedure for each new dataset.

\begin{table}[htbp]
\centering
\footnotesize
\setlength{\tabcolsep}{5pt}
\renewcommand{\arraystretch}{1.2}

\resizebox{\textwidth}{!}{
\begin{tabular}{cccccccccccc}
\toprule
\multicolumn{4}{c}{\textbf{True Parameters}} &
\multicolumn{4}{c}{\textbf{Posterior Estimates}} &
\multicolumn{4}{c}{\textbf{Accuracy and Coverages}} \\
\cmidrule(lr){1-4} \cmidrule(lr){5-8} \cmidrule(lr){9-12}
$\mu$ & $\phi$ & $\alpha$ & $\beta$ &
$\hat{\mu}$ & $\hat{\phi}$ & $\hat{\alpha}$ & $\hat{\beta}$ &
RMSE & 0.75 & 0.90 & 0.95 \\
\midrule
0 & 0.95 & 1.10 & 0 &
\est{-0.182}{-0.421}{0.153} &
\est{0.940}{0.922}{0.961} &
\est{1.122}{1.052}{1.234} &
\est{0.002}{-0.018}{0.027} &
0.722 & 0.709 & 0.868 & 0.926 \\
0 & 0.95 & 1.30 & 0.90 &
\est{0.150}{-0.109}{0.448} &
\est{0.942}{0.919}{0.962} &
\est{1.302}{1.220}{1.410} &
\est{0.716}{0.552}{0.863} &
0.505 & 0.735 & 0.890 & 0.944 \\
0 & 0.95 & 1.50 & -0.50 &
\est{0.036}{-0.195}{0.334} &
\est{0.936}{0.911}{0.958} &
\est{1.525}{1.433}{1.644} &
\est{-0.479}{-0.624}{-0.315} &
0.553 & 0.726 & 0.881 & 0.936 \\
0 & 0.95 & 1.75 & 0.50 &
\est{0.010}{-0.254}{0.289} &
\est{0.934}{0.906}{0.957} &
\est{1.746}{1.665}{1.835} &
\est{0.398}{0.181}{0.593} &
0.530 & 0.730 & 0.885 & 0.939 \\
0 & 0.95 & 1.95 & 0 &
\est{-0.025}{-0.269}{0.260} &
\est{0.932}{0.903}{0.955} &
\est{1.905}{1.860}{1.951} &
\est{0.035}{-0.340}{0.429} &
0.510 & 0.730 & 0.885 & 0.940 \\
\bottomrule
\end{tabular}}
\caption{\footnotesize
Estimated parameters ($\hat{\mu}$, $\hat{\phi}$, $\hat{\alpha}$, $\hat{\beta}$) with 95\% credible intervals below each estimate. 
The first block shows true parameter values, followed by estimates, and finally the RMSE and coverage metrics for the latent states sequence $(X_t)$ at levels 0.75, 0.90, and 0.95. 
All values are averages across 100 simulations.
}
\label{tab:true_vs_est_with_x}
\end{table}

\section{Empirical Study}\label{sec:Empirical Study}

Empirical studies of financial time series have consistently identified a set of recurring patterns, commonly referred to as \emph{stylized facts}, which any realistic asset pricing model should strive to replicate. These empirical regularities have been documented across a wide range of assets, asset classes, and markets, posing significant challenges to the classical assumptions of homoskedasticity and normally distributed returns that underpin traditional financial models, such as the Black-Scholes framework for option pricing \citep{black1973pricing}. In a seminal contribution, \citet{Cont01022001} systematically cataloged eleven such features. An $\alpha$-stable SV model is capable of reproducing several of these stylized facts, including: the absence of linear autocorrelation, conditional and unconditional heavy tails, gain/loss asymmetry, volatility clustering, and slow decay of autocorrelation in absolute returns. Moreover, the GBF framework introduced in this paper provides a foundation for developing and estimating more sophisticated models capable of capturing the remaining stylized facts identified by \citet{Cont01022001}.

As an example of a financial time series exhibiting these features, we consider the Short VIX Short-Term Futures ETF, issued by ProShares\footnote{https://www.proshares.com/our-etfs/strategic/svxy} and commonly referred to as SVXY. This product is designed to provide inverse exposure to the S\&P 500 VIX Short-Term Futures Index, a benchmark tracking a continuously rolled position in short-term VIX futures. As a result, SVXY delivers positive returns when market volatility declines and the VIX futures curve remains in contango.

In reality, market volatility is prone to abrupt fluctuations, which translate into sharp jumps in VIX futures prices and correspondingly large losses for inverse-volatility products such as SVXY. While periods of sustained calm in equity markets may result in smooth, positive returns, episodes of market stress can lead to rapid increases in volatility and significant losses. An episode of this type occurred in February 2018 during the so-called \emph{Volmageddon}, when a sudden spike in equity market volatility precipitated an unprecedented surge in VIX futures. In a single trading session, several short-volatility products experienced severe drawdowns, with some of them ultimately being liquidated (see, e.g., the closure of Credit Suisse's XIV ETN). In the aftermath of this event, numerous volatility-linked exchange-traded products, including SVXY, underwent substantial restructuring. In particular, ProShares reduced the fund’s exposure, such it passed from $-1\times$ to $-0.5\times$ leverage, with the objective of mitigating tail-risk. As a consequence of this adjustment, which we expect to have modified the return-generating process, we focus our analysis to the period from March 2014 through April 2018 (1,000 trading days), a window that includes the February 2018 shock while excluding the post-restructuring regime.

\begin{figure}[ht]
    \centering
\includegraphics[width=\textwidth]{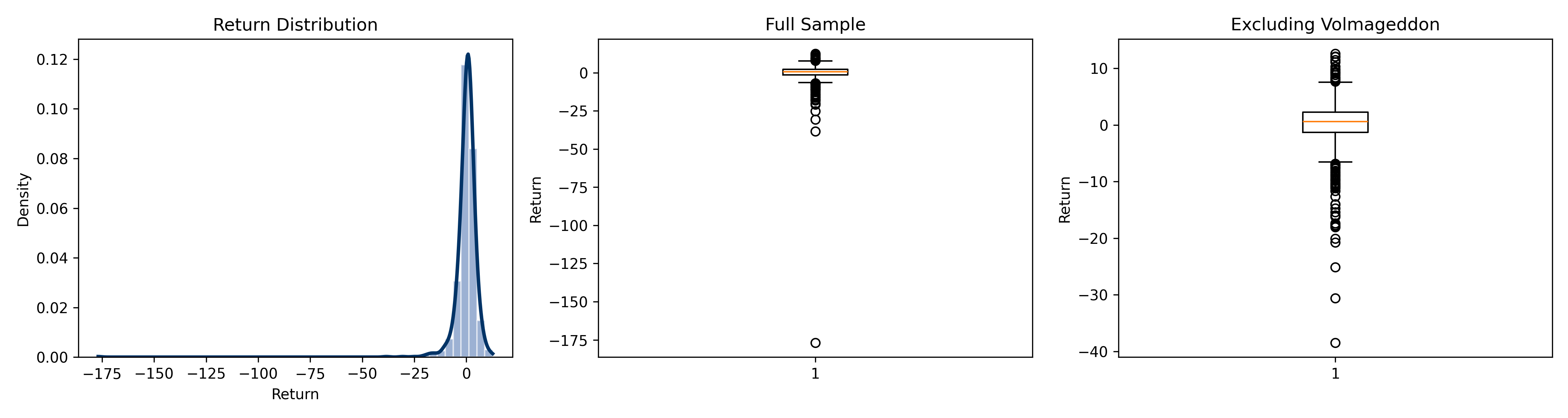}
    \caption{\footnotesize \textbf{SVXY}. Distribution of daily returns. The left panel shows the empirical histogram and kernel 
density estimate; the middle and right panels report boxplots for the full sample and for the 
sample excluding the observation associated with the 2018 ``Volmageddon'' shock.
}    \label{fig:SVXY_return_distribution}
\end{figure}

\begin{figure}[ht]
    \centering
\includegraphics[width=\textwidth]{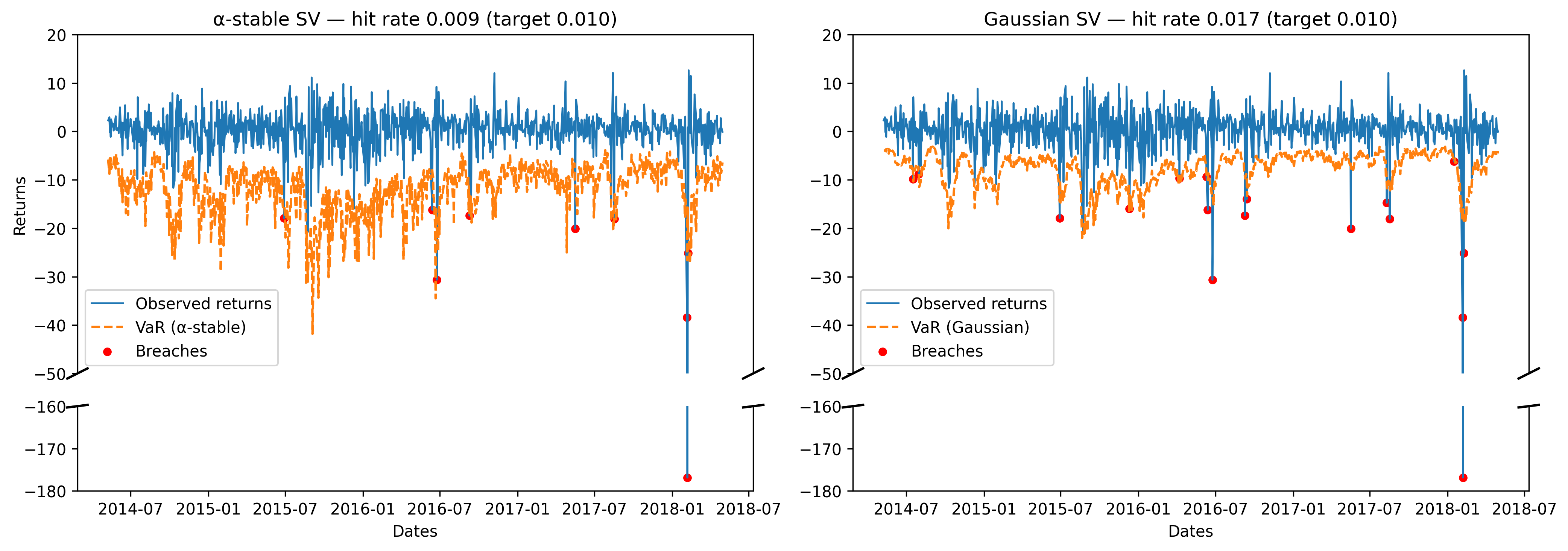}
    \caption{\footnotesize \textbf{SVXY}. VaR backtesting results at the 1\% tail probability. 
The figure compares the $\alpha$-stable and Gaussian specifications, showing 
observed returns, estimated VaR levels, and breach events. The $\alpha$-stable specification provides tighter alignment with observed 
tail losses and fewer violations than the Gaussian benchmark.
}    \label{fig:SVXY_VaR_plot_q001}
\end{figure}

We transform the original price series into demeaned daily log returns as follows:
\begin{equation*}
y_t = 100 \left( \log\left(\frac{P_t}{P_{t-1}}\right) 
- \frac{1}{T} \sum_{t=1}^{T} \log\left(\frac{P_t}{P_{t-1}}\right) \right),
\end{equation*}
where $P_t$ denotes the price of the asset at time $t$. 
Visual inspection of the resulting series reveals clear evidence of volatility clustering (Figure \ref{fig:SVXY_VaR_plot_q001}). In addition, the return distribution appears negatively skewed and exhibits pronounced
heavy tails, as illustrated in Figure \ref{fig:SVXY_return_distribution}.
These features are corroborated by standard diagnostic tests. To avoid distortions arising from the extreme 2018 volatility episode, we exclude that observation when computing test statistics. The Ljung-Box test \citep{ljung1978onameasure} provides no evidence of linear dependence in returns (20 lags, $\text{p-value}=0.89$). In contrast, when applied to absolute returns, the test yields $\text{p-value} \ll 0.001$, consistent with persistent volatility. Normality is strongly rejected by the Jarque-Bera test \citep{jarque1980efficiency} ($\text{p-value} \ll 0.001$), indicating substantial departure from Gaussian behavior. Additionally, the full sample exhibits extreme excess kurtosis ($410$) and pronounced negative skewness ($-16$), and these values remain elevated at $13$ and $-2$, respectively, even after excluding the 2018 shock.

We fit an $\alpha$-stable SV model to the return data, adopting the same priors, summaries, and simulation scheme as in the previous simulation study, and benchmark the results against a Gaussian-SV model. As reported in Table \ref{tab:empirical_results}, the posterior estimates of $\mu$ and $\phi$ are similar under both models. On the other hand, the estimated parameters of the $\alpha$-stable distribution, $\alpha$ and $\beta$, reveal pronounced  tail heaviness and negative skewness, capturing distributional characteristics that the Gaussian-SV model is structurally unable to accommodate. To assess the quality of these estimates, we use the smoothed latent-state distribution together with the joint posterior of the model parameters to generate artificial return sequences.
Specifically, we draw $500$
predictive samples from the posterior predictive distribution
\[
\tilde{y}_t \sim \int p\!\left(\tilde{y}_t \mid h_t,\alpha,\beta\right)\,
       p\!\left(h_t \mid y_{1:T}, \mu, \phi\right)\,
       p\!\left(\mu,\phi,\alpha,\beta \mid y_{1:T}\right)\,
       dh_t\, d\mu\, d\phi\, d\alpha\, d\beta .
\]
In practice, under the Gen-Gibbs sampler, $y_{1:T}$ is replaced by summary statistics $S(y_{1:T})$ in the conditioning distributions, as discussed in Section \ref{sec:Gen-Gibbs}. Using these simulated predictive draws, we begin evaluating model performance by computing the predictive coverage values reported in Table \ref{tab:empirical_results}.
This check is important because it ensures that the superior tail performance of the $\alpha$-stable specification is not achieved merely by producing overly diffuse predictive distributions. If the $\alpha$-stable model were simply inflating uncertainty, its empirical coverage rates would significantly exceed their nominal levels. Instead, the predictive coverage results confirm that both models are, on average, well calibrated.

\begin{table}[htbp]
\centering
\footnotesize
\setlength{\tabcolsep}{5pt}
\renewcommand{\arraystretch}{1.2}

%\resizebox{\textwidth}{!}{
\begin{tabular}{lccccccc}
\toprule
\multicolumn{1}{c}{\textbf{Model}} &
\multicolumn{4}{c}{\textbf{Posterior Estimates}} &
\multicolumn{3}{c}{\textbf{Predictive Coverage}} \\
\cmidrule(lr){2-5} \cmidrule(lr){6-8}
 & $\hat{\mu}$ & $\hat{\phi}$ & $\hat{\alpha}$ & $\hat{\beta}$ &
0.75 & 0.90 & 0.95 \\
\midrule
Gaussian-SV &
\est{0.265}{-0.359}{0.889} &
\est{0.975}{0.961}{0.988} &
\est{2.000}{2.000}{2.000} &
\est{0.000}{0.000}{0.000} &
0.704 & 0.888 & 0.950 \\
\midrule
$\alpha$-Stable-SV &
\est{0.331}{-0.285}{0.937} &
\est{0.970}{0.953}{0.985} &
\est{1.746}{1.647}{1.840} &
\est{-0.735}{-0.974}{-0.507} &
0.696 & 0.894 & 0.962 \\
\bottomrule
\end{tabular}%}
\caption{\footnotesize Posterior parameters' estimates and empirical predictive coverage for the return series.
Credible intervals reported in parentheses.}
\label{tab:empirical_results}
\end{table}

Coverage alone cannot reveal tail behavior. To properly assess models' tail performance, we focus on the Monte Carlo estimates of the Value at Risk (VaR) and Expected Shortfall (ES) at level $q$:
\[
\widehat{\text{VaR}}_t(q)
= \widehat{F}^{-1}_t(q), \quad \widehat{\text{ES}}_t(q)= \mathbb{E}_{\widehat{F}_t}[\tilde{y}_t\mid \tilde{y}_t\leq \widehat{\text{VaR}}_t(q)]
\]
where $\widehat{F}_t(x)=M^{-1} \sum_{j=1}^{M}
\mathbb{I}\{ \tilde{y}_t^{(j)} \le x \}$. For each confidence level $q\in\{0.05,0.01,0.005,0.001\}$, we evaluate the adequacy of the VaR estimates using standard backtesting metrics (see \cite{kupiec1995}, and \cite{christoffersen1998}). In particular, we compute the hit rate, given by $\text{hit rate}=T^{-1}\sum_{t=1}^{T}\mathbb{I}\{y_t\leq\widehat{\text{VaR}}_t(q)\} $, measuring the empirical frequency of VaR violations (also known as breaches). To evaluate whether the observed frequency of violations is consistent with the expected frequency, we employ the Unconditional Coverage Likelihood Ratio ($LR_{uc}$) test. We also verify that exceedances occur independently over time via a Independence Likelihood Ratio ($LR_{ind}$) test, which detects serial dependence in violations. Lastly, we consider the Conditional Coverage test, $LR_{cc}=LR_{un}+LR_{ind}$, for a joint assessment of both correct violation frequency and temporal independence. Backtesting results are reported in Table \ref{tab:Var_backtesting}.
\begin{table}[t!]
\centering
\setlength{\tabcolsep}{6pt}
\begin{tabular}{c l c c c c c}
\toprule
$q$ & Model & Hit Rate & $LR_{\text{uc}}$ & $LR_{\text{ind}}$ & $LR_{\text{cc}}$ & ES \\
\midrule
\multirow{2}{*}{0.050}
 & GaussianSV     & 0.049 & 0.023 & 0.999 & 1.022 & $-6.764$ \\
 & $\alpha$-StableSV & 0.048 & 0.090 & 1.152 & 1.241 & $-11.392$ \\
\midrule
\multirow{2}{*}{0.010}
 & GaussianSV     & 0.017 & 4.077 & 4.682 & 8.759 & $-9.303$ \\
 & $\alpha$-StableSV & 0.009 & 0.107 & 3.386 & 3.492 & $-25.974$ \\
\midrule
\multirow{2}{*}{0.005}
 & GaussianSV     & 0.013 & 8.892 & 2.004 & 10.896 & $-10.049$ \\
 & $\alpha$-StableSV & 0.005 & 0.000 & 5.839 & 5.839  & $-33.758$ \\
\midrule
\multirow{2}{*}{0.001}
 & GaussianSV     & 0.008 & 19.306 & 3.857 & 23.163 & $-11.406$ \\
 & $\alpha$-StableSV & 0.002 & 0.772  & 10.271 & 11.043 & $-58.034$ \\
\bottomrule
\end{tabular}
\caption{\footnotesize VaR Backtesting Results Across Confidence Levels. For each model, the table reports empirical hit rates, Unconditional Coverage LR test,
Independence LR test, Conditional 
Coverage test, and average Expected Shortfall over time. 
More negative ES values indicates larger tail losses.}\label{tab:Var_backtesting}
\end{table}

At moderate quantiles ($q=0.05$), the models perform similarly; however, as  the quantile become more extreme, the Gaussian SV model increasingly underestimates downside risk. This is evident from systematically higher hit rates relative to $q$ and the rapid inflation in the $LR_{uc}$ and $LR_{ind}$ statistics, which indicate both miscalibrated tail probabilities and clustering of violations. In contrast, the $\alpha$-stable specification maintains hit rates that remain close to the nominal levels across all quantiles, and its corresponding test statistics remain comparatively small, suggesting superior calibration of extreme events \footnote{We note that the elevated $LR_{ind}$
 statistic at $q=0.001$ is mainly driven by the very small number of breaches (only two) that happen to occur close together. This artificial clustering mechanically inflates the test statistic and does not reflect true dependence in violations.}. Moreover,  
when extreme losses occur, the $\alpha$-stable SV model anticipates more pronounced drawdowns than the Gaussian benchmark, as evidenced by its substantially more negative ES values. This reflects a more realistic assessment of the severity of tail losses. Such enhanced sensitivity to extreme downside risk is particularly relevant in light of recent market stress episodes, including the volatility shock of February 2018. 

The backtesting results are illustrated in Figure \ref{fig:SVXY_VaR_plot_q001}, where observed returns are shown together with the model-implied VaR values at $q=0.01$ (equivalently, the 99\% VaR). The highlighted points indicate instances where realized returns breached the VaR threshold. The $\alpha$-stable SV model records 9 breaches over $T=1000$ observations (hit rate 0.009), with violations appearing relatively isolated over time. On the other hand, the Gaussian SV model exhibits 17 breaches (hit rate 0.017), with violations that tend to occur in cluster, reflecting its poorer ability in modelling downside risk.

Since the returns generated from the posterior predictive distribution successfully
capture the dynamics of actual returns, the results provide strong evidence that the proposed estimation framework delivers reliable inference for latent states and parameters even in environments where the likelihood is analytically intractable and standard computational Bayesian techniques cannot be applied. By overcoming these limitations, our approach significantly expands the class of state-space models that can be estimated within the Bayesian paradigm. This, in turn, enables the use of richer and more realistic specifications that incorporate features often neglected for computational convenience, such as abrupt negative market movements and other forms of extreme behavior in financial returns.

\section{Discussion}

In this article, we have presented a novel framework for filtering and parameter learning in state-space models.
Our methodology proves particularly valuable in situations where the model specification induces complex systems of priors and likelihoods that make conventional MCMC and SMC methods difficult or even impossible to apply, such as in the case of intractable state-space models. We demonstrate that as long as simulation from the model is feasible, estimation of the latent states remains possible regardless of the noise distributions or the functional forms of the transition and observation equations, through our Gen-Filter procedure. A Pre-Trained variant is also provided, offering an efficient alternative that we recommend  for applications requiring rapid updates of the filtering distribution -- such as in real-time object tracking and high-frequency volatility monitoring -- when the latent process can be reasonably assumed to be stationary and the emission distribution time-homogeneous.
 Both approaches demonstrated superior performance compared to the benchmark ABC-PF, achieving higher accuracy, better coverage, and closer proximity to the true posterior.

For scenarios in which model parameters are unknown and must be inferred jointly with the latent states, we develop the Gen-Gibbs sampler. This method provides a fully density-free sampling scheme that enables Bayesian inference in models with complex hierarchical architectures and intractable densities, among which some classes of state-space models are a particular instance. When standard MCMC techniques can be applied, the Gen-Gibbs sampler achieves comparable results, demonstrating its validity and robustness as a general inference tool.

Our GBF framework is broadly applicable across scientific domains, as it can be utilized for any model that admits a state-space representation. In this work, we focused on financial applications, specifically on volatility estimation, an enduring challenge in the filtering literature. The complex dynamics observed in financial returns, as documented by \citet{cont2023tailganlearningsimulatetail}, are difficult to capture using simple models with restrictive assumptions, though such models remain useful benchmarks. Our framework opens the door to more flexible and realistic modeling. In particular, we showed that adopting $\alpha$-stable distributions can capture some well-known stylized facts of financial returns and enhance volatility estimation. We invite economists and quant researchers to further explore our framework and extend our analysis to account for richer dynamics such as jumps, leverage effects, and other nonlinearities. All the materials required to replicate our results are available in the first author’s GitHub repository\footnote{The repository will be published once the Arxiv is submitted}.

Researchers adopting our framework should be aware that the methodology may involve a considerable computational cost during the training of the deep learning models. However, this burden can be greatly alleviated by employing high-performance computing resources -- such as GPUs -- and by exploiting parallel processing. Nonetheless, our results show that strong performance can still be achieved using standard computational setups. Moreover, in the context of the Pre-Trained Gen Filter and Gen-Gibbs sampling, this computational cost is incurred only once; after training, both filtering and parameter learning proceed at speeds comparable to those of classical PF and PMCMC methods. In particular, the maps learned during the training phase can be easily reused to estimate an entire class of state-space models simply by supplying new data. This represents a particularly appealing advantage over conventional ABC approaches, in which the estimation procedure must be reinitialized whenever the dataset changes.

Another relevant point is that our implementation of the GBF framework mainly relies on QNNs for learning the inverse CDF maps used to generate samples from target distributions. As such, it is subject to the intrinsic limitations of this technique. Although the use of QNNs is not strictly necessary and other implicit quantile methods could be adopted, the training of neural networks in general requires careful tuning and validation.

In this paper, we have focused on the univariate case. As part of future research, we aim to extend the proposed framework to multidimensional state-space models, where both $Y_t$
 and 
$X_t$  are vector-valued. This direction is motivated by the recent work of \citet{kim2025deepgenerativequantilebayes}, which generalizes the GBC approach to multivariate settings.

\printbibliography

\newpage
\appendix

\section{Non-stationary data}

We now present the version of the Pre-Trained Gen-Filter which constructs summaries using information from the previous filtering step. 
At each time step, we compute the filtering distribution and extract its mean and variance, denoted by $m_{t-1}$ and $v_{t-1}$, respectively. 
These statistics are then used as inputs to a learned mapping that generates samples from the posterior distribution 
$p(x_t \mid y_t, m_{t-1}, v_{t-1})$, providing an approximation of the current filtering distribution. We visually illustrate the performance of the proposed method in comparison with the Kalman filter, 
considering both stationary and non-stationary latent processes.

\begin{figure}[H]
    \centering
    \begin{subfigure}{1\textwidth}
        \centering
        \includegraphics[width=0.7\textwidth]{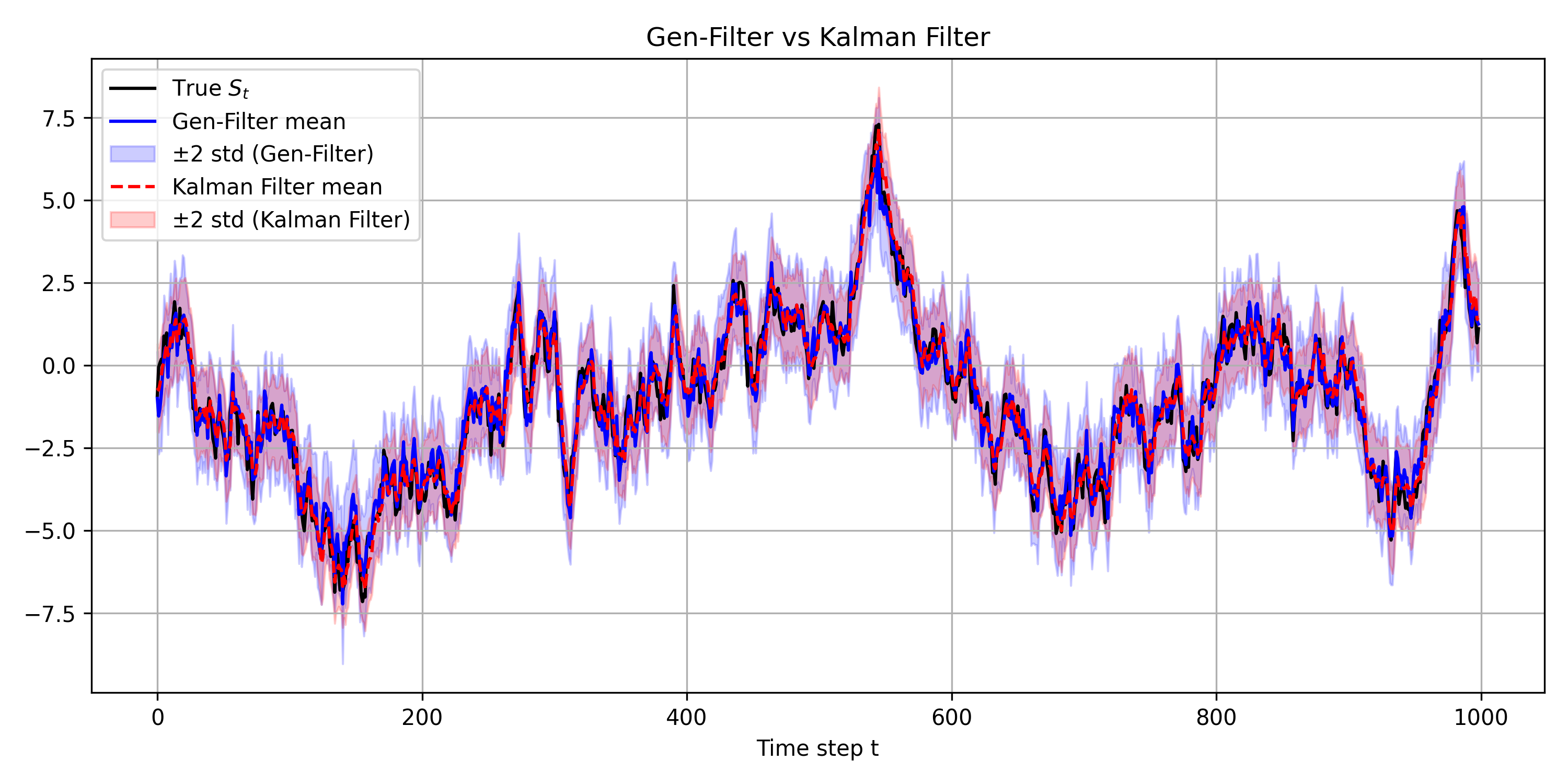}
        \caption{\footnotesize Pre-Trained Gen-Filter applied to a       
        \emph{stationary} latent process.}
        \label{fig:AI-vs-Kalman-a}
    \end{subfigure}
    
    \vspace{1em}
    
    \begin{subfigure}{1\textwidth}
        \centering
        \includegraphics[width=0.7\textwidth]{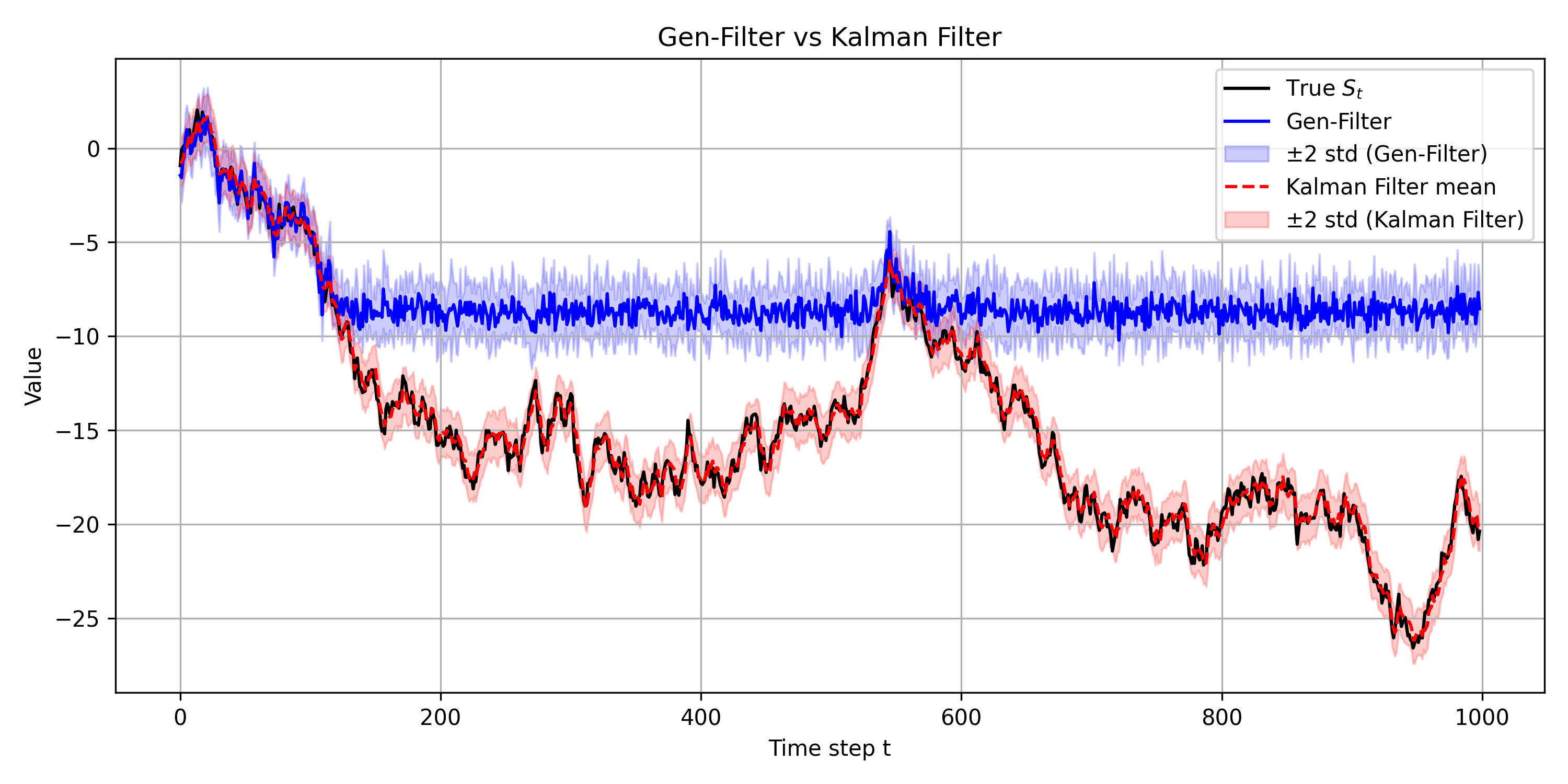}
        \caption{ \footnotesize Pre-Trained Gen-Filter applied to a       
        \emph{non stationary} latent process.}
        \label{fig:AI-vs-Kalman-b}
    \end{subfigure}
    
    \caption{\footnotesize \textbf{LG model}. Comparison between the Pre-trained Gen-Filter and the Kalman Filter. 
    True latent states are shown in black. 
    The upper and lower panels illustrate that while the Gen-Filter performs comparably to the Kalman Filter within the training range, it fails to generalize to non-stationary latent dynamics, 
    highlighting its limited ability to generalize outside the support of the training set. }
    \label{fig:AI-vs-Kalman-combined}
\end{figure}

\section{Simulation Study -- Additional Results }

\subsection{Linear Gaussian Model}

\begin{figure}[H]
    \centering
    \includegraphics[width=\textwidth]{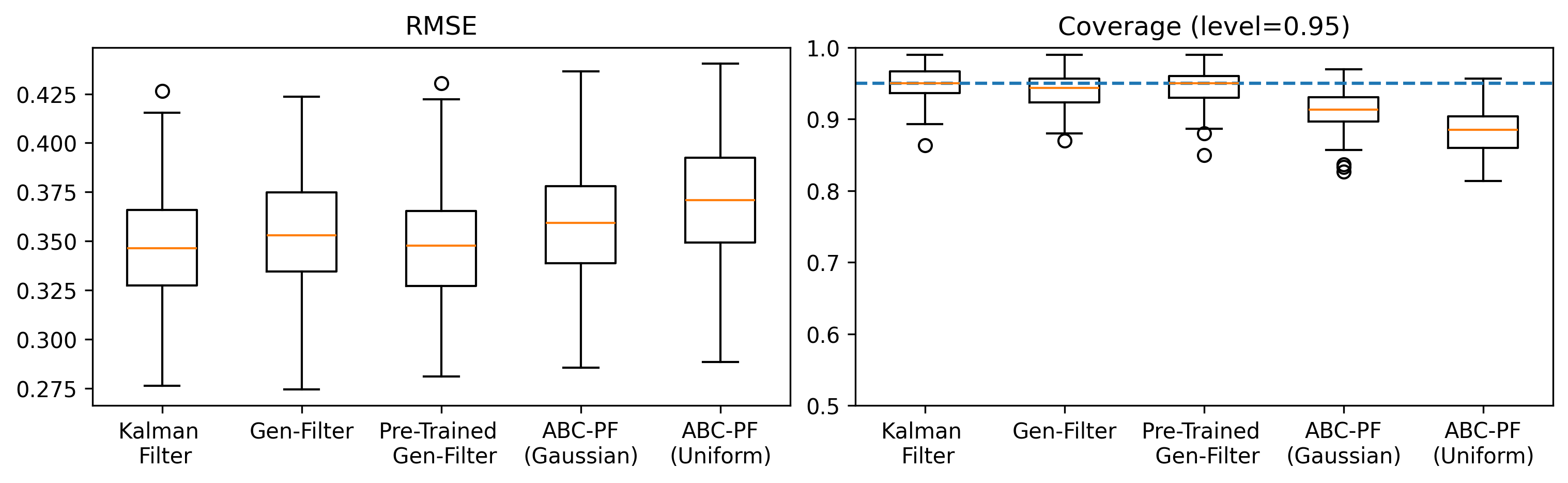}
    \caption{\footnotesize \textbf{Linear Gaussian model.} RMSE and 95\% coverage over 100 simulations for the Gen-Filter. 
    Averages are reported in Table~\ref{tab:RMSE_coverage_means}.}
    \label{fig:plot_RMSE_coverage}
\end{figure}

\begin{table}[H]
\centering
\resizebox{\textwidth}{!}{
\begin{tabular}{lrrrr}
\toprule
Method & RMSE & Coverage (0.75) & Coverage (0.90) & Coverage (0.95) \\
\midrule
Kalman Filter & 0.347 & 0.750 & 0.898 & 0.949 \\
Gen-Filter (30 lags) & 0.347 & 0.741 & 0.903 & 0.948 \\
Gen-Filter (20 lags) & 0.347 & 0.750 & 0.902 & 0.951 \\
Gen-Filter (10 lags) & 0.348 & 0.737 & 0.893 & 0.945 \\
ABC-PF (Gaussian) & 0.360 & 0.709 & 0.859 & 0.911 \\
ABC-PF (Uniform) & 0.370 & 0.691 & 0.831 & 0.883 \\
\bottomrule
\end{tabular}
}
\caption{\footnotesize \textbf{Linear Gaussian model}. RMSE and Coverage at levels 0.75, 0.90, and 0.95 are reported as averages across 100 simulations. The table highlights the performance of the Pre-Trained Gen-Filter for different lag specifications. }
\label{tab:RMSE_coverage_means_PT_Appendix}
\end{table}

\begin{table}[H]
\centering
\resizebox{\textwidth}{!}{
\begin{tabular}{lccccc}
\toprule
Distance from Ground Truth & Wasserstein & MMD  & Energy & Mean Diff & Std Diff \\ 
\midrule
Gen-Filter (30 lags)       & 0.040       & 0.001 & 0.004  & 0.033     & 0.015    \\
Gen-Filter (20 lags)       & 0.034       & 0.001 & 0.003  & 0.028     & 0.012    \\
Gen-Filter (10 lags)       & 0.036 & 0.001 & 0.003 & 0.030 & 0.013    \\
ABC-PF (Gaussian)          & 0.085 & 0.017 & 0.039 & 0.068 & 0.035    \\
ABC-PF (Uniform)           & 0.112 & 0.021 & 0.048 & 0.090 & 0.047    \\ 
\bottomrule
\end{tabular}
}
\caption{\footnotesize \textbf{Linear Gaussian model}. Average distance metrics between filtering distributions of each method and the Kalman Filter (ground truth), computed over 100 simulations. The table highlights the performance of the Pre-Trained Gen-Filter for different lag specifications.}
\label{tab:RMSE_coverage_means_PT_Appendix}
\end{table}

\subsection{Stochastic Volatility Model}

\begin{table}[H]
\centering
\begin{subtable}{\textwidth}
 \centering
 \caption{Gen-Filter}
        \resizebox{\textwidth}{!}{
        \begin{tabular}{llrrrr}
\toprule
Model & Method & RMSE & Coverage (0.75) & Coverage (0.90) & Coverage (0.95) \\
\midrule
\multirow{3}{*}{GaussianSV}
& PF & 0.488 & 0.757 & 0.904 & 0.952 \\
& Gen-Filter & 0.508 & 0.766 & 0.903 & 0.941 \\
& ABC-PF (Gaussian) & 0.524 & 0.684 & 0.831 & 0.886 \\
& ABC-PF (Uniform) & 0.572 & 0.633 & 0.773 & 0.825 \\
\midrule
\multirow{3}{*}{CauchySV}
& PF & 0.654 & 0.746 & 0.896 & 0.945 \\
 & Gen-Filter & 0.687 & 0.585 & 0.783 & 0.854 \\
& ABC-PF (Gaussian) & 0.912 & 0.426 & 0.541 & 0.595 \\
& ABC-PF (Uniform) & 0.923 & 0.418 & 0.525 & 0.567 \\
\midrule
\multirow{4}{*}{$\alpha$-StableSV}
& Ground  Truth & 0.540 & 0.740 & 0.889 & 0.937 \\
& Gen-Filter ($N=10^5$) & 0.578 & 0.705 & 0.853 & 0.900 \\
& Gen-Filter & 0.576 & 0.648 & 0.844 & 0.911 \\
& ABC-PF(Gaussian) & 0.617 & 0.615 & 0.758 & 0.816 \\
& ABC-PF (Uniform) & 0.657 & 0.567 & 0.698 & 0.749 \\
\bottomrule
\end{tabular}
        }
        \label{tab:RMSE_coverage_means_SV}
\end{subtable}
\begin{subtable}{\textwidth}
\centering
\caption{Pre-Trained Gen-Filter}
\resizebox{\textwidth}{!}{
\begin{tabular}{llrrrr}
\toprule
Model & Method & RMSE & Coverage (0.75) & Coverage (0.90) & Coverage (0.95) \\
\midrule
\multirow{6}{*}{GaussianSV}
 & PF & 0.488 & 0.752 & 0.903 & 0.952 \\
 & Gen-Filter (30 lags) & 0.494 & 0.746 & 0.890 & 0.941 \\
 & Gen-Filter (20 lags) & 0.493 & 0.742 & 0.891 & 0.945 \\
 & Gen-Filter (10 lags) & 0.510 & 0.750 & 0.889 & 0.940 \\
 & ABC-PF (Gaussian)    & 0.532 & 0.680 & 0.828 & 0.880 \\
 & ABC-PF (Uniform)     & 0.574 & 0.628 & 0.770 & 0.823 \\
\midrule
\multirow{6}{*}{CauchySV}
 & PF & 0.641 & 0.751 & 0.899 & 0.951 \\
 & Gen-Filter (30 lags) & 0.729 & 0.745 & 0.898 & 0.942 \\
 & Gen-Filter (20 lags) & 0.743 & 0.753 & 0.896 & 0.943 \\
 & Gen-Filter (10 lags) & 0.753 & 0.738 & 0.893 & 0.946 \\
 & ABC-PF (Gaussian)    & 0.882 & 0.433 & 0.554 & 0.607 \\
 & ABC-PF (Uniform)     & 0.959 & 0.393 & 0.497 & 0.542 \\
\midrule
\multirow{6}{*}{$\alpha$-StableSV}
& Ground  Truth & 0.536 & 0.742 & 0.890 & 0.938 \\
& Gen-Filter (30 lags) & 0.540 & 0.754 & 0.897 & 0.947 \\
& Gen-Filter (20 lags) & 0.540 & 0.745 & 0.899 & 0.954 \\
& Gen-Filter (10 lags) & 0.562 & 0.750 & 0.898 & 0.951 \\
& ABC-PF (Gaussian) & 0.627 & 0.606 & 0.749 & 0.804 \\
& ABC-PF (Uniform) & 0.657 & 0.563 & 0.697 & 0.748 \\
\bottomrule
\end{tabular}
}
\label{tab:RMSE_coverage_means_PT_SV}
\end{subtable}
\caption{\footnotesize RMSE and Coverage at levels 0.75, 0.90, and 0.95 are reported as averages across 100 simulations. The number of samples $N$ is equal to $1000$ when not specified.}
\label{tab:RMSE_coverage_means_panels_SV}
\end{table}

\begin{table}[H]
\centering
\begin{subtable}{\textwidth}
 \centering
 \caption{Gen-Filter}
        \resizebox{\textwidth}{!}{
        \begin{tabular}{llccccc}
\toprule
Model & Method & Wasserstein & MMD & Energy & Mean Diff & Std Diff \\
\midrule
\multirow{2}{*}{GaussianSV} 
& Gen-Filter  & 0.140 & 0.067 & 0.142 & 0.122 & 0.042 \\
& ABC-PF (Gaussian) & 0.167 & 0.045 & 0.094 & 0.140 & 0.069 \\
& ABC-PF (Uniform) & 0.240 & 0.069 & 0.145 & 0.210 & 0.090 \\
\midrule
\multirow{3}{*}{CauchySV} 
% & Gen-Filter ($N=10^5$) & 0.145 & 0.044 & 0.098 & 0.118 & 0.046 \\
& Gen-Filter & 0.225 & 0.065 & 0.133 & 0.169 & 0.144 \\
& ABC-PF (Gaussian) & 0.543 & 0.168 & 0.371 & 0.486 & 0.244 \\
& ABC-PF (Uniform) & 0.580 & 0.183 & 0.408 & 0.521 & 0.256 \\
\midrule
\multirow{5}{*}{$\alpha$-StableSV}
& Gen-Filter ($N=10^5$) & 0.213 & 0.034 & 0.074 & 0.207 & 0.040 \\
& Gen-Filter & 0.203 & 0.031 & 0.070 & 0.176 & 0.078 \\
& ABC-PF (Gaussian) & 0.264 & 0.069 & 0.148 & 0.230 & 0.107 \\
& ABC-PF (Uniform) & 0.334 & 0.096 & 0.207 & 0.295 & 0.130 \\
\bottomrule
\end{tabular}
        }
        \label{tab:PF_vs_methods}
\end{subtable}
\begin{subtable}{\textwidth}
\centering
\caption{Pre-Trained Gen-Filter}
\resizebox{\textwidth}{!}{
\begin{tabular}{llccccc}
\toprule
Model & Method & Wasserstein & MMD & Energy & Mean Diff & Std Diff \\
\midrule
\multirow{5}{*}{GaussianSV} 
 & Gen-Filter (10 lags) & 0.120 & 0.056 & 0.118 & 0.115 & 0.023 \\
 & Gen-Filter (20 lags) & 0.069 & 0.045 & 0.093 & 0.060 & 0.023 \\
 & Gen-Filter (30 lags) & 0.069 & 0.045 & 0.093 & 0.060 & 0.024 \\
 & ABC-PF (Gaussian)   & 0.172 & 0.045 & 0.096 & 0.147 & 0.068 \\
 & ABC-PF (Uniform)    & 0.244 & 0.070 & 0.149 & 0.214 & 0.089 \\
\midrule
\multirow{5}{*}{CauchySV} 
 & Gen-Filter (10 lags) & 0.324 & 0.078 & 0.191 & 0.309 & 0.102 \\
 & Gen-Filter (20 lags) & 0.302 & 0.068 & 0.165 & 0.288 & 0.093 \\
 & Gen-Filter (30 lags) & 0.284 & 0.067 & 0.160 & 0.272 & 0.079 \\
 & ABC-PF (Gaussian)   & 0.535 & 0.165 & 0.364 & 0.480 & 0.239 \\
 & ABC-PF (Uniform)    & 0.607 & 0.198 & 0.445 & 0.550 & 0.257 \\
\midrule
\multirow{5}{*}{$\alpha$-StableSV}
& Gen-Filter (10 lags) & 0.169 & 0.026 & 0.056 & 0.161 & 0.040 \\
& Gen-Filter (20 lags) & 0.113 & 0.015 & 0.033 & 0.102 & 0.035 \\
& Gen-Filter (30 lags) & 0.113 & 0.015 & 0.032 & 0.103 & 0.033 \\
& ABC-PF (Gaussian) & 0.269 & 0.065 & 0.139 & 0.235 & 0.109 \\
& ABC-PF (Uniform) & 0.334 & 0.088 & 0.189 & 0.295 & 0.132 \\
\bottomrule
\end{tabular}
}
\label{tab:PF_vs_methods_PT}
\end{subtable}
\caption{\footnotesize Average distance metrics between filtering distributions of each method and the PF (ground truth), computed over 100 simulations. For the $\alpha$-stable SV model, the ground truth is given by the ABC-PF with $10^5$ particles. The number of samples $N$ is equal to $1000$ when not specified.}
\label{tab:PF_vs_methods_panels}
\end{table}

\begin{figure}[H]
  \centering

  \begin{subfigure}{\linewidth}
    \centering
    \includegraphics[width=0.9\linewidth]{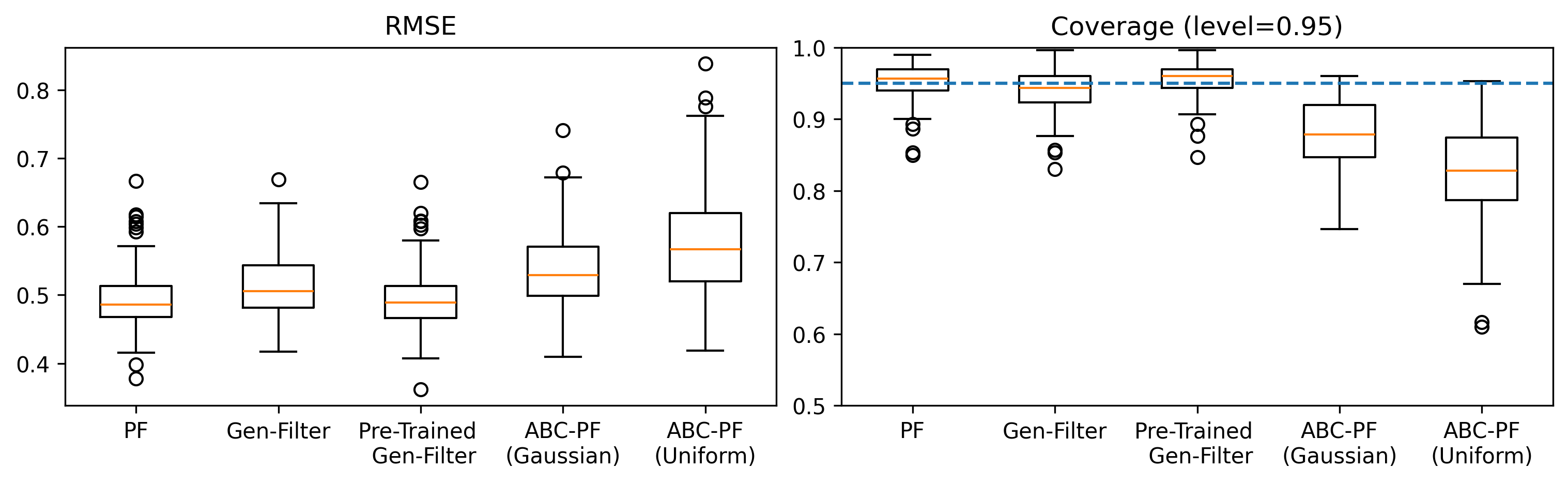} 
    \caption{Gaussian SV}
    \label{fig:Gaussian_rmse_coverage_95_SV}
  \end{subfigure}\par\medskip

  \begin{subfigure}{\linewidth}
    \centering
    \includegraphics[width=0.9\linewidth]{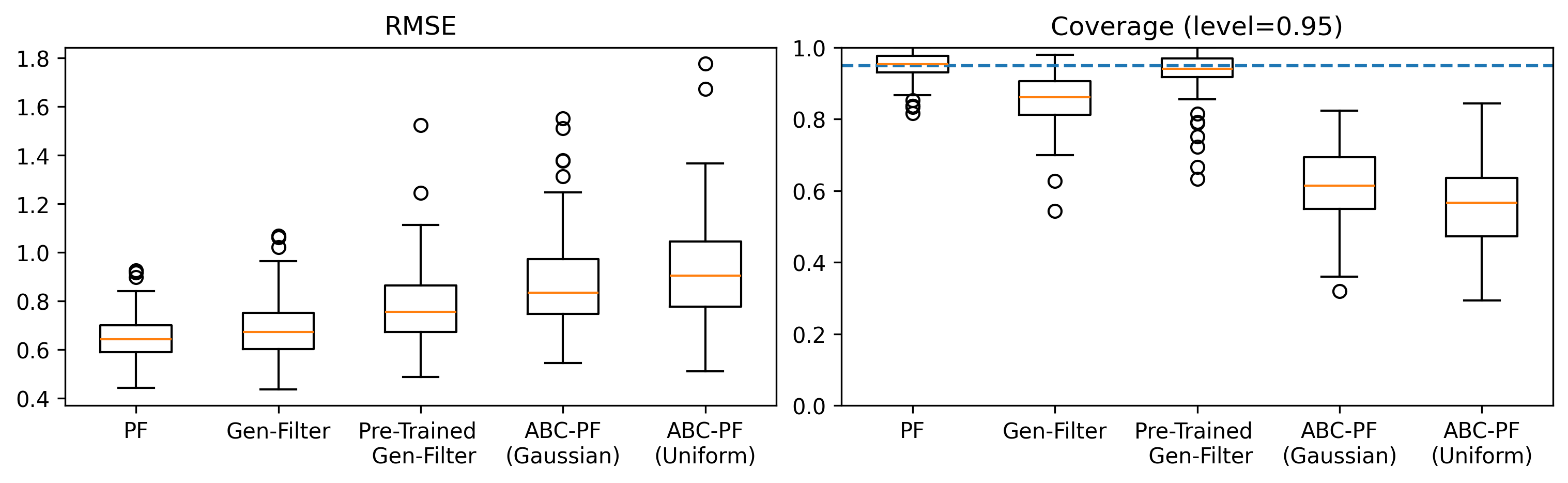}
    \caption{Cauchy SV}
    \label{fig:Cauchy_rmse_coverage_95_SV}
  \end{subfigure}\par\medskip

  \begin{subfigure}{\linewidth}
    \centering
    \includegraphics[width=0.9\linewidth]{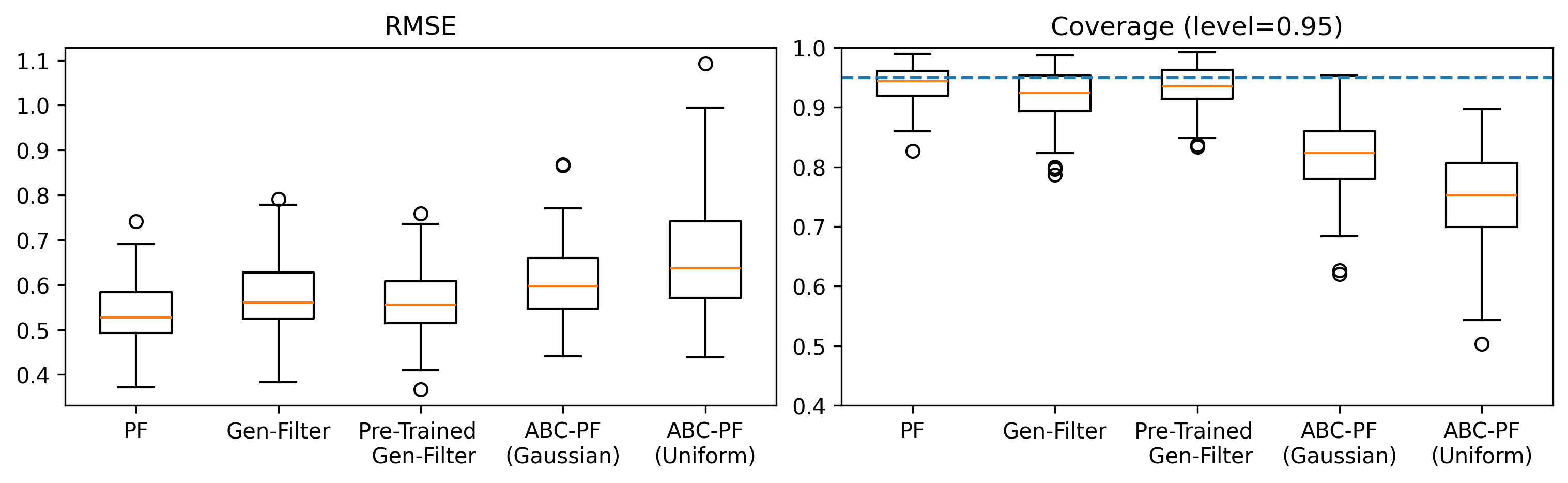}
    \caption{$\alpha$-Stable SV}
    \label{fig:AlphaStable_rmse_coverage_95_SV}
  \end{subfigure}

  \caption{\footnotesize \textbf{SV model}. RMSE and 95\% coverage over $100$ simulations. 
    Averages are reported in Table \ref{tab:RMSE_coverage_means_SV}. }
  \label{fig:three-by-one}
\end{figure}

\begin{figure}[H]
  \centering

  \begin{subfigure}{\linewidth}
    \centering
    \includegraphics[width=0.9\linewidth]{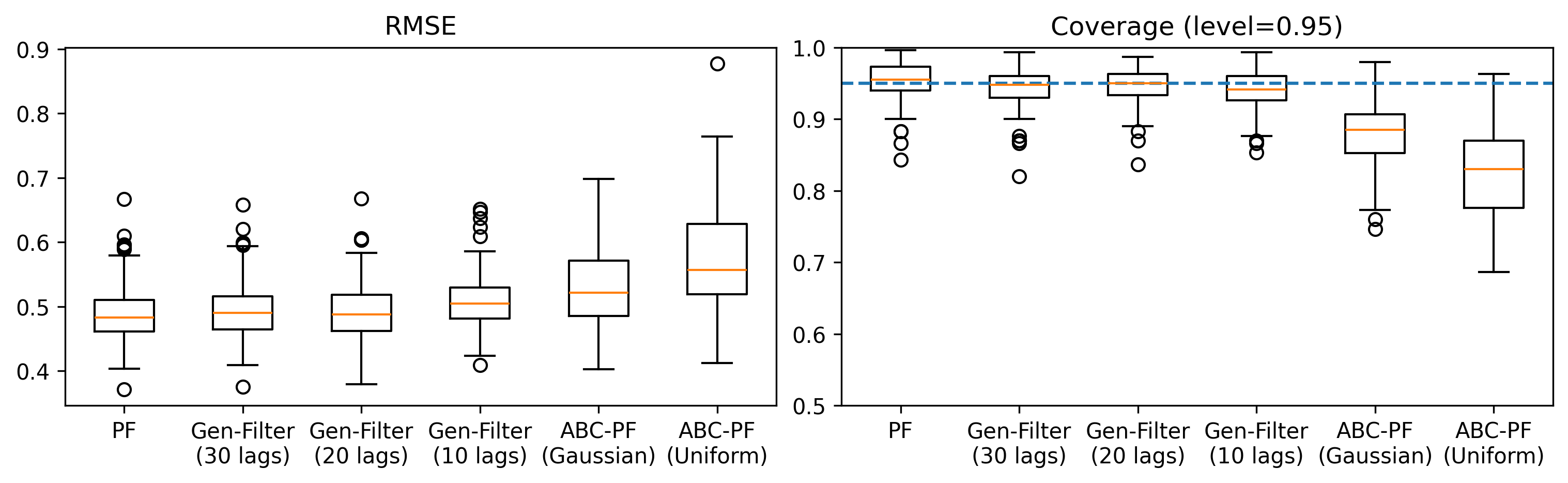} 
    \caption{Gaussian SV}
    \label{fig:Gaussian_rmse_coverage_95}
  \end{subfigure}\par\medskip

  \begin{subfigure}{\linewidth}
    \centering
    \includegraphics[width=0.9\linewidth]{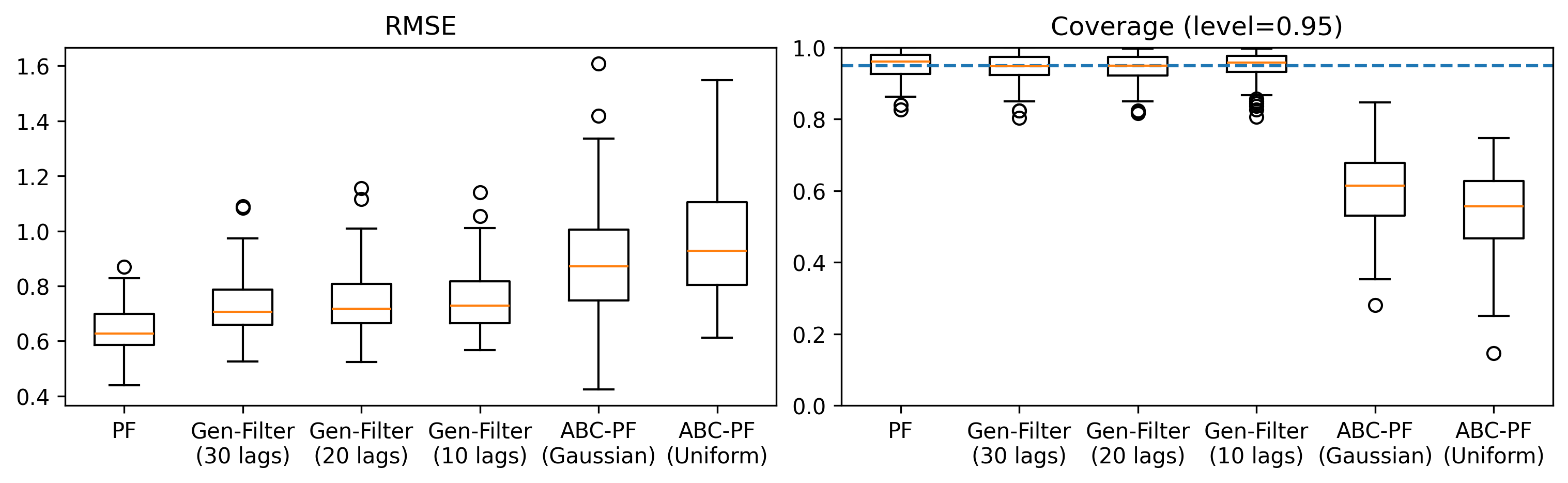}
    \caption{Cauchy SV}
    \label{fig:Cauchy_rmse_coverage_95}
  \end{subfigure}\par\medskip

  \begin{subfigure}{\linewidth}
    \centering
    \includegraphics[width=0.9\linewidth]{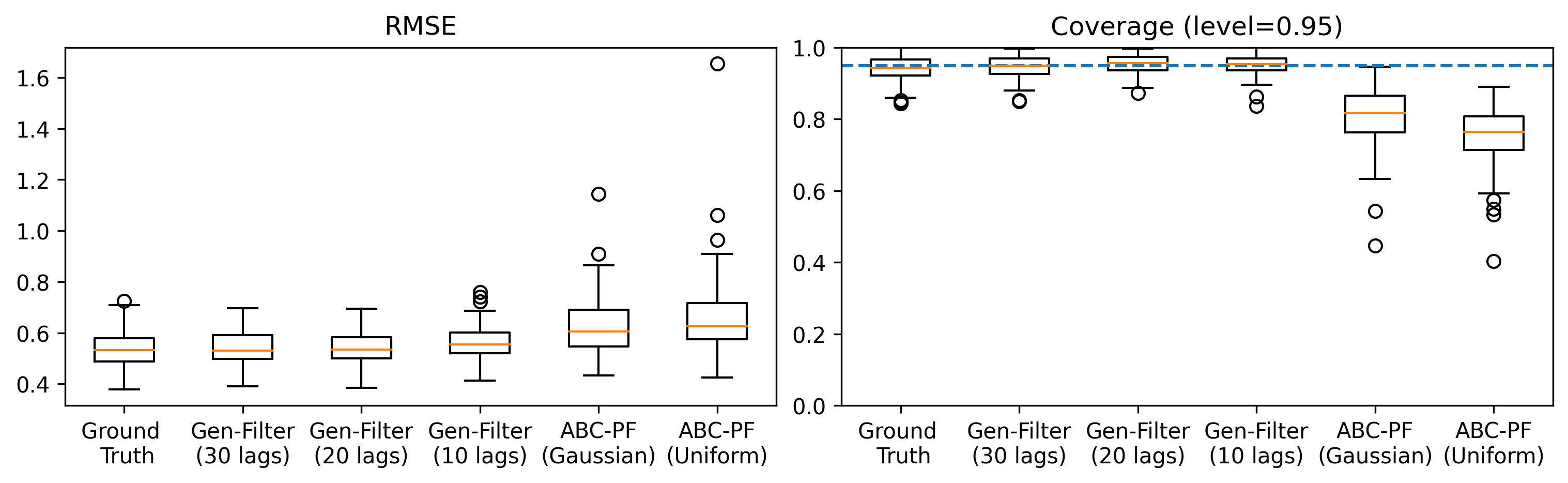}
    \caption{$\alpha$-Stable SV}
    \label{fig:}
  \end{subfigure}

  \caption{\footnotesize \textbf{Pre-Trained Gen-Filter}. RMSE and 95\% coverage over $100$ simulations. 
    Averages are reported in Table \ref{tab:RMSE_coverage_means_PT_SV}.}
  \label{fig:three-by-one}
\end{figure}

\section{Generative Gibbs Sampling}

\begin{algorithm}[H]
\caption{Generative Gibbs Sampling}
\label{alg:gibbs-training}
\begin{algorithmic}[1]
\STATE \textbf{\underline{Training Phase}}
\STATE \textbf{Input:}
Parameter blocks $\theta=(\theta_1,\ldots,\theta_B)$, prior $p(\theta)$, probabilistic model $p(y\mid \theta)$
\STATE \textbf{Output:} 
$B$ trained maps $\{H_{b\mid -b}(\cdot)\}_{b=1}^{B}$ approximating the inverse CDFs $F^{-1}_{\theta_b\mid \theta_{-b},y}(\cdot)$ for each $b=1,\dots,B$
\FOR{$i=1$ to $N$}
\STATE Sample $\theta^{(i)}\sim p(\theta)$
\STATE Sample $\tilde{y}^{(i)}\sim p(y\mid \theta)$
\STATE Sample $u^{(i)}\sim \mathcal{U}(0,1)$
\ENDFOR
\STATE Use the synthetic dataset $\{\theta_b^{(i)},\theta_{-b}^{(i)},\tilde{y}^{(i)},u^{(i)}\}_{1=}^{N}$ to learn the map
$\theta_b\overset{d}{=}H_{b\mid -b}(\theta_{-b},y, u)$ 
\STATE \textbf{\underline{Gibbs Sampling}}
\STATE \textbf{Input:} Data $y$, trained maps $\{\hat{H}_{b\mid -b}(\cdot)\}_{b=1}^{B}$ for each $b=1,\dots,B$, initial value $\theta^{(0)}$
\STATE \textbf{Output:} Chain of draws $\{\theta^{(i)}\}_{i=1}^{N}$ approximating $p(\theta\mid y)$
\STATE Initialize $\theta^{(0)}$ 
\FOR{$i = 1$ to $N$}
\STATE $\theta^{(i)}\leftarrow \theta^{(i-1)}$
  \FOR{$b = 1$ to $B$}
    \STATE Sample
    $u^{(b)}\sim \mathcal{U}(0,1)$
    \STATE Set $      \theta_b^{(i)} \leftarrow \hat{H}( \theta_{-b}^{(i)},y,u^{(b)}) $
  \ENDFOR
\ENDFOR

\STATE \textbf{return} $\{\theta^{(i)}\}_{i=1}^{N}$
\end{algorithmic}
\end{algorithm}

\section{MCMC for Linear Gaussian State-Space Models}
\begin{algorithm}[H]
\caption{FFBS for the LG-SSM with Gamma precision priors}
\label{alg:ffbs-gibbs-ar1}
{\footnotesize
\begin{algorithmic}[1]
\STATE \textbf{Input:}
Observations $\{y_t\}_{t=1}^T$, hyperparameters $(\phi,a_0,b_0,m_0, C_0)$.
\STATE \textbf{Output:}
Draws $\{x_{1:T}^{(s)}, \psi_x^{(s)}, \psi_y^{(s)}\}_{s=1}^S$ from the posterior

\STATE
\STATE \textbf{\underline{Initialization}}
\STATE Set initial precisions $\psi_x^{(0)}, \psi_y^{(0)}$ (e.g., method-of-moments or rough guesses)
\STATE For each iteration $s=1,\dots,S$:
\STATE \quad Let $\sigma_x^2 \gets 1/\psi_x^{(s-1)}$, \ $\sigma_y^2 \gets 1/\psi_y^{(s-1)}$
\STATE \quad  set $m_0 \gets 0$, $C_0 \gets \sigma_x^2/(1-\phi^2)$

\STATE
\STATE \textbf{\underline{Forward Filtering}}
\STATE \quad Set $m_0, C_0$
\FOR{$t=1$ to $T$}
  \STATE \quad Prediction: $a_t \gets \phi\, m_{t-1}$,\;\; $R_t \gets \phi^2 C_{t-1} + \sigma_x^2$
  \STATE \quad One-step forecast: $f_t \gets a_t$,\;\; $S_t \gets R_t + \sigma_y^2$
  \STATE \quad Kalman gain: $K_t \gets R_t/S_t$
  \STATE \quad Update: $m_t \gets a_t + K_t (y_t - f_t)$,\;\; $C_t \gets (1-K_t) R_t$
  \STATE \quad \textbf{Store} $a_t, R_t, m_t, C_t$
\ENDFOR

\STATE
\STATE \textbf{\underline{Backward Sampling (FFBS)}}
\STATE \quad Sample $x_T^{(s)} \sim \mathcal{N}(m_T, C_T)$
\FOR{$t=T-1$ down to $1$}
  \STATE \quad Smoother gain: $J_t \gets C_t \phi / R_{t+1}$
  \STATE \quad Conditional mean: $\tilde m_t \gets m_t + J_t \big(x_{t+1}^{(s)} - a_{t+1}\big)$
  \STATE \quad Conditional var.: $\tilde C_t \gets C_t - J_t^2 R_{t+1}$
  \STATE \quad Sample $x_t^{(s)} \sim \mathcal{N}(\tilde m_t, \tilde C_t)$
\ENDFOR

\STATE
\STATE \textbf{\underline{Gibbs updates for precisions}}
\STATE \quad Compute residual sums:
\STATE \quad \quad Observation residuals: $SS_y \gets \sum_{t=1}^T (y_t - x_t^{(s)})^2$
\STATE \quad \quad State residuals: $SS_x \gets \sum_{t=1}^T \big(x_t^{(s)} - \phi x_{t-1}^{(s)}\big)^2$
\STATE \quad Sample
\STATE \quad \quad $\tau_y^{(s)} \sim \mathrm{Gamma}\!\left(a_0 + \tfrac{T}{2},\; b_0 + \tfrac{1}{2} SS_y\right)$
\STATE \quad \quad $\tau_x^{(s)} \sim \mathrm{Gamma}\!\left(a_0 + \tfrac{T}{2},\; b_0 + \tfrac{1}{2} SS_x\right)$

\STATE
\STATE \textbf{Return} $\{x_{1:T}^{(s)}, \tau_x^{(s)}, \tau_y^{(s)}\}_{s=1}^S$
\end{algorithmic}
}
\end{algorithm}

\section{Summary Statistics for the SV Model}
\subsection*{1. Observation-based summaries}
Given observations $\{y_t\}_{t=1}^T$:

- Sample mean
\[
\bar{y} = \frac{1}{T}\sum_{t=1}^T y_t
\]

- Sample variance
\[
s_y^2 = \frac{1}{T-1}\sum_{t=1}^T (y_t - \bar{y})^2
\]

- Sample autocovariance at lag $k$ ($k=1,3,5$)
\[
\gamma_k = \frac{1}{T-k}\sum_{t=k+1}^T (y_t - \bar{y})(y_{t-k} - \bar{y})
\]

- Sample quantile at level $\tau$
\[
Q_\tau(y) = \inf\{q \in \mathbb{R} : F_T(q) \geq \tau\}, \quad 
F_T(q) = \frac{1}{T}\sum_{t=1}^T \mathbb{I}\{y_t \leq q\}
\]

\subsection*{2. States-based summaries ($\tilde{x}_t$)}
Given latent states $\{\tilde{x}_t\}_{t=1}^T$:

- Sample mean of latent states
\[
\bar{\tilde{x}} = \frac{1}{T}\sum_{t=1}^T \tilde{x}_t
\]

- Autoregressive coefficient estimate
\[
\hat{\phi} = \frac{\sum_{t=2}^T (\tilde{x}_t - \bar{\tilde{x}})(\tilde{x}_{t-1} - \bar{\tilde{x}})}{\sum_{t=2}^T (\tilde{x}_{t-1} - \bar{\tilde{x}})^2}
\]

- Innovation variance estimate
\[
\hat{\sigma}^2_\eta = \frac{1}{T-1} \sum_{t=2}^T \Big( \tilde{x}_t - \bar{\tilde{x}} - \hat{\phi}(\tilde{x}_{t-1} - \bar{\tilde{x}}) \Big)^2
\]

\section{Gen-Gibbs for $\alpha$-Stable SV Model}

In this appendix, we provide a detailed discussion of the Gen-Gibbs sampling strategy employed to generate samples from the posterior distribution of the parameters in the SV model with an $\alpha$-stable distribution. 

The parameters $\gamma=\mu/\sigma$ and $\phi$ are samples using a their full conditionals and MH step as discussed in \citet{Vankov_2019}. 
It can easily shown that
\begin{equation*}
\begin{aligned}
p(\gamma \mid x_{0:T}, \phi)
\propto \exp\Bigg\{-\frac12 \Big[\, &
\gamma^2 \Big( (1-\phi^2) + T(1-\phi)^2 + \tfrac{1}{\tau_0^2} \Big) \\
&- 2\gamma \Big( x_0(1-\phi^2)
+ (1-\phi)\!\sum_{t=1}^{T}\!(x_t - \phi x_{t-1}) \Big)
\Big] \Bigg\}.
\end{aligned}
\end{equation*}
which is the kernel of a Normal distribution. On the other hand, for $\phi$ we can assume an alternative prior $\tilde{p}(\phi)$ which is analytically convenient i.e. $\mathcal{N}(0, \sigma_\phi^2)$. Thus we have
\begin{equation*}
\begin{aligned}
p(\phi \mid x_{0:T}, \gamma)
\propto & p(x_0 \mid \gamma, \phi) 
\exp\Bigg\{
-\frac{1}{2}
\Bigg(
\phi^2
\Bigg(
\sum_{t=1}^{T} (x_{t-1} - \gamma)^2 
+ \frac{1}{\sigma_\phi^2}
\Bigg)
\\& - 2\phi
\Bigg(
\sum_{t=1}^{T} (x_t - \gamma)(x_{t-1} - \gamma)
\Bigg)
\Bigg)
\Bigg\}.
\end{aligned}
\end{equation*}
which is again a kernel of a a Normal distribution. Given that we are using a simpler, auxiliary prior  to construct the proposal distribution, a MH step is applied to ensure convergence to the correct posterior:
\begin{align*}
\alpha(\phi, \phi^{*}) 
&= \min\left\{
1,\,
\frac{p(x_{0:T} \mid \phi^{*}, \gamma)p(\phi^{*})}
     {p(x_{0:T} \mid \phi, \gamma)p(\phi)}
\times
\frac{q(\phi \mid x_{0:T}, \gamma)}
     {q(\phi^{*} \mid x_{0:T}, \gamma)}
\right\} \nonumber \\[1em]
&= \min\left\{
1,\,
\frac{p(x_{0:T} \mid \phi^{*}, \gamma)p(\phi^{*})}
     {p(x_{0:T} \mid \phi, \gamma)p(\phi)}
\times
\frac{p(x_{1:T} \mid \phi, \gamma)\tilde{p}(\phi)}
     {p(x_{1:T} \mid \phi^{*}, \gamma)\tilde{p}(\phi^{*})}
\right\} \nonumber \\[1em]
&= \min\left\{
1,\,
\frac{p(x_0 \mid \phi^{*}, \gamma)p(\phi^{*})}
     {p(x_0 \mid \phi, \gamma)p(\phi)}
\times
\frac{\tilde{p}(\phi)}{\tilde{p}(\phi^{*})}
\right\},
\end{align*}
The proposed value $\phi^{*}$ is then accepted with probability $\alpha(\phi, \phi^{*})$; otherwise, the current value $\phi$ is retained.

The $\alpha_y$ and $\beta_y$ parameters of the observation noise distribution are sampled using a Gen-Gibbs step. In particular, for each draw of the latent volatility sequence $x_{1:T}^{(i)}$, $i=1,\ldots,N$, we obtain the standardized residuals
\[
\hat{\varepsilon_t} = y_t\exp\left(-0.5x_t\right)
\]
and compute summaries statistics. Then we sample from the corresponding conditional posteriors
\[
p(\alpha_y \mid S_{\alpha_y}(\hat{\varepsilon}_{1:T}), \beta_y)
\quad \text{and} \quad
p(\beta_y \mid S_{\beta_y}(\hat{\varepsilon}_{1:T}), \alpha_y),
\]
where $S_{\alpha_y}(\hat{\varepsilon}_{1:T})$ and $S_{\beta_y}(\hat{\varepsilon}_{1:T})$ denote the summary statistics for the respective parameters. The list of summaries is given in the following Table \ref{tab:summaries_alpha_stable}.

\begin{table}[H]
\centering
\setlength{\tabcolsep}{8pt}
\renewcommand{\arraystretch}{1.15}
\begin{tabularx}{\linewidth}{l|>{\raggedright\arraybackslash}X|>{\raggedright\arraybackslash}X}
\toprule
Summary & $\alpha_y$–focused & $\beta_y$–focused \\
\midrule
S$_1$ &
\makecell[tl]{\textbf{ECF Slope}\\[2pt]
$\displaystyle \log(-\log|\phi(t)|)=a_0+\alpha_{\mathrm{ECF}}\log t$} &
\makecell[tl]{\textbf{Quantile Asymmetry}\\[2pt]
$\displaystyle \frac{q_{0.95}+q_{0.05}-2q_{0.50}}{q_{0.95}-q_{0.05}}$} \\
\midrule
S$_2$ &
\makecell[tl]{\textbf{Phase Slope}\\[2pt]
$\displaystyle \frac{\theta(t)}{t}=b_0+\beta_{\mathrm{phase}}\,t$} &
\makecell[tl]{\textbf{Sign Imbalance}\\[2pt]
$\displaystyle \frac{1}{T}\sum_{t=1}^{T}\mathbb{I}\{\hat{\varepsilon}_t>0\}$} \\
\midrule
S$_3$ &
\makecell[tl]{\textbf{Hill Tail Index}\\[2pt]
$\displaystyle 
\alpha_{\mathrm{Hill}}
=\left[\frac{1}{k}\sum_{i=1}^{k}\bigl(\ln|\hat{\varepsilon}_{(i)}|-\ln|\hat{\varepsilon}_{(k+1)}|\bigr)\right]^{-1}$} &
\makecell[tl]{\textbf{Tail Ratio}\\[2pt]
$\displaystyle \frac{|q_{0.95}-q_{0.50}|}{|q_{0.05}-q_{0.50}|}$} \\
\midrule
S$_4$ &
\makecell[tl]{\textbf{Outer/Inner Spread Ratio}\\[2pt]
$\displaystyle \frac{q_{0.975}-q_{0.025}}{q_{0.75}-q_{0.25}}$} &
\makecell[tl]{\textbf{Extreme Quantile Skew}\\[2pt]
$\displaystyle \frac{q_{0.99}+q_{0.01}-2q_{0.50}}{q_{0.99}-q_{0.01}}$} \\
\bottomrule
\end{tabularx}
\caption{Summary statistics for $\alpha_y$– and $\beta_y$–focused measures in the $\alpha$–stable SV model. 
Here, $\hat{\varepsilon}_t$ denotes the standardized residuals, and $\hat{\varepsilon}_{(i)}$ the $i$th largest in absolute value.}
\label{tab:summaries_alpha_stable}
\end{table}

These summaries were chosen to provide complementary diagnostics of the $\alpha$-stable stochastic volatility process, capturing both tail thickness and skewness in a compact form. 
The \textit{ECF Slope} and \textit{Hill Tail Index} arise from classical tail index estimation: the ECF slope exploits the power-law decay of the empirical characteristic function, while the Hill estimator uses upper-order statistics of the absolute innovations to quantify heavy-tailedness. 
The \textit{Phase Slope} extends the ECF approach by examining the phase of the characteristic function, whose approximately linear dependence on $t^{\alpha}$ reflects the skewness parameter $\beta$ in the L\'evy--Khintchine representation; estimating its slope thus provides a $\beta$-sensitive measure consistent with stable-law theory. 
The \textit{Outer/Inner Spread Ratio}, in contrast, compares dispersion across extreme and central quantiles of the noise distribution; it is motivated by the fact that heavy tails inflate outer quantile ranges faster than inner ones, making this ratio monotonic in $\alpha$ even when moments are undefined. 

Regarding the $\beta$-focused summaries, these are moment-free statistics that directly measure asymmetry in the empirical distribution. 
Together, these $\alpha$- and $\beta$-sensitive summaries balance theoretical grounding (via characteristic-function and tail-behavior properties) with practical robustness, providing an interpretable framework for inference and comparison in $\alpha$-stable stochastic volatility models.

\begin{figure}[ht]
  \centering

  \begin{subfigure}{\linewidth}
    \centering
    \includegraphics[width=0.7\linewidth]{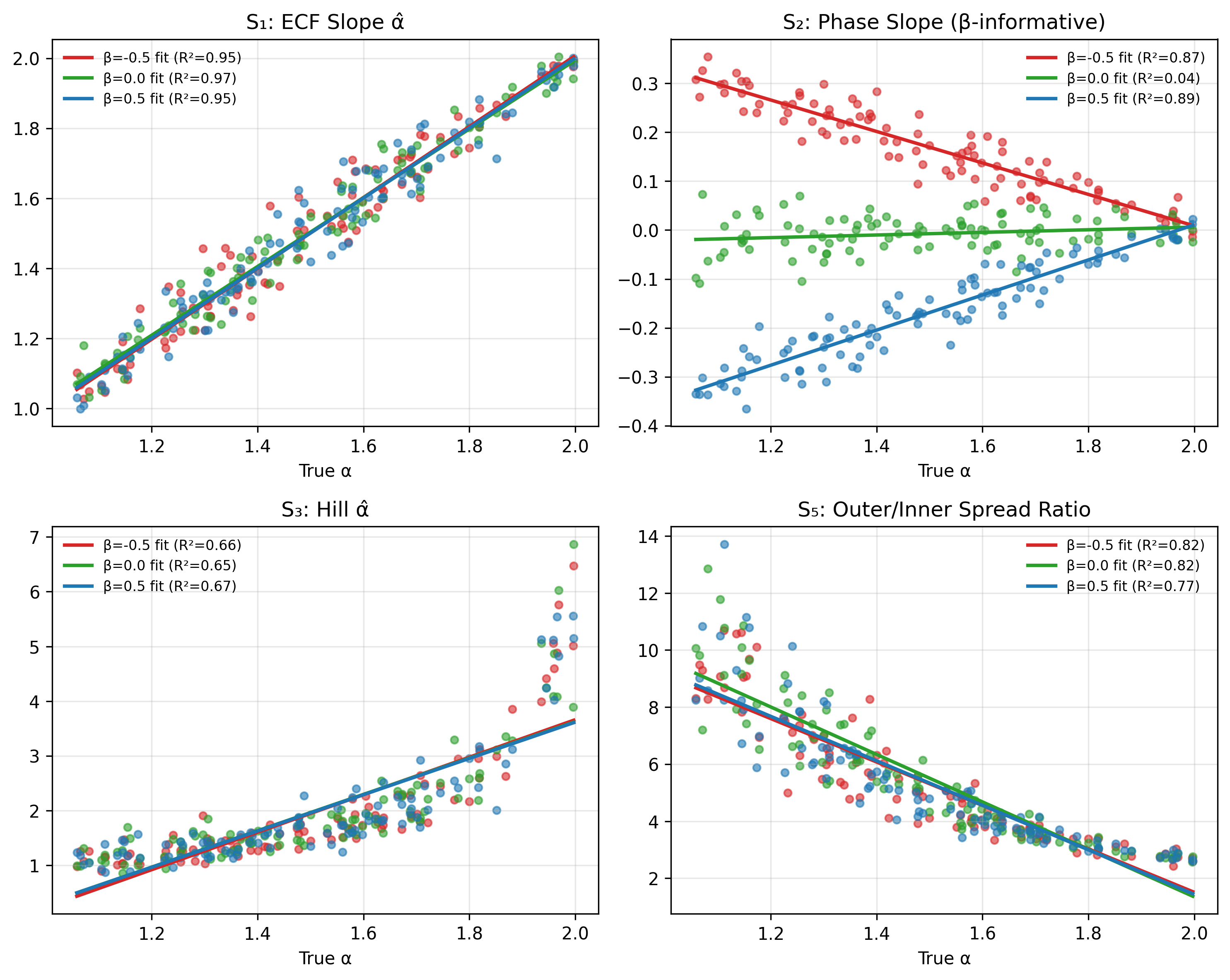} 
    \caption{\footnotesize Summary statistics for $\alpha$ parameter}
    \label{fig:alpha_stats}
  \end{subfigure}\par\medskip

  \begin{subfigure}{\linewidth}
    \centering
    \includegraphics[width=0.7\linewidth]{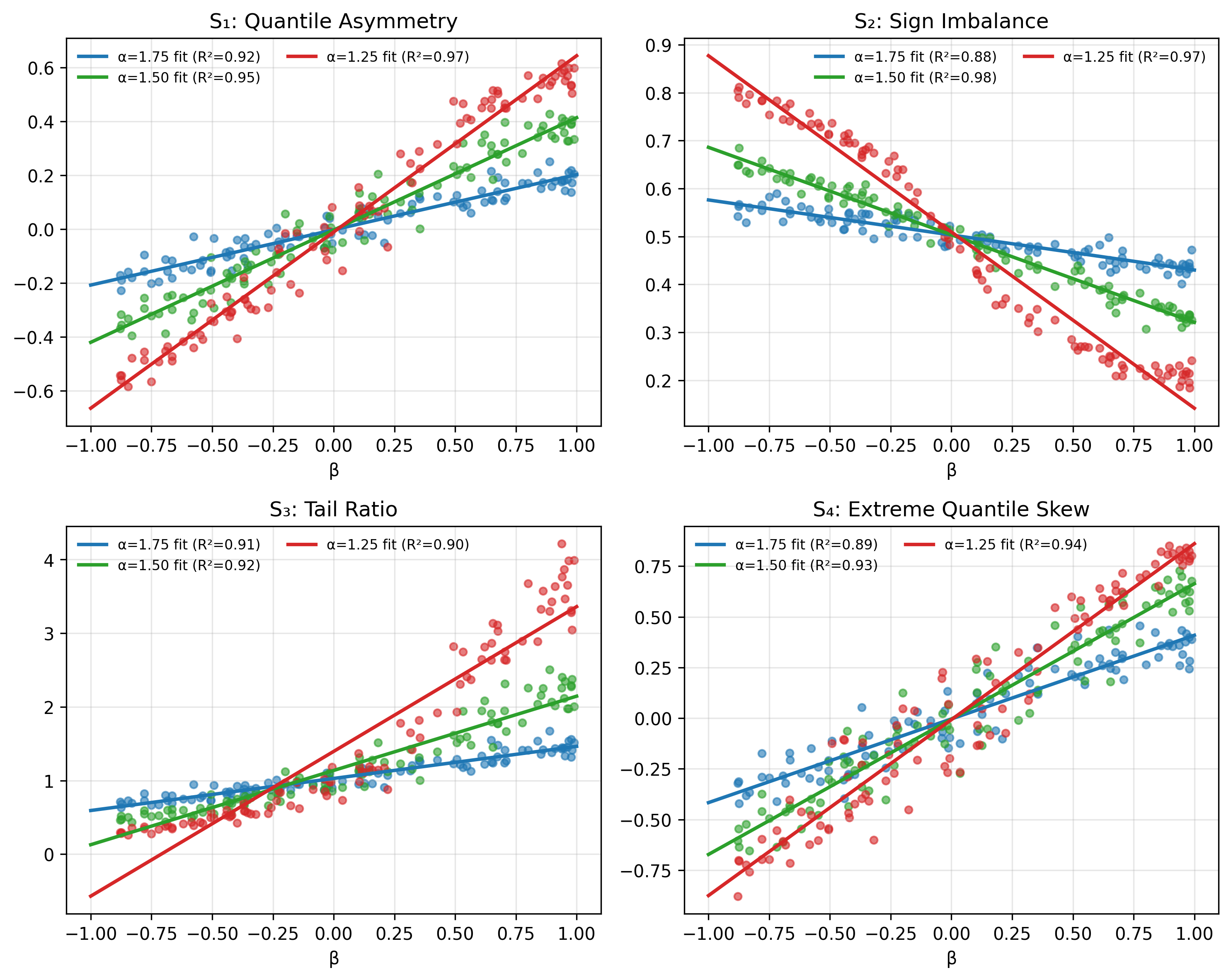}
    \caption{\footnotesize Summary statistics for $\beta$ parameter}
    \label{fig:beta_stats}
  
  \end{subfigure}

  \caption{\footnotesize \textbf{$\alpha$-Stable SV Model}. Evaluation of the informativeness of the summary statistics employed in the Gen–Gibbs sampling scheme. High correlation among summaries and parameter values suggests that the chosen summaries effectively capture information relevant for parameter inference. }
  \label{fig:alpha_beta_stats}
\end{figure}

We assess the adequacy of the selected summary statistics through visual inspection.
Figure \ref{fig:alpha_beta_stats} shows that the chosen summaries respond sensitively to changes in the parameter values, indicating that they provide informative signals for parameter inference.

\end{document}